\documentclass[namedreferences,10pt,onecolumn,fleqn]{article}
\usepackage{parskip}
\usepackage{graphicx}
\usepackage{flafter}
\def\arcsec{\hbox{$^{\prime\prime}$}}
\def\arcmin{\hbox{$^{\prime}$}}
\def\deg{\hbox{$^{\rm o}$}}
\begin{document}

{\bf \large Morphology, dynamics and plasma parameters of plumes and \\
inter-plume regions in solar coronal holes}

{\bf \small K.~Wilhelm $\cdot$ L.~Abbo $\cdot$ F.~Auch\`ere
$\cdot$ N.~Barbey $\cdot$ L.~Feng $\cdot$
A.H.~Gabriel $\cdot$ S.~Giordano~$\cdot$ S.~Imada $\cdot$
A.~Llebaria $\cdot$ W.H.~Matthaeus $\cdot$
G.~Poletto $\cdot$ N.-E.~Raouafi $\cdot$ S.T.~Suess~$\cdot$
L.~Teriaca $\cdot$ Y.-M.~Wang}

\begin{footnotesize}
Received: 18 March 20101\\
DOI 10.1007/s00159-01100035-7\\
\end{footnotesize}

{\bf Abstract}~~Coronal plumes, which extend from solar coronal holes (CH)
into the high corona and\,---\,possibly\,---\,into the solar wind (SW), can
now continuously be studied with
modern telescopes and spectrometers on spacecraft, in addition to
investigations from the ground, in particular, during total eclipses.
Despite the large amount of data available on these prominent features and
related phenomena, many questions remained unanswered as to their generation and
relative contributions to the high-speed streams emanating from CHs.
An understanding of the processes of plume formation and evolution requires
a better knowledge of the physical conditions at the base of CHs,
in plumes and in the surrounding inter-plume regions (IPR).
More specifically, information is needed on
the magnetic field configuration, the electron densities
and temperatures, effective ion temperatures, non-thermal motions,
plume cross-sections relative to the size of a CH,
the plasma bulk speeds, as well as any plume signatures in the SW.
In spring 2007, the authors proposed a study on
``Structure and dynamics of coronal plumes and
inter-plume regions in solar coronal holes''
to the International Space Science Institute (ISSI) in Bern to clarify some of
these aspects by considering relevant observations and the extensive literature.
This review summarizes the results and conclusions of the study. Stereoscopic
observations allowed us to include three-dimensional reconstructions of
plumes. Multi-instrument investigations carried out during several campaigns
led to progress in some areas, such as plasma densities, temperatures, plume
structure and the relation to other solar phenomena,
but not all questions could be answered concerning the details
of plume generation process(es) and interaction with the SW.


------------------------------------\\
\begin{small}
K. Wilhelm (corresponding author), L. Feng$^*$, L. Teriaca\\
Max-Planck-Institut f\"ur Son\-nen\-sy\-stem\-for\-schung\\
37191 Katlenburg-Lindau, Germany\\
e-mail: wilhelm@msp.mpg.de; fax: +49 5556 979 240; tel.: +49 5556 979 423 \\
$^*$ also at Purple Mountain Observatory,
Chinese Academy of Sciences\\
210008 Nanjing, China, e-mail: lfeng@pmo.ac.cn

L. Abbo, S. Giordano\\
INAF -- Osservatorio Astronomico di Torino\\
via Osservatorio 20,
10025 Pino Torinese, Italy

F. Auch\`ere, N. Barbey, A.H. Gabriel\\
Institut d'Astrophysique Spatiale\\
Universit\'e Paris XI, b\^atiment 121,
91405 Orsay, France

S. Imada\\
Institute of Space and Astronautical Science,
Japan Aerospace Exploration Agency\\
3-1-1 Yoshinodai, Sagamihara-shi, Kanagawa, 229-8510, Japan

A. Llebaria\\
Observatoire Astronomique de Marseille-Provence,
Laboratoire d'Astrophysique de Marseille\\
P$\hat{\rm o}$le de l'\'Etoile Site de Ch$\hat{\rm a}$teau-Gombert\\
38, rue Fr\'ed\'eric Joliot--Curie,
13388 Marseille Cedex 13, France

W.H. Matthaeus\\
Bartol Research Institute and Department of Physics and Astronomy \\
University of Delaware,
Newark, DE 19716, USA

G. Poletto\\
Osservatorio Astrofisico di Arcetri\\
Largo Enrico Fermi 5,
50125 Firenze, Italy

N.-E. Raouafi\\
Johns Hopkins University,
Applied Physics Laboratory\\
11100 Johns Hopkins Road,
Laurel, MD 20723-6099, USA

S.T. Suess\\
National Space Science and Technology Center\\
320 Sparkman Drive,
Huntsville, AL 35805, USA

Y.-M. Wang\\
Code 7672,
E. O. Hulburt Center for Space Research\\
Naval Research Laboratory\\
Washington, DC 20375-5352, USA

\end{small}

\begin{scriptsize}
\tableofcontents
\end{scriptsize}
\section{\small Introduction}
% 1
\label{Introduction}

\begin{figure}[!t]
% Figure 1
\centering
\includegraphics[width=\textwidth]{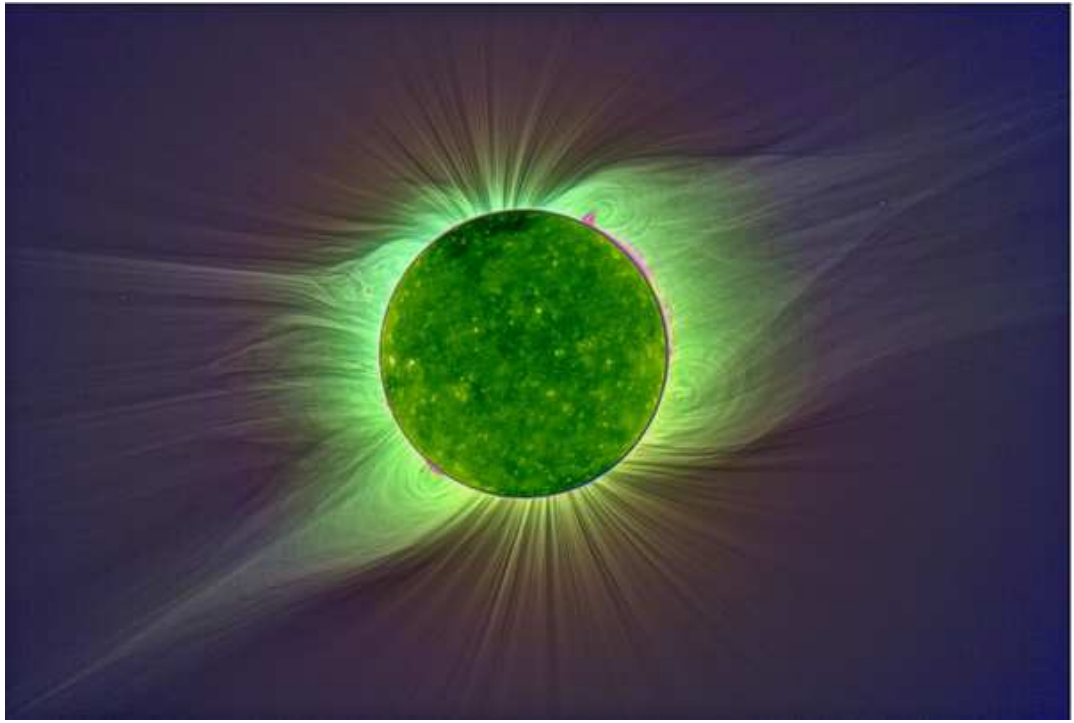} % {PASACHOFF_EIT.EPS}
\caption{\label{Figcor} \small The solar corona during the total eclipse
on 1 August 2008 observed from
Mongolia. The corona at solar minimum conditions has wide PCHs
with reduced radiation, open magnetic field lines and
many plume structures. At lower latitudes
closed field-line regions dominate the corona and extend into coronal streamers
(from Pasachoff et al. 2009; composite eclipse image by M. Druckm\"uller,
P. Aniol and V. Ru\v{s}in). An image in 19.5~nm of the solar disk taken from
EIT/SOHO at the time of the eclipse has been inserted into the shadow
of the Moon.}
\end{figure}

Coronal plumes, extending as bright, narrow structures from
the solar chromosphere into the high corona,
have long been seen as fascinating phenomenon
during total eclipses (cf., e.g., van de Hulst 1950a, b),
and can now be observed with
telescopes and spectrometers on spacecraft without interruption.
They are prominent features of the solar corona, both in visible and
ultraviolet
(UV)\footnote{A list of acronyms and abbreviations is compiled in Appendix~A.}
radiation, and are rooted in coronal holes (CH).
A spectacular image of the solar corona during an eclipse is shown in
Fig.~\ref{Figcor}.
Notice\,---\,in the context of our study\,---\,the plumes at the N
and S poles as well as the bright coronal material in the N that would
interfere with any line-of-sight (LOS) observations of the plume configuration.
In order to demonstrate the relation between coronal plumes and the northern and
southern polar coronal holes (PCH),
the occulted disk of the Sun is filled with an extreme-ultraviolet (EUV)
image taken by the EUV Imaging Telescope (EIT) (cf., Sect.~\ref{Remote}).

Van de Hulst (1950b) confirmed Alfv\'en's conclusion
that polar coronal plumes (rays in the old terminology,
cf., Sect.~\ref{Classification}) coincided with ``open''
magnetic lines of force
and thus outline the general magnetic field of the Sun.
PCHs are best developed during the minimum of the solar
activity. Consequently, many plume and PCH studies were carried out
after the launch of the Solar and Heliospheric Observatory (SOHO)
under very quiet conditions of the Sun in 1996 and 1997,
followed by additional observations during the recent minimum of the 11 year
sunspot cycle. Plumes are also observed in non-polar CHs
(Del Zanna and Bromage 1999; Del Zanna et al. 2003; Wang and Muglach 2008).
Most of the past observations have, however, been related to polar
plumes, and they will be the main topic of this study proposed to
the International Space Science Institute (ISSI), Bern, in March 2007.
It was motivated by the
fact that no undisputed theoretical concept was available for the formation of
plumes, and even many observational facts appeared to be in conflict with each
other. In particular, the three-dimensional (3D) structure of plumes and their
dynamical properties, along with those of
the inter-plume regions (IPR) had been under discussion.

We, the members of the study team on ``Structure and dynamics of
coronal plumes and inter-plume regions'' reviewed,
without any claim to be exhaustive,
the wealth of observational data available on
coronal plumes and their environment.\footnote{Taking advantage of the
prevailing solar minimum conditions at the beginning of the proposed study,
some additional plume observations had also been suggested and were carried out
in 2007 and 2008.
The analysis of these data sets is not yet completed, but some campaign
information and results are included in this report.}
This, together with the analyses
carried out in the past, allowed us to answer a number of questions formulated
in the proposal phase of the study. These questions will be repeated in the
appropriate sections, and we will restrict the discussion to these specific
topics as a general review on coronal plumes taking all aspects into
account will appear in Living Reviews (Poletto, to be submitted).
We can also refer the reader to earlier plume studies,
e.g., by Saito (1965a), Newkirk and Harvey (1968), Ahmad and Withbroe (1977),
Del Zanna et al. (1997), DeForest et al. (1997) and Koutschmy and
Bocchialini (1998).
Reference can also be made to reviews on the extended corona of the Sun
(Kohl et al. 2006) and to solar UV spectroscopy
(Wilhelm et al. 2004, 2007). The interest in coronal plumes led to two
special sessions in the past, namely, ``Solar jets and coronal plumes'',
Guadeloupe, 23 -- 26 February 1998 (ESA SP-421, 1998, T-D Guyenne, ed.) and
``Solar polar plumes'' at the 2nd Asia Oceania Geoscience (AOGS) Conference,
Singapore, 20 -- 24 June 2005.

It might be appropriate to mention from the
outset that not all questions could be answered conclusively. Future
investigations utilizing multi-instrument, high-resolution observations
will be needed to complete the task.

\section{\small Instrumentation}
% 2
\label{Instrumentation}

The study of coronal plumes and IPRs in CHs of the Sun requires many
observational facts obtained with the help of ground-based and space
instruments. It is beyond the scope of this review to provide detailed
descriptions of these devices, but short characterizations of some of
the instruments mentioned and the cooperation in observational campaigns
might be useful for appreciating the corresponding investigations.

% 2.1
\subsection{\small Ground-based systems}
\label{Ground}

The main advantage of ground-based observations is related to the large
telescope apertures available permitting high spatial and temporal resolutions,
including polarization measurements. However, detailed plume investigations
can only be conducted in the optical window of the
terrestrial atmosphere and, in general, during total solar eclipse periods
with temporary campaign installations.\footnote{The chances that
the totality path of an eclipse crosses a site of a very large permanent
telescope are low, however
on 11 July 1991 an eclipse could be observed with the 3.6~m
Canada-France-Hawai Telescope on Mauna Kea
(see, e.g., November and Koutchmy 1996).}
Exceptions are, for instance, the white-light (WL) plume observations
with the Mk~III K-coronameter of the
Mauna Loa Solar Observatory (MLSO)
(DeForest et al. 2001a), and the determination of plume lifetimes
between 10~h and 20~h with the help of the Fe\,{\sc x} 637.4~nm
($T_{\rm F} = 0.98$~MK)
line\footnote{The formation temperatures, $T_{\rm F}$, of ionic emission lines
(cf., Sect.~\ref{Electron}) are given in parentheses.}
during the solar minimum 1954 (Waldmeier 1955).

The Synoptic Optical Long-term Investigations of the Sun (SOLIS) project
(Keller et al. 2003) employs a 50~cm aperture Ritchey--Chr\'etien
telescope and a
vector spectro-magnetograph (VSM)
for investigations of solar magnetic fields. It is designed to help
understand the origin of the solar cycle (complementing helioseismic studies)
through the study of different aspects of the Sun's magnetic activity
related to the cycle at different scales (dynamo, turbulent magnetic fields,
irradiance changes, differential rotation). One of the main goals is to
develop methods and techniques for solar activity forecast (e.g., flares,
coronal mass ejections). VSM provides vector magnetic fields and
the LOS field using spectral lines
characterized by their Zeeman-induced polarization,
in addition to a chromospheric line that serves as a proxy for
coronal structures ensuring observational continuity at different heights in the
solar atmosphere.
The LOS magnetograms obtained from chromospheric lines
benefit from the canopy structure of the field yielding strong
signals everywhere on the solar disk, in particular close to the limb.

%2.2
\subsection{\small Space systems}
\label{systems}

\begin{table}[!t]
% Table~1
\noindent
\caption{Some characteristics of space-borne remote sensing instrumentation}
\vspace{0.3cm}
\begin{small}
\begin{tabular*}{14.2cm}{lcccc}
\hline
Mission (Time) & Wavelength &
\multicolumn{2}{c}{Resolutions:} & FOV \\
~~~Instrument & ranges$^{\rm a}$, $\lambda$/nm &
spectral, $\lambda/\Delta \lambda$ & angular & \\
\hline
\hline
\multicolumn{2}{l}{Spartan 201 (1993 to 1998)} & & &  \\
~~~UCL & 103.2, 103.8, & $\approx 5000$ & $\geq 30\arcsec \times 150\arcsec$
& $4\arcmin \times 5\arcmin$ \\
 & 121.6, 124.2 & & & corona\\
\cline{2-5}
~~~WLC &  480, $pB$ & $\approx 10$  & 22.5\arcsec &
1.25~$R_\odot$ to 6~$R_\odot$ \\
\hline
\multicolumn{2}{l}{SOHO (1995 to ---)} & & &  \\
~~~CDS & $15 \ldots 80$ & 1000 to  & $\approx 3\arcsec$ & Slit, disk, \\
~~~(NI, GI) &             & 10\,000  & (2\arcsec~slit width)   & low corona \\
\cline{2-5}
~~~EIT & 17.1, 19.5, & Filter bands: &$2.6\arcsec \times 2.6\arcsec$ &
$45\arcmin \times 45\arcmin$ \\
&  28.4, 30.4 & $12 \ldots 20$ & pixel & \\
\cline{2-5}
~~~LASCO & WL, 530.3 & C1: $\approx 8000$ &
5.6\arcsec, 11.4\arcsec, & $1.1~R_\odot$ to \\
~~~(C1, C2, C3) & & Filter bands & 56\arcsec~pixels & $30~R_\odot$ \\
\cline{2-5}
~~~MDI & WL, 676.78 & 70\,000 & 2\arcsec~pixel &
$34\arcmin \times 34\arcmin$ \\
& & & 0.6\arcsec~pixel & $11\arcmin \times 11\arcmin$ \\
\cline{2-5}
~~~SUMER & $78 \ldots 161$, WL  &  20\,000  &
0.3\arcsec, 1\arcsec, 4\arcsec & Slits, disk,  \\
& ($46.5 \ldots 80.5$) & (40\,000) & slit widths & low corona \\
\cline{2-5}
~~~UVCS & $103.2 \ldots 124.2$, WL  & 5000
& 7\arcsec~pixel & $40\arcmin \times 141\arcmin$ \\
& ($49.9 \ldots 62.5$) & (7000)  & & corona \\
\hline
TRACE & WL, 17.1, 19.5, 28.4,  & Filter bands & 0.5\arcsec~pixel
& $8.5\arcmin \times 8.5\arcmin$ \\
(1998 to 2010) &121.6, 155, 160, 170 &&& disk, low corona\\
\hline
\multicolumn{2}{l}{Hinode (2006 to ---)} & & &  \\
~~~SOT & $380 \ldots 657$ & Narrow bands & 0.08\arcsec~pixel &
$328\arcsec \times 164\arcsec$\\
       & $380 \ldots 657$ & Broad bands & 0.053\arcsec~pixel &
$218\arcsec \times 109\arcsec$\\
~~~~~SP & 630.08 \ldots 630.32 & 30\,000 & 0.16\arcsec~pixel
& $320\arcsec \times 151\arcsec$\\
\cline{2-5}
~~~EIS & $17 \ldots 21$  & $\approx 10\,000$  &
1\arcsec~pixel & $590\arcsec \times 512\arcsec$ \\
& $25 \ldots 29$ & $\approx 14\,000$ & & disk, low corona \\
\cline{2-5}
~~~XRT & $0.2 \ldots 20, 430.5 $ &Filter bands & 1\arcsec~pixel &
$2048\arcsec \times 2048\arcsec$\\
\hline
\multicolumn{2}{l}{STEREO (2006 to ---)} & & &  \\
~~~EUVI & 17.1, 19.5, & Filter bands: & 1.59\arcsec~pixel &
$0 \ldots 1.7~R_\odot$ \\
& 28.4, 30.4 & 12, 12, 14, 10 & & \\
\hline
\end{tabular*}

\vspace{0.3cm}
$^{\rm a}$ ranges in the second-order of diffraction in parentheses\\
\label{Tabinst}
\end{small}
\end{table}
The restrictions on mass and size in the case of space instrumentation
are to a large extent compensated by the wide energy ranges accessible both
for photons and charged particles combined with the flexibility in selecting the
spacecraft position during the observations.

% 2.2.1
\subsubsection{\small Ulysses}
\label{Ulysses}
Launched in October 1990, Ulysses travelled outwards to Jupiter,
where it used
a gravitational assist to turn the orbit away from the ecliptic plane.
The final orbit has an 80.4\deg~inclination, 1.34~ua perihelion, 5.4~ua
aphelion, and a period of 6.2~years. Three polar passes were completed, two
of them near
sunspot minimum, before the mission was terminated in June 2009. The
spacecraft carried a complete compliment of fields and particles instruments.
For plume studies, the relevant instruments are
the Solar Wind Observations Over the Poles of the Sun (SWOOPS)
(Bame et al. 1992), a thermal ion and
electron spectrometer with energy ranges for electrons from 0.81~eV to
862~keV and ions from 255~eV/$Z$ to 34.4~keV/$Z$,
the Solar Wind Ionization state and Composition Spectrometer (SWICS)
(Gloeckler et al. 1992), and
the Vector Helium and Fluxgate Magnetometers (VHM/FGM)
(Balogh et al. 1992).

SWOOPS returned the temperatures, densities and vector speeds of protons
(H$^{+}$, p), $\alpha$ particles (He$^{2+}$) and electrons.
Speed changes of a few kilometres per second over a few seconds could be
resolved. SWICS recorded the speed and density of He$^{2+}$
and the densities, ionization states  and
speeds of several minor ions of the elements C, O, Ne, Mg, Si and Fe.
The data rate of Ulysses
was limited by the transmitter power available and the distance from the Earth
so that expected SWICS plume signatures are at the level of
detectability. VH-FGM measured the vector magnetic field at a far higher
cadence than SWOOPS.

\begin{figure}[!t]
% Figure 2
\includegraphics[width=\textwidth]{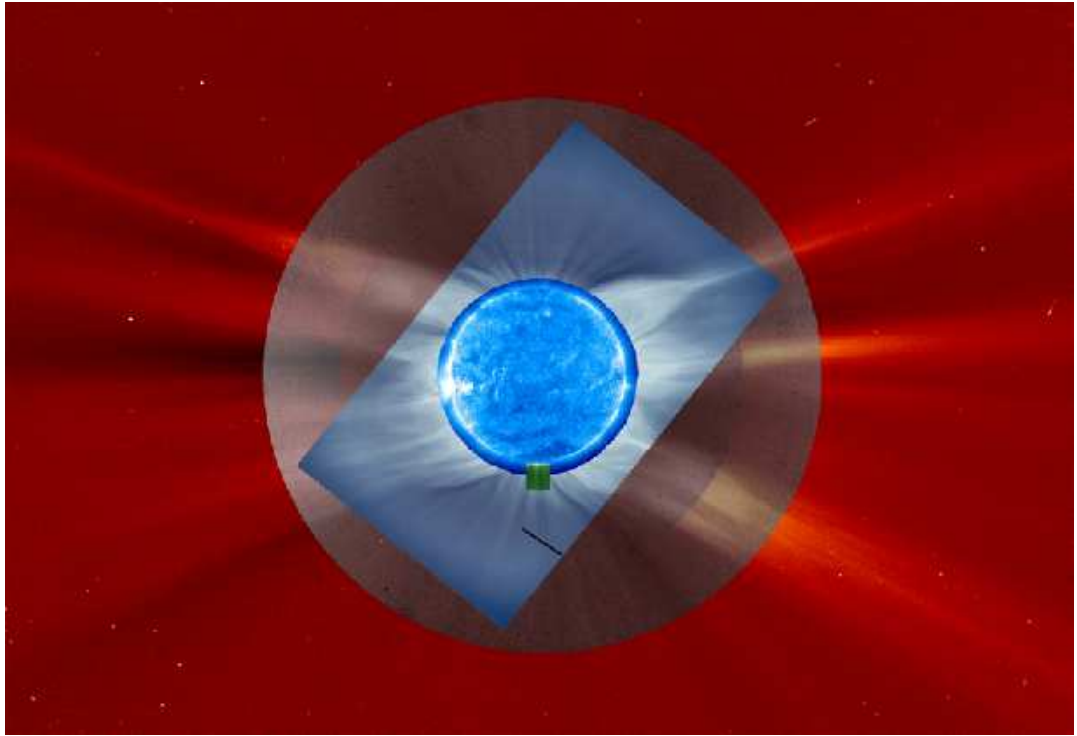} %{ECLIPSE06_2.EPS}
\caption{\label{Lybia} \small
Composite image of the corona during the eclipse of 29 March
2006 built up from SOHO data (outer frame from LASCO, green rectangle
near south pole from SUMER), polarized radiation from the EKPol experiment
(circular portion from Abbo et al. 2008) and WL ground-based
observations (rectangular portion from Koutchmy et al. 2006; obs.: J.
Mouette). The solar disk in the 17.1 nm band of EIT has been inserted
into the shadow of the Moon. The black solid line represents the
operational UVCS slit during the eclipse.}
\end{figure}
%

% 2.2.2
\subsubsection{\small Remote-sensing instrumentation}
\label{Remote}

Some of the relevant instrument characteristics are
listed in Table~\ref{Tabinst} in order to present a compact and coherent
overview of the following missions and their operational periods: \\
-- Spartan was a satellite system launched and retrieved by the Space Shuttle
on four occasions.
It carried the Ultraviolet Coronal Spectrometer (UCS),
an externally and internally occulted coronagraph with a dual spectrograph
as well as the White Light Coronagraph (WLC)
(Kohl et al. 1995; Guhathakurta and Fisher 1995), a coronagraph with
polarimeter for polarized brightness, $pB$, measurements
(cf., Sect.~\ref{Densities}). \\
-- SOHO was launched on 2 December 1995 and injected into a halo orbit
around the Sun-Earth Lagrange point L1 ($\approx 0.01$~ua sunward of the Earth)
on 14 February 1996 (see Domingo et al. 1995 and references therein).
The following instruments are of importance in our context: (1)~The
Coronal Diagnostic Spectrometer (CDS) with normal-
and grazing-incidence (NI/GI) spectrometers in the EUV wavelength range.
(2)~EIT,
a full-disk solar imager for
observations in the emission lines
Fe\,{\sc ix} 17.1~nm (0.71~MK),
(Fe\,{\sc x} 17.5~nm);
Fe\,{\sc xii} 19.5~nm (1.38~MK);
Fe\,{\sc xv} 28.4~nm (2.08~MK);
He\,{\sc ii} 30.4~nm (81\,000~K).
(3)~The Large Angle and Spectrometer Coronagraph (LASCO),
a triple coronagraph (C1, C2, C3) in the visible wavelength regime
with nested FOVs out to heliocentric distances $R = 30~R_\odot$
(1~$R_\odot$ = 696~Mm, the radius of the Sun\footnote{$1\,R_\odot$ is
seen from Earth under an angle of $961\arcsec \pm 15\arcsec$, depending on the
season. The angle is $\approx$\,10\arcsec~larger from SOHO.
A distance of $\approx$\,720~km in the solar photosphere corresponds to
1\arcsec.}).
C1 observed with a Fabry--Perot interferometer, among other lines,
Fe\,{\sc xiv} 530.3~nm (1.82~MK). C2 and C3 had a set of wideband filters.
Most of images were obtained with the orange filter of C2: (540 to 640)~nm,
and the clear filter of C3: (400 to 850)~nm;
both used a set of three polarizers (--\,60\deg, 0\deg, 60\deg)
on specific sequences.
(4)~The Michelson Doppler Imager (MDI)
observes solar oscillations and LOS magnetic fields in the
Ni\,{\sc i} 678.8 line with a tunable interferometer.
(5)~The Solar Ultraviolet Measurements of Emitted Radiation spectrometer (SUMER)
measures radiation in the
vacuum-ultraviolet (VUV) wavelength range with slit lengths
of 120\arcsec~and 300\arcsec~ and spatial rasters.
(6)~ The Ultraviolet Coronagraph Spectrometer (UVCS)
spectrometers are fed by three occulted telescopes for
observations of the extended solar corona.\\
-- The Transition Region and Coronal Explorer (TRACE),
was launched by a Pegasus
vehicle into a Sun-synchronous Earth orbit on 2 April 1998. The spacecraft
carried  a 30~cm Cassegrain telescope. A typical temporal resolution was 5~s
(Handy et al. 1999). \\
-- Hinode was launched on 22 September 2006 (Kosugi et al. 2007).
The three instruments on board are:
(1)~The Solar Optical Telescope (SOT),
a 50~cm diffraction-limited Gregorian
telescope (Suematsu et al. 2008) feeding narrow-band and broad-band
filter imagers and a spectro-polarimeter (SP).
Polarization
spectra of the Fe\,{\sc i} 630.15~nm and 630.20~nm lines are obtained
for high-precision Stokes polarimetry and measurements of the
three components of the magnetic field in the photosphere
(Tsuneta et al. 2008a).
(2)~The EUV Imaging Spectrometer (EIS),
a NI stigmatic spectrometer fed by a multi-layer telescope
(Culhane et al. 2007), observes
prominent emission lines, e.g., Fe\,{\sc viii} 18.52~nm (0.44~MK),
Fe\,{\sc xii}, Fe\,{\sc xiii} 20.2~nm (1.58~MK) and
Fe\,{\sc xxiv} 19.20~nm (17.0~MK), with
four slit or slot widths from 1\arcsec~to 266\arcsec.
(3)~The X-Ray Telescope (XRT), a 30 cm aperture GI
telescope with analysis filters, provides nine X-ray
wavelength bands with different lower cut-off energies (Golup et al. 2007).\\
-- The Solar Terrestrial Relations Observatory (STEREO),
was launched in
October 2006 near a minimum of solar activity.
Two identical
spacecraft (STEREO~A and B) are drifting apart along the Earth's orbit and
observe the Sun almost simultaneously with
the Extreme Ultraviolet Imager (EUVI) telescopes (Wuelser et al. 2004)
of the instrument package Sun Earth Connection Coronal and
Heliospheric Investigation (Secchi) (Howard et al. 2008).
For plume observations, long exposure times and low compression rates are
applied to increase the signal-to-noise ratio.
%

% 2.3
\subsection{\small Eclipse and other campaigns}
\label{campaigns}

In particular during total eclipse periods, but also at other times, special
plume observing
campaigns have been organized involving many instruments on the ground and
in space. Examples are:

\subsubsection{\small Eclipse campaign 2006}
\label{Eclipse}

Total solar eclipses offer great opportunities to observe the faint solar
corona, especially in its inner portions, which are not easily accessible
by coronagraphic telescopes, owing to the instrumentally scattered light. This
background is significantly reduced during an eclipse.

During the total eclipse of 29 March 2006, the greatest eclipse point
was at Waw an Namous, Lybia, in the Sahara Desert
at 10:11:18~UTC with a duration of 4~min 7~s.
An Italian scientific expedition was organized to reach this site
and measure the linearly polarized radiation of the corona, with the help of
the EKPol experiment, a liquid crystal K-corona imaging polarimeter
(Zangrilli et al. 2006).
These and other ground-based observations during the eclipse
were coordinated with those of the space instruments EIT, CDS, LASCO, SUMER
and UVCS. The composite image in Fig.~\ref{Lybia} shows some of the data
obtained and the position of the operational UVCS slit during
the eclipse.
The polar angle of the slit centre was 147\deg~at
a heliocentric distance of 1.63~$R_\odot$.
UVCS observed plume and IPR structures during the time interval from 06:16
to 18:40~UTC in the O\,{\sc vi} 103.2~nm, 103.8~nm doublet (0.3~MK),
and\,---\,immediately before the eclipse\,---\,in the
H\,{\sc i} 121.6~nm Ly\,$\alpha$ line.

\subsubsection{\small Multi-instrument campaign 2008}
\label{Multi}

Based on the experience gained during further cooperative efforts in 2007,
the southern PCH was the target of a multi-instrument campaign
from 22 June to 3 July 2008.
The purpose of the observations was to get morphological information on plumes
as well as on the physical parameters of the plume and IPR plasmas. We
summarize the campaign characteristics here and present preliminary
results from XRT in Sects.~\ref{Electron} and \ref{Jets}.
SUMER recorded spectra over 8~h time intervals covering the TRACE and Hinode
observing times on a daily basis (with a gap on 26 June).
Included in the spectra are the H\,{\sc i} Ly\,$\alpha$ and 102.6~nm
Ly\,$\beta$ lines,
as well as lines from O\,{\sc vi} 103.2~nm, 103.8~nm,
N\,{\sc v} 123.9~nm (0.18~MK),
Si\,{\sc viii} (144.0, 144.6)\,nm (0.81~MK), Ne\,{\sc viii} 77.0~nm
(0.62~MK), Mg\,{\sc ix} (70.6, 75.0)\,nm (0.95~MK),
Mg\,{\sc x} 62.5~nm (1.12~MK) and Fe\,{\sc xii}.
UVCS took data from $R = 1.8~R_\odot$ to $3~R_\odot$.
Ly\,$\alpha$ radiances and spectral profiles
have been obtained.
TRACE images in the 17.1~nm and 160~nm channels
have been acquired daily, at
times when Hinode was taking data and also at extra times, including a 20~h
continuous observation run on 1 July.
Hinode data are available with gaps on 23, 27, 30~June and 3~July. A typical
observing sequence lasted 2~h to 4~h. XRT images have been acquired with
three filters (namely, Al\_poly, Al\_mesh and C\_poly)
in a FOV of $526.6\arcsec \times 526.6\arcsec$ and a cadence of 35~s.
EIS performed calibration-related studies
over most of this campaign, but, in addition, observed the
He\,{\sc ii} 25.6~nm (87\,000~K), Mg\,{\sc vi} 27.0~nm (0.44~MK),
Mg\,{\sc vii} 27.7~nm (0.63~MK),
Si\,{\sc vii} 27.5~nm (0.59~MK) lines,
and emissions from a number of Fe ions with ionization stages from 10+ to 16+.

\begin{figure}[!t]
% Figure 3
\begin{minipage}[b]{14.2cm}
\begin{minipage}[b]{10.2cm}
\includegraphics[width=10.2cm]{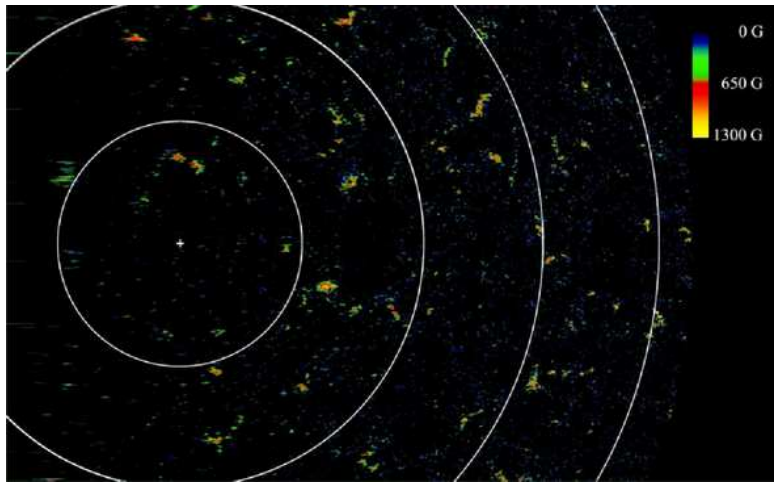} %{Tsuneta-fg3.eps}
\end{minipage}\hfill
\parbox[b]{3.5cm}{\caption{\label{landscape} \small
S polar view of the magnetic field observed between 12:02:19 and
14:55:48~UTC on 16~March 2007. Heliospheric latitudes are shown for --\,85\deg,
--\,80\deg, --\,75\deg~and --\,70\deg. The FOV is about 327\arcsec~(E--W; shown
in the vertical direction) by 473\arcsec~(N--S along the LOS).
(from Tsuneta et al. 2008b, reproduced by
permission of the American Astronomical Society, AAS).}}
\end{minipage}
\end{figure}

% 3
\section{\small Morphology of plumes in coronal holes}
\label{Morphology}

Coronal plumes are obviously 3D structures in solar CHs,
and the following questions have to be answered:
-- Can the fore- and background problems caused by closed
magnetic field regions around CHs be solved?
-- Can \emph{standard} magnetic field configurations for CHs,
plumes and IPRs be determined?
-- Does the plume assembly expand super-radially?
-- What is the cross-section of plumes, and, in general, their
3D structure?
-- Do SOHO and STEREO instruments
 observe single plumes or bright features
as combinations of structures along the LOS ?
-- What are the prospects of tomographic methods, and how
could they evolve in the STEREO era?
-- What is the ratio of the total plume area to the IPR in CHs at the base
of the corona?
-- Is the apparent plume width of $\approx 30$~Mm
in the low corona an agreed value?
-- Can plumes be describe as fractal structures?
-- How can the rotation speed of coronal plumes be evaluated?

% 3.1
\subsection{\small Magnetic field configuration}
\label{Magnetic}

\begin{figure}[!t]
% Figure 4
\centering
\includegraphics[width=0.352\textwidth]{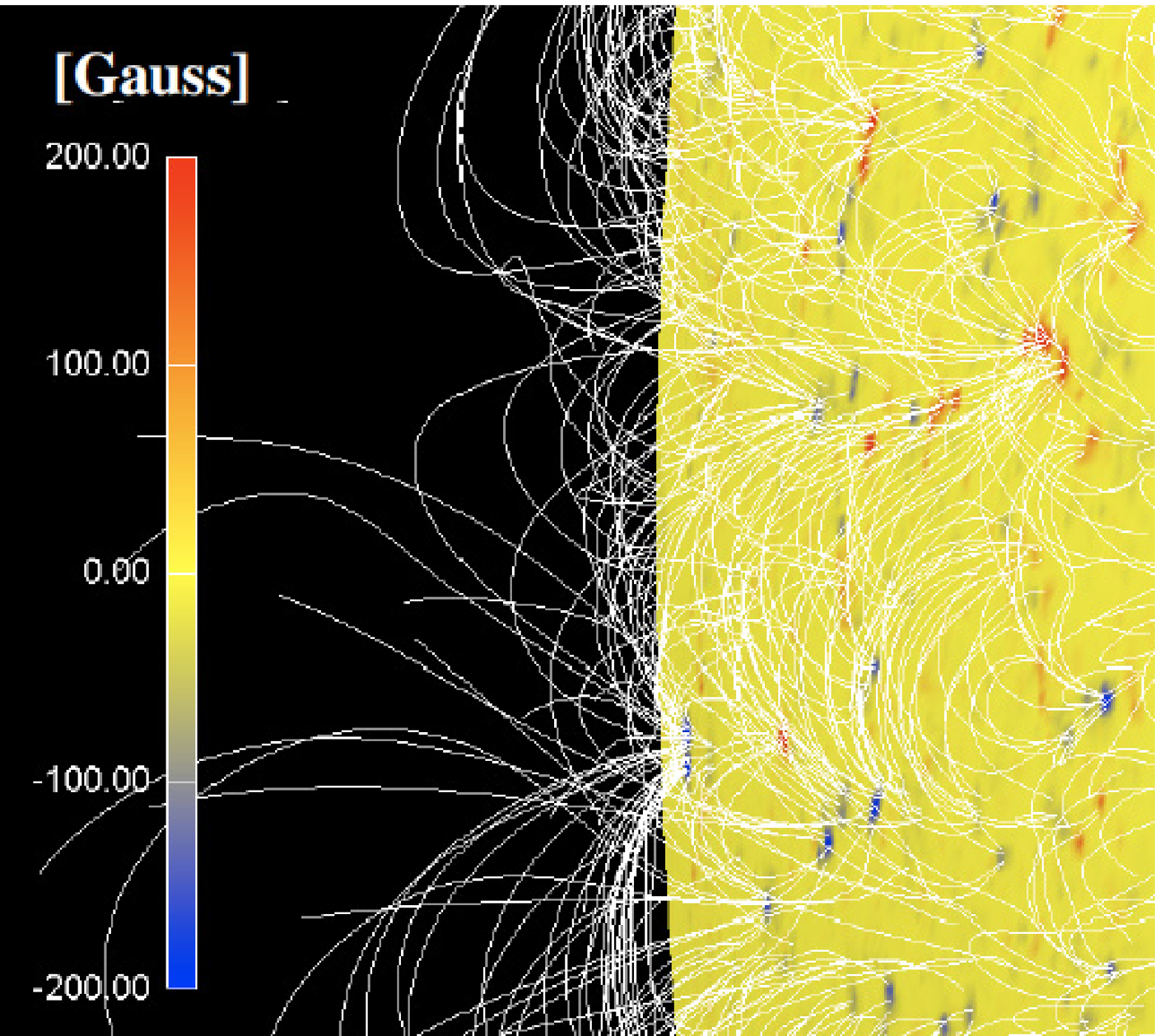} %{Ito_B.eps}
\includegraphics[width=0.63\textwidth]{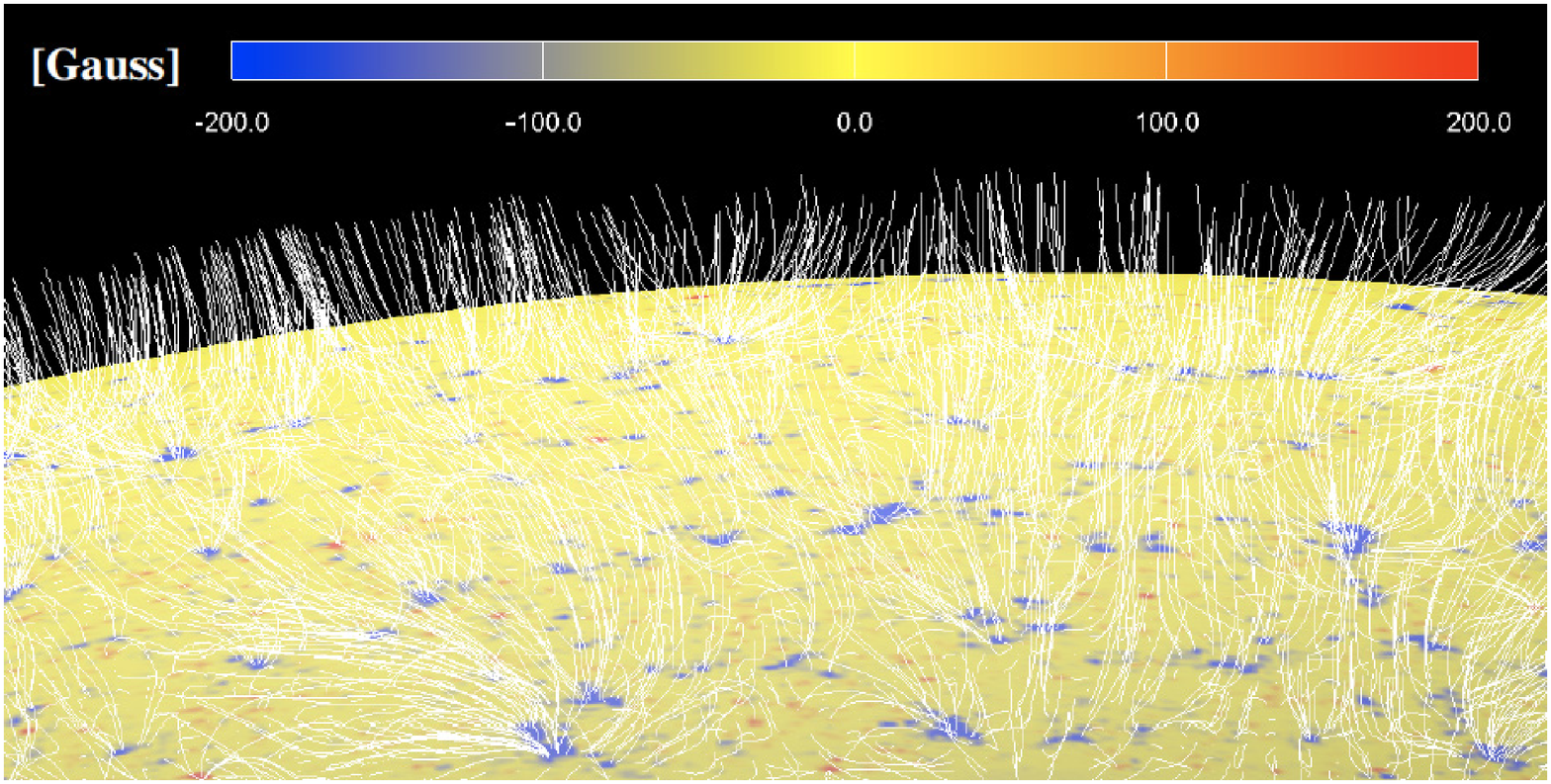} %{Ito_A.eps}
\caption{\label{CH_QS_mag} \small
Magnetic field structures extrapolated from SOT/SP observations in 2007.
A fraction of a PCH (280~Mm $\times$ 140~Mm) on 16~March is shown in the right
panel with many open field lines extrapolated to a height of $\approx 20$~Mm.
The left side
depicts an equatorial region (160~Mm $\times$ 140~Mm)
with many closed field lines
extrapolated to a height of $\approx 70$~Mm
on 28~November (from Ito et al. 2010, reproduced by permission of the AAS).}
\end{figure}

Coronal plumes are rooted in solar CHs, and consequently the magnetic
field configuration of CHs is of major importance. CHs have first been
identified as dark regions of the corona by Waldmeier (1951, 1957) and later as
sources of the fast solar wind (SW) streams by Krieger et al. (1973). CHs are
characterized by open magnetic field lines of the majority magnetic flux
(cf., Bohlin 1977; Bohlin and Sheeley 1978)
interacting with chromospheric network loops (cf., Wang and Sheeley 1995;
Wang et al. 1997).
The loop heights in CHs are, in general, smaller than in quiet-Sun (QS) regions
(Wilhelm 2000; Tian et al. 2008)\,---\,about half as high as in
QS regions (Wiegelmann and Solanki 2004). Despite this different
magnetic loop configuration, CHs and QS regions are difficult to
distinguish in most of the chromospheric and
transition-region (TR) lines
(Huber et al. 1974).

The difficulty in measuring magnetic fields in polar regions
with ground-based instruments stems from
the noise in the magnetograms caused by a strong radiance
gradient and the foreshortening effect near the solar limb. Moreover,
most of the observations in polar regions yield
the LOS component of the magnetic field. Full Stokes polarimetry
without seeing concerns can now be carried out with SOT.
Tsuneta et al. (2008b) studied the magnetic landscape of the S polar region
of the Sun shown in Fig.~\ref{landscape}.
Many patchy magnetic flux concentrations
with field strengths of more than 0.1~T (1~kG) were found
at heliographic latitudes between --\,70\deg~and --\,90\deg.
The correction of the foreshortening effect can be
noticed on the left-hand side of the figure. In general,
the strong vertical magnetic fields have the same polarity,
consistent with the global polarity of the polar region.
A comparative study of the magnetic-field structures in CHs and equatorial
QS regions by Ito et al. (2010) confirmed that the positive and negative
vertical magnetic fluxes in equatorial regions are balanced,
whereas the field is dominated by a single polarity in CHs.
Potential field extrapolations in Fig.~\ref{CH_QS_mag} show in an impressive way
that most of the field lines in QS regions are closed and
the majority of the magnetic field lines from the flux concentrations
in CHs are open fanning out with height.
Such funnel-type geometries extending from strong unipolar flux concentrations
have been derived in many studies using magnetic-field extrapolations from
photospheric observations into the corona (cf., Gabriel 1976;
Dowdy et al. 1986; Suess 1998; Tu et al. 2005; Wiegelmann et al. 2005).
Although these funnels
are mainly considered as source regions of the fast SW, they can also
describe coronal plumes. In fact, one funnel analysed by Tu et al. at
$x = 50\arcsec$~and $y = 175\arcsec$~does not show any outflow speed
(cf., Sect.~\ref{Outflow}).
This feature had earlier been identified as a plume (Wilhelm et al. 2000).
We will expand the discussion on the presence of outflows
in plumes and IPRs in Sect.~\ref{Outflow}.

``Rosettes'' in the magnetic field structure related to flux
concentrations have been described by Beckers (1963).
A connection between plumes and rosettes was suggested
by Newkirk and Harvey (1968), who estimated that $\approx 10$ plumes
would be present at each polar cap at a time\,---\,only about 1/30 of the number
of rosettes.
The filling factor of plumes in CHs is $\approx 0.1$ according to
Ahmad and Withbroe (1977),
however, even smaller factors must be expected from results
discussed in the next section.
Axford and McKenzie (1992, 1996) argued that small-scale reconnection
events\,---\,so-called microflares; in analogy to the nanoflare concept
(Parker 1988)\,---\,in the chromospheric network
(with field strengths of 20~mT and more)
would produce waves, shocks, energetic
particles and hot plasma jets that should suffice to create the fast SW
on open field-line structures (McKenzie et al. 1995).
\begin{figure}[!t]
% Figure 5
\begin{minipage}[b]{14.2cm}
\begin{minipage}[b]{9cm}
\includegraphics[width=9cm]{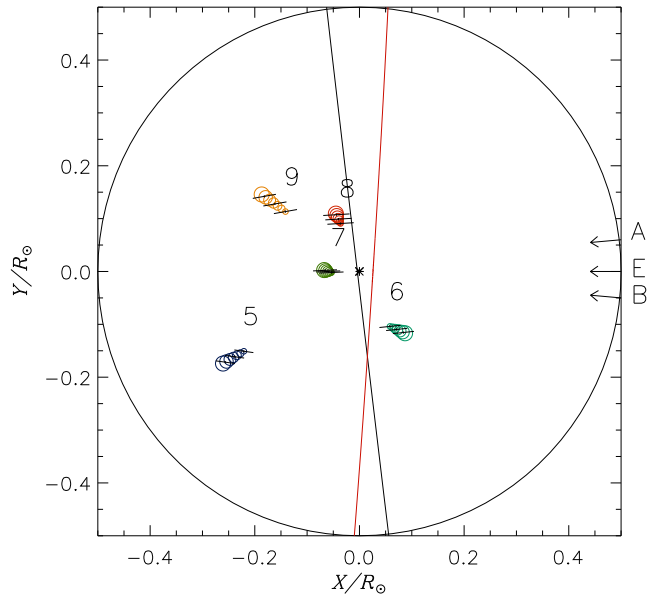} %{AA_f2_new.eps}
\end{minipage}\hfill
\parbox[b]{4.2cm}{\caption{\label{Figrec} \small
Projections onto the solar equatorial plane of reconstructed
plumes observed in the northern CH on 1 June 2007 by EUVI.
The LOS directions from the Earth (E) and STEREO~A and B are marked
together with the corresponding limb projections
from A (black) and B (red) (after Feng et al. 2009).
The numbers~5 to 9 refer to the plumes identified in
Fig.~\ref{FigEUVI}.}}
\end{minipage}
\end{figure}

From WL observations, plumes appear to expand super-radially together with the
CHs with altitude (Saito 1965a; Ahmad and Withbroe 1977;
Munro and Jackson 1977; Fisher and Guhathakurta 1995; Koutchmy and
Bocchialini 1998).
DeForest et al. (1997) found from SOHO observations that plumes rapidly expand
(super-radially with a half-cone angle of 45\deg)
in their lowest height range $h = R - 1~R_\odot < 30$~Mm to diameters of
20~Mm to 30~Mm, and more slowly above;
the linear expansion ratios of plumes
seen in the plane of the sky
were 1, 3 and 6 at heights of
$h$ = (0.05, 4, 14)~$R_\odot$, respectively, and
1, 3 and 3 for the background CH.
The expansion factor of a plume is also a weak function of the
footpoint location in the CH (cf., Goldstein et al. 1996;
DeForest et al. 2001b).
The CH expansion factor, $f(R)$, for a flux tube with cross-section $A$
is often defined (cf., e.g.,
Kopp and Holzer 1976) as
$A(R)/A(R_\odot) = (R/R_{\odot})^2\,f(R)$, where $f(R_{\odot}) = 1$ and
$f(R)$ depends upon the parameters $f_{\rm max}$,
$R_1$ as well as $\sigma$ ($f_{\rm max}$ is the net non-radial divergence,
$R_1$ the distance of the most rapid expansion and $\sigma$ the range
over which it occurs).
In this framework, the CH expansion published by Cranmer et
al. (1999) can be parametrized by the values 6.5, $1.5~R_\odot$ and
$0.6~R_\odot$ of the above parameters,
whereas the expansion of DeForest et al. (2001b) can be fitted reasonably
well by the values 5.65, $1.53~R_\odot$ and $0.65~R_\odot$\,---\,after
noticing that the quantity $\sqrt{f(R)}$ is shown in their Fig.~8.
The conclusion was reached that a radial
expansion claimed by some authors (cf., e.g., Woo and Habbal 2000)
is inconsistent with observations in PCHs.
The plumes subtend a solar latitude angle
of 2\deg~to 2.5\deg~near the limb, which is in
agreement with eclipse observations
(Newkirk and Harvey 1968).
Observations during four eclipse periods showed an average width of 31~Mm
(corresponding to 2.6\deg) at $h = 0.05~R_\odot$ (Hiei et al. 2000).
The fraction of plumes wider
than 4\deg~or narrower than 2\deg~was $\approx 20~\%$, each.

As mentioned in Sect.~\ref{Introduction}, the best observing conditions of
coronal plumes exist near the minimum of the activity
cycle when the PCHs have their maximum extension from the poles
to $\pm\,60^\circ$ solar latitudes (Wang and Sheeley 1990; Banaszkiewicz et al.
1998). In this phase, it is unlikely that
neighbouring streamers significantly contaminate the
PCH observations.\footnote{The LOS geometry from the Earth
will also be influenced by the
tilt angle of the solar rotation axes of 7.25\deg~with respect to the
ecliptic plane.}
The southern CH in Fig.~\ref{Figcor}
shows an example of such an optimal
condition. It is obvious from this figure that the plumes diverge
super-radially with altitude in the PCHs. Plumes near the limb
appear to converge to
points on the solar rotation axis more than half the way between the centre of
the Sun and the poles
(Saito 1965a; Marsch et al. 1997; Boursier and Llebaria 2008). The
super-radial expansion could also be confirmed in the 3D
reconstructions using EUVI data in 2007 (Feng et al. 2009).
An example of such a reconstruction is shown in Fig.~\ref{Figrec}.
In addition to the divergence of the plume assembly, the cross-sectional area of
the plumes expands as well (cf., Casalbuoni et al. 1999;
and the discussion in the previous paragraph). According to model
calculations, most of this expansion
occurs at the base below $\approx 35$~Mm, where the plumes grow
in diameter from $\approx 3$~Mm by nearly a factor of ten. In the regime of low
$\beta$ (the ratio of the plasma pressure to the magnetic pressure) up to at
least 5~$R_\odot$, the geometric spreading factors in plumes and
IPRs vary together (Suess et al. 1998; Suess 2000).

In the photosphere, the footpoints of plumes\,---\,more specific
\emph{beam plumes} as defined in the next section\,---\,lie near
unipolar flux concentrations, and in the corona the plumes
follow, as mentioned before, magnetic field lines, although Lamy et
al. (1997) and Zhukov et al. (2001) detected twisted and
helicoidal shapes in LASCO-C2 images,
and a ``doublet'' fine structure was seen in eclipse images (cf.,
Koutchmy and Bocchialini 1998). Observations with many SOHO
instruments and MLSO Mk~III allowed DeForest et
al. (1997, 2001a) to trace plumes to $R = 15~R_\odot$,
where the brightness observed by LASCO-C3 decreased significantly,
although the brightest structures could be identified out to
30~$R_\odot$.

The boundaries of CHs rotate more rigidly
than the underlying photospheric plasma (Timothy et al. 1975; Wang et al. 1996),
and hence apparently do the ensemble of plumes within the CH boundaries
(Llebaria et al. 1998).
This does not necessarily imply that individual plumes
rotate rigidly, only that closed loops must be continually converted into open
field lines and vice versa at the CH boundaries, presumably via interchange
reconnection (Crooker et al. 2002; Wang and Sheeley 2004).
For the
boot-shaped CH in August 1996, Zhao et al. (1999) give a nearly rigid rotation
rate of 13.25\deg/d. Smaller CHs during more active periods of the Sun
do, however, show differential rotation (cf., Wang et al. 1996; Wang 2009). In
these CHs at lower solar latitudes, coronal plumes have also been observed with
properties similar to those in polar regions (Wang and Muglach 2008).

The distribution
of magnetic flux concentrations in CH has been studied with
chromospheric magnetograms from SOLIS/VSM (Raouafi et al. 2007a).
The monthly averaged distribution of polar flux elements as a function of
latitude (normalized to the surface area)
The distribution is relatively flat up
to $\approx 75\deg$~and drops to higher latitudes. The CH
boundary varied between $\approx$\,60\deg~and 70\deg~with time and longitude.
Larger flux elements fall off more sharply than smaller ones. If they would be
required for the generation of prominent plumes, this result is
consistent with the forward
modeling by Raouafi et al. (2007b) demonstrating that plumes near the
poles do not yield the O\,{\sc vi} profiles observed by UVCS
(cf., Sect.~\ref{Outflow}), and also with Saito's (1965a)
observations that plumes
are rooted preferably in a ring at latitudes between 70\deg~and 80\deg.

% 3.2
\subsection{\small Plume geometry and dimensions}
\label{Dimension}
\begin{figure}[!t]
% Figure 6
\begin{minipage}[b]{14.2cm}
\begin{minipage}[b]{10cm}
\includegraphics[width=10cm]{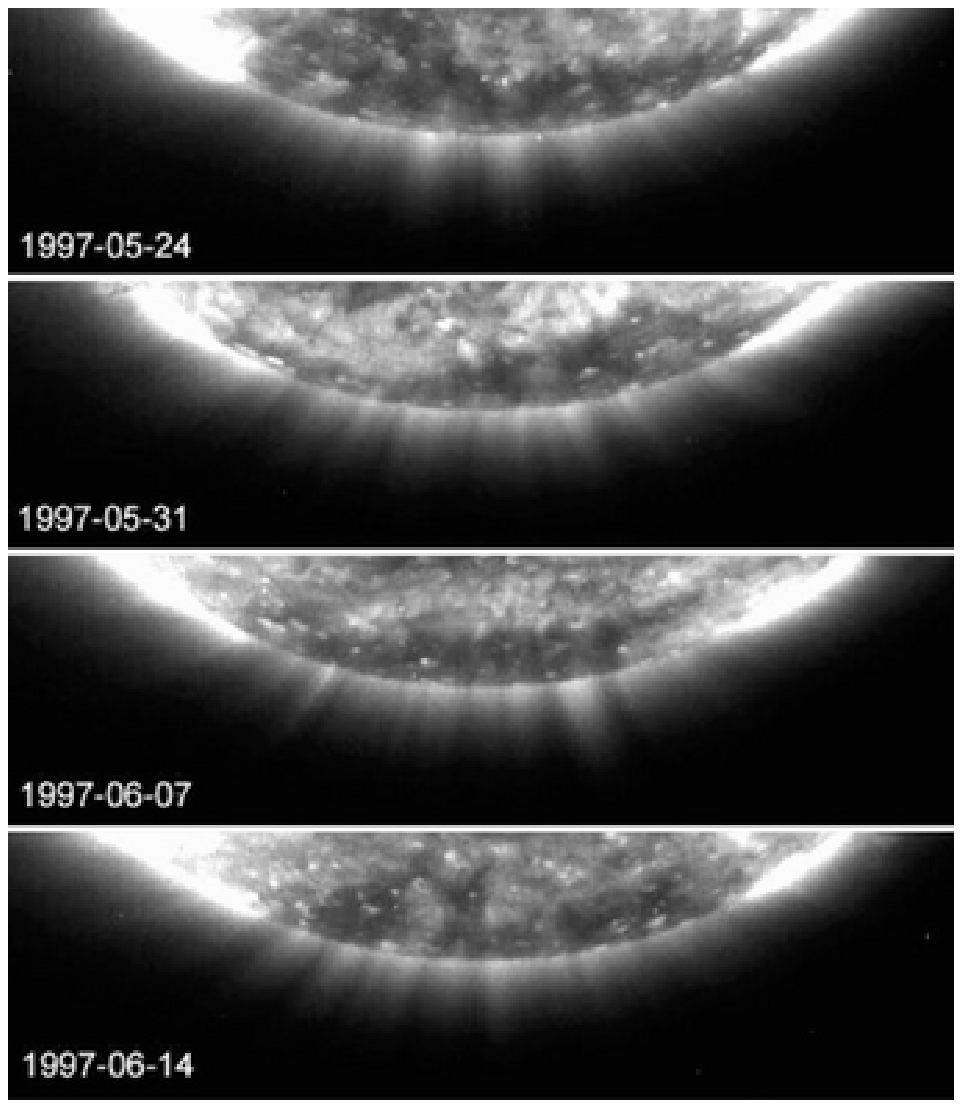} %{Gabriel_2009_fig2.eps}
\end{minipage}\hfill
\parbox[b]{3.7cm}{\caption{\label{FigEIT} \small
Four EIT images of the southern PCH in the 17.1~nm
wavelength band, recorded at 7~d intervals corresponding to about a
quarter solar rotation (from Gabriel et al. 2009).}}
\end{minipage}
\end{figure}
\begin{figure}[!t]
% Figure 7
\centering
\includegraphics[width=\textwidth]{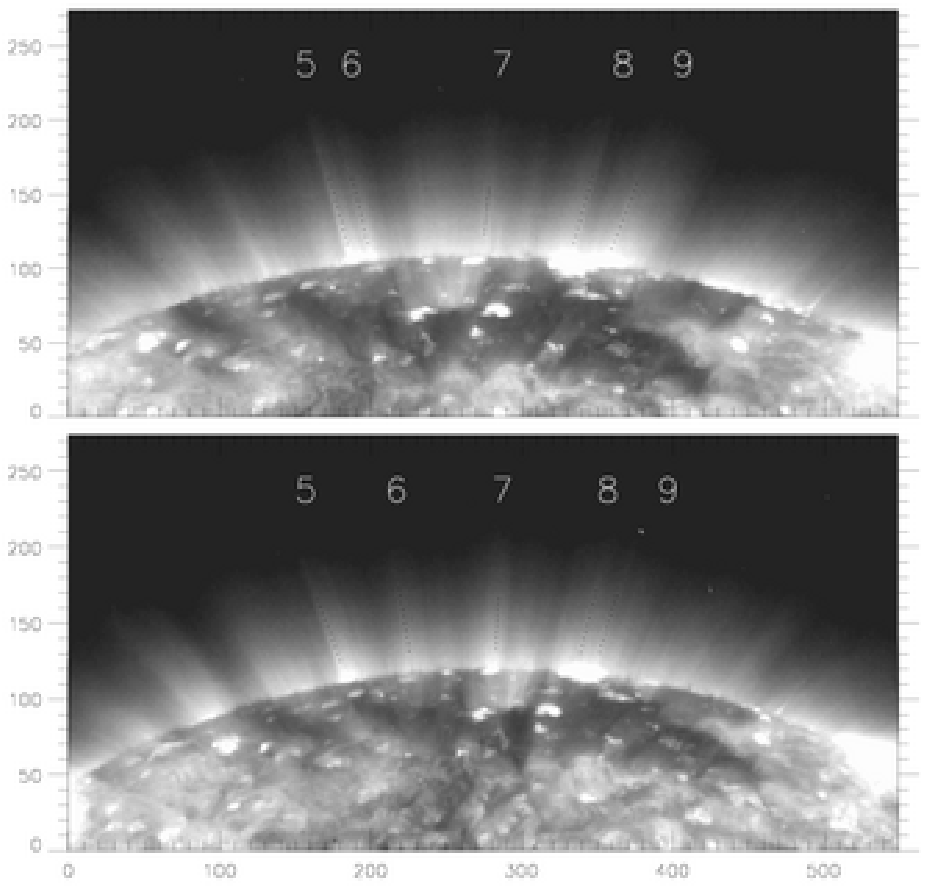} %{AA_F8_NEW3.eps}
\caption{\small \label{FigEUVI} EUVI images in the 17.1~nm
band of the northern CH on 1 June 2007 seen
from STEREO~A (top panel) and B (bottom panel).
Strong (beam) plumes are marked by dark dotted lines (after Feng et al. 2009).
The LOS geometries and the reconstructions of the plumes labelled 5 to 9
are shown in Fig.~2. The scales of the axes are in pixels with a size
of 1.59\arcsec.}
\end{figure}

Some aspects of the plume dimensions have already been mentioned in relation to
the magnetic field configuration in the previous section. It is not
directly possible to determine the 3D geometry of the plume
structures from 2D observations integrated along the LOS, and
thus not all plumes might be of cylindrical shape with diameters
of $\approx 30$~Mm in the low corona.
It has been suggested by Gabriel et al. (2003) that
``curtain'' plumes would only become visible when the curtain will be aligned
with the LOS. These features would be a second type of plumes as there can be no
doubt that near-cylindrical plumes do exist, for instance, those
related to bright points (BPs) (cf. Sect.~\ref{BP}).
The main argument for non-cylindrical plume
structures stems from EIT images in the 17.1~nm band, such as those in
Fig.~\ref{FigEIT}, giving the impression that most of the
plumes in the CH are located
on the far side of the Sun. The same effect is evident in EUVI images as can be
seen from
Fig.~\ref{FigEUVI}.
Since there is no reason for such an asymmetry, the
increase of the LOS length through a curtain plume above the limb was thought to
create the effect.
A separation into two types of plumes could also be made by considering
the relative brightness profiles of individual plumes versus heliocentric
distance. Similar curtain- or sheet-like plumes had been identified
by Wang and Sheeley (1995) extending over several network cells.

Electron density measurements, discussed in detail in Sect.~\ref{Densities},
provide strong evidence that the CH plasma consists of two distinct density
regimes. If identified with plume and IPR plasmas, the plumes
occupy a maximum of $\approx 10$~\% of the length of the LOS through the CH
(Wilhelm 2006). These findings are inconsistent with a
conterminous curtain plume, but might be compatible with microplumes
aligned in a certain fashion. Such a scenario has been
proposed by Gabriel et al. (2009). Simulations produce realistic
images of plume assemblies under the assumption that the footpoints of
microplumes are
aligned along lanes of the chromospheric network. This population of plumes is
therefore called ``network plumes'' in contrast to ``beam plumes'' that are,
in general, related to BPs at some stage of their life (cf., Sect.~\ref{Life}).
However, it is quite possible that beam plumes are
also composed of microplumes in a more compact arrangement.
Such an option might shed some light on the findings of Newkirk and Harvey
(1968) that a \emph{typical} plume with cylindrical symmetry has a
core (electron) density of $\approx 10^8\,{\rm cm}^{-3}$ and a radial density
profile dependent on the distance from the plume axis.
The apparent density profile could, however, also be attributed to a varying
LOS length through the plume cross-section with more or less constant density.
Loops and plumes composed of multiple strands below the resolution of
present-day coronal imagers are considered by DeForest (2007).

One method of disentangling the LOS integration in the optically thin coronal
plasma is rotational tomography (Frazin 2000) using image
sequences taken by LASCO and EIT.
In traditional
rotational tomography, two major assumptions are made. First it is
assumed that the solar rotation rate is the same at all latitudes and
altitudes. Second, it is assumed that the coronal structures are stable
over the acquisition time, i.e. about two weeks. Both assumptions induce
artifacts in the reconstructed emissions, and modern tomographic
inversion codes aim at avoiding these problems. Barbey et
al. (2008) have addressed the effect of temporal variation in the case
of polar plumes. They have proposed a reformulation of the inversion
problem in order to obtain both the 3D structure of
plumes and their temporal evolution. Results by DeForest et al. (2001b)
suggest that even though plumes seem to be short-lived, their magnetic
skeleton may be stable, at least in some cases, for several days. Based
on this observation, the code developed by Barbey et al. (2008) assumes
that temporal evolution occurs only in a limited number of volume
regions. In the reconstructed volumes, one can identify both tube-like
structures similar to the intuitive idea of plumes, but also more
elongated features and a structuration in cells.

\begin{figure}[!b]
% Figure 8
\includegraphics[width=0.98\textwidth]{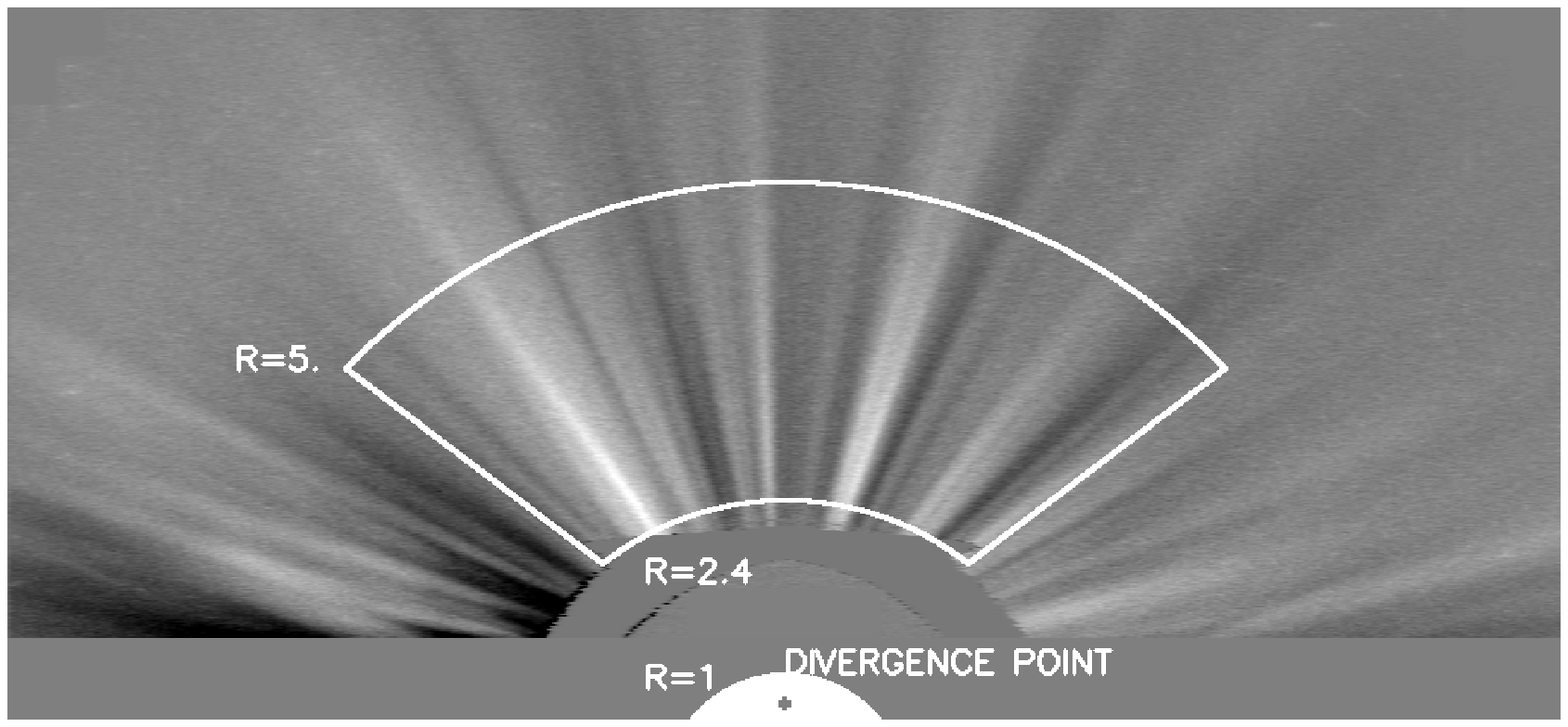} %{antoine_1.ps}
\caption{\label{fig:plumes} \small
A processed image of plumes and IPRs at time, $t_0$, 21:31~UTC
on 23 March 1997, i.e. 47~h into
the 66~h sequence taken with LASCO-C2 from 21 to 24 March 1997.
The F-corona, the stray light and the K-continuum are removed from this image.
It is not corrected for vignetting in order to keep a uniform visual contrast.
Notice the thin IPRs relative to the broad plumes. The intense ray
at $\approx -\,50\deg$ is a high-latitude streamer crossing the FOV, in which
the TID integration is performed
along radial directions centred on the divergence point
($R_{\rm div} = 0.843~R_\odot$) from 2.4~$R_\odot$ to 5~$R_\odot$
in the angular range between
$\delta = -\,60\deg$ and +\,60\deg. This yields the integrated
radiance, $L_{\rm I}(\delta,t_0)$, as function of angle at time $t_0$.
}
\end{figure}

WL plumes could continuously be observed with LASCO-C2 in the
polar areas during the periods of low solar activity
(for instance in the years 1996 to 1998
and 2007 to 2010).
Examples are shown in Fig.~\ref{fig:plumes}. Note that in the FOV of
LASCO-C2 the plumes diverge more or less radially from a point on the polar axis
not very far from the pole (Llebaria et al. 2001), whereas closer to the Sun
the plume geometry looks totally different (cf., Fig.~\ref{Figcor}).
Due to the intrinsic differences between VUV and WL photo-emission
mechanisms (cf., Sect.~\ref{Electron})
as well as temporal variations, it is difficult to demonstrate the continuity
of plumes in the different wavelength regimes. The correlation between both
types of plumes was studied by comparing EUV and WL sinograms
derived from a large set of EIT, LASCO-C1 and C2 images
(DeForest et al. 2001b). Sinograms are 2D images which display
the evolution with time of an evolving profile (frequently resulting
from a projection).
The correlation between sinograms obtained in different wavelengths and
radial distances required an angular correction to compensate for the
super-radial expansion. The resulting correlation coefficient was significant
and established a direct link between EUV and WL plumes.
This demonstration based on the co-evolution of both features
is more robust than image-to-image comparisons of local details.
The CH expansion factor deduced is 2.25 at $R = 3~R_\odot$ in good agreement
with other determinations (DeForest et al. 1997; Cranmer et al. 1999).

\begin{figure}[!t]
% Figure 9
\begin{minipage}[b]{14.2cm}
\begin{minipage}[b]{8.5cm}
\includegraphics[width=8.5cm]{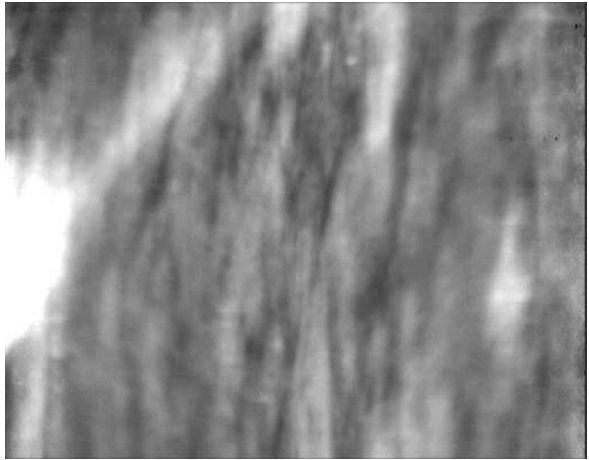} %{tid5.eps}
\end{minipage}\hfill
\parbox[b]{5.2cm}{\caption{\label{tidfive} \small
A TID, $L_{\rm I}(\delta,t)$, of 21 to 24 March 1997.
Horizontal axis:
$\delta = -\,60\deg$ to 60\deg. Vertical axis: time, $t$, with 66~h full scale
(cf., Fig.~\ref{dtid} for labelled coordinate axes).
Each horizontal line shows, at a specific time, the radially integrated radiance
of the FOV indicated in Fig.~\ref{fig:plumes}.
The global fluffy aspect is characteristic of fractal images.
IPRs (in dark) are sharper than bright plumes.
On the left side, a streamer crosses the FOV.
The overall texture shows an inclination of 7\deg~caused by the polar tilt
angle relative to the plane of the sky at that time
(from Llebaria et al. 2002a).}}
\end{minipage}
\end{figure}

With LASCO-C2 it was possible to restitute simultaneously images of the
polarized K-corona and the unpolarized F-corona from polarization
measurements (Llebaria et al. 2010). In circular profiles of the
K-corona centred on the divergence point, plumes and IPRs appear
as random oscillations with a relative amplitude of $\approx 1\,\%$.
From a spectral analysis of such oscillations,
Llebaria et al. (2002a) concluded that the angular size distribution
is fractal, and thus the concept of a characteristic angular size is
inapplicable to images of WL plumes. A superposition of multiple strands
of different sizes leads to a smooth variation undulating the background.
The low level of high-frequency oscillations relative to the background
is indicative of
a large number of tiny plumes along the LOS.
Support for this concept has been obtained
by a forward modeling approach (Boursier and Llebaria 2005;
cf., Sect.~\ref{Forward}).
Since the observed plumes are projections
of 3D structures, the fractal 2D structure provides a strong clue for
assuming a fractal structure also for the physical plumes,
i.e., the 3D electron density distribution over the CH domain
is probably fractal. Its dimension
must be  $ D = 2.9$ over the CH, in order to obtain a fractal
dimension of $ D = 1.5$ in the transverse profile of WL plumes.
The high fractal dimension required by the density distribution of plumes
looks surprising, but is understandable because of the strong integration effect
along the LOS. The implication that CHs have a fibrous structure might
indicate that the beam plumes mentioned earlier should indeed
not be thought of as
compact entities.

\begin{figure}[!t]
% Figure 10
\centering
\includegraphics[width=10cm]{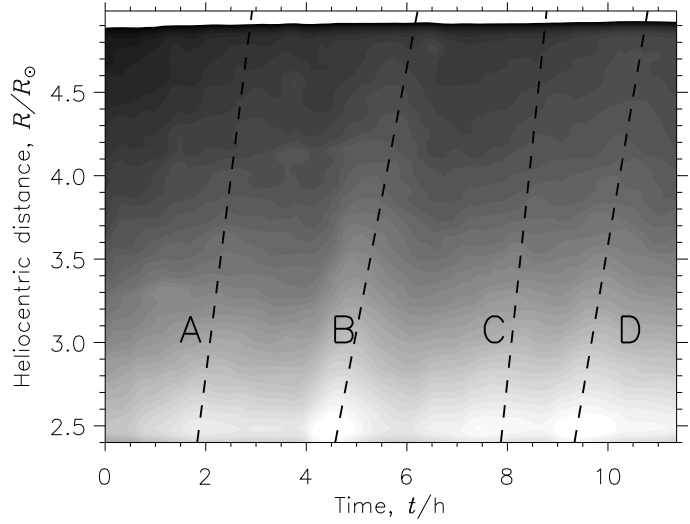} %{antoine_3a2.ps}
\caption{\label{traces} \small
Brightness profiles along the trajectory of the plume which starts at a
colatitude of 11.7\deg~at 52.7~h in time, dimming at 11.5\deg, 64.2~h.
The graph displays the logarithmic brightness in heliocentric
distance from 2.4~$R_\odot$ to 5~$R_\odot$ versus time.
In this example, the persistence was 11.5~h.
The ejection speeds of the events~A and C were $\approx 450$~km\,s$^{-1}$,
higher than the mean speeds of $\approx 300$~km\,s$^{-1}$ for B and D.}
\end{figure}

\begin{figure}[!t]
% Figure 11
\includegraphics[width=\textwidth]{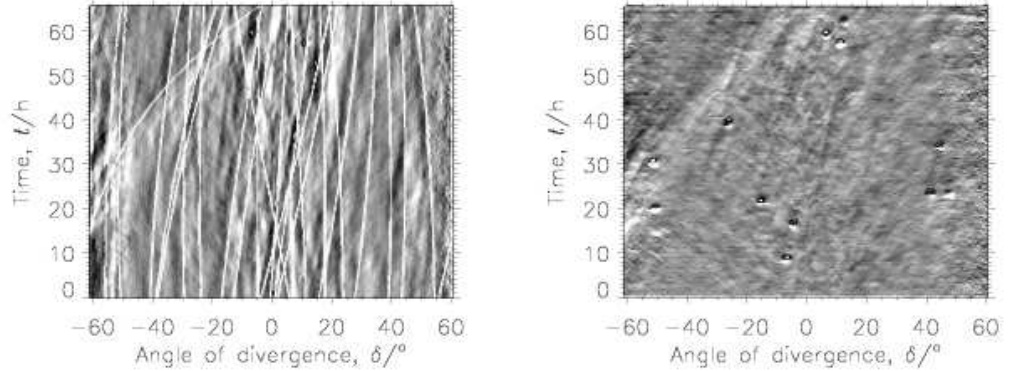} %{antoine_4a.ps}
\caption{\label{dtid} \small
(Left panel) The TID derivative in the angular direction,
$\partial L_{\rm I}(\delta,t)/\partial \delta$,
with plume structures superimposed as black
lines. (Right panel) The temporal derivative,
$\partial L_{\rm I}(\delta,t)/\partial t$,
indicating some jets as circular areas centred around each jet have been
emphasized by a factor of two. A total of 12 jets were detected in the 66~h
interval (from Llebaria et al. 2002b).}
\end{figure}

In order to analyse the spatio-temporal evolution of WL plumes,
special sequences of
LASCO-C2 images were obtained in 1997, 2007 and 2010. These sequences employed
a high cadence (one image every 9.9~min) and high signal-to-noise
ratios (through an overexposure by a factor of four).
This increased the visibility of changes in structural details.
A time intensity diagram (TID) has been constructed in Fig.~\ref{tidfive},
using a variant of the sinogram technique (Lamy et al. 1997),
by compiling 402 images taken with LASCO-C2
in March 1997. It shows that not only the
spatial distribution is of a fractal type, but also the temporal
evolution (Llebaria et al. 2002b).
Plume trajectories could be tracked for many hours (see also Fig.~\ref{dtid})
and statistics obtained on duration and intermittence along the
trajectories as well as on the distribution of projected speeds of
plume propagation into the corona.
The apparent speeds peaked at $300~{\rm km\,s}^{-1}$ with
a median value of $\approx 400~{\rm km\,s}^{-1}$.
The speeds are deduced by fitting straight
lines to the local profiles of brightness in Fig.~\ref{traces}.
The high variability
of the estimates seem to indicate real outflows,
rather than wave phenomena, for which the
propagation speed would be constraint by the local plasma conditions.

The first derivatives of the TID are also very informative.
The derivative in the angular direction facilitates the detection of plume
trajectories and determines the relative position of each one on
the CHs ``plane'' as shown in the left panel of Fig.~\ref{dtid}.
The derivative relative to time enhances the onset and extinction of plumes,
but also unambiguously reveals in the right panel the appearance of jets
with angular widths of less than 2\deg~at $3.5~R_\odot$ and
lifetimes of $\approx 200$~min (for more details on
\emph{jets} see Sect.~\ref{Jets}).


% 4
\section{\small Dynamics}
\label{Dynamics}

The importance of wave phenomena for the generation and support
of plumes is unclear and their life cycle is rather elusive as is
the relation to the fast SW streams. In this context, the following
questions have been considered and are discussed in this section.
-- Are the characteristics of plumes changing within their
life cycle?
-- Do plumes show a tendency of recurrence?
-- Is there wave activity that is characteristic of plumes
and IPRs and how are they related to heating processes and the
SW acceleration?
-- It is established that the fast SW streams
emanate from CHs: what are the outflow speeds in plumes and IPRs?
-- Are plumes or IPRs the main contributors to the fast
SW streams?

% 4.1
\subsection{\small Life cycle of plumes}
\label{Life}

DeForest et al. (1997) observed temporal changes of plumes
on time scales of less than 10~min as brightenings of small
filamentary structures of $\approx 5\arcsec$~width with relative radiance
variations of $\approx 10~\%$ and outward propagation speeds of
300~km\,s$^{-1}$
to 500~km\,s$^{-1}$. On the other hand, the shapes of the plumes on spatial
scales of $\approx$\,30\arcsec~(also typical for the network super-granulation,
cf., Sect.~\ref{Network}) appeared to be nearly constant
for hours to days (cf., Waldmeier 1955).
Sinogram analyses of EIT and LASCO image sequences obtained in December 1996
indicated plume lifetimes from 0.5~d to 2~d
(and longer for beam plumes, cf., Sect.~\ref{Dimension})
with a pronounced recurrence
tendency for weeks at the same locations (Lamy et al. 1997; DeForest et al.
2001b). A certain magnetic flux tube therefore is only occasionally filled with
high-density plume plasma. Plumes are episodic in nature,
and are both transient and persistent.

\begin{figure}[!t]
% Figure 12
\centering
\includegraphics[width=\textwidth]{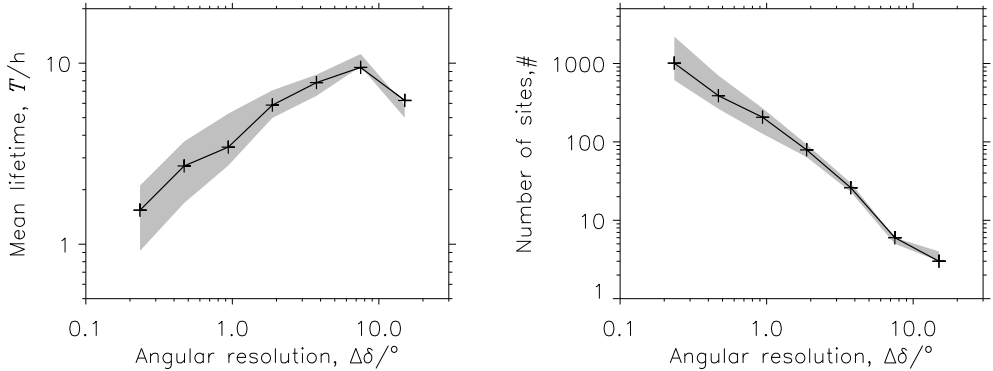} %{antoine_5.ps}
\caption{\label{fig:statres} \small
(Left panel) The distribution of the mean lifetime of WL plumes as function of
the spatial resolution. (Right panel)  Number of plumes identified in the
TID of 66~h at
the corresponding resolution. Grey areas shows the variability depending on the
minimal gap (between 30 and 50~pixels) chosen to dissociate two successive
plumes. The steep increase of the number of plumes with resolution speaks
for a (fractal-like) fine structure of WL plumes
(from Llebaria et al. 2004).}
\end{figure}

The question of
how long plumes last is difficult to
answer, because plumes and IPRs cannot easily be separated
in images over long time intervals. The transverse profile is
characterized by a continuous succession of local minima and maxima.
The apparent structure of plume profiles led
Llebaria et al. (2002b) to the definition of an
multi-resolution analysis.
It deploys the initial profile in a set of diagrams with
spatial scales of increasing size, and will thus progressively
smooth the profiles. It turns out that there is a strong
correlation between the size level and the time duration as can be seen
in Fig.~\ref{fig:statres}. However, because plumes are intermittent phenomena,
these results are to some degree dependent on the
choice of the activity threshold.

The formation of a plume above a BP has been observed in the 17.1~nm band of EIT
followed by the disappearance of the BP before the
plume faded away within $\approx 28$~h in May 1997 (Wang 1998).
The chromospheric evaporation time scale of $\approx 6$~h appears to
control the formation
of the plume and the radiative cooling time of $\approx 4$~h its decay.
Del Zanna et al. (2003) observed diffuse plumes without BPs at their base and
a near isothermal temperature of $\approx 0.8$~MK. The authors suggest that the
plumes with BPs might represent an early stage of plume evolution; and
Newkirk and Harvey (1968) speculated that the most long-lived flux
concentrations might underlie plumes. It must, therefore, be concluded that the
magnetic configuration of a plume is relatively stable, but that the processes
generating the actual plume plasma operate only intermittently.

Raouafi et al. (2008) studied the relationship between polar coronal
jets and plumes (see also Sect.~\ref{Jets}). Multiple occurrences of
short-lived, jet-like event were found at the base of long-lived plumes,
suggesting that these brightenings might be the result of
magnetic reconnection of continuously emerging flux and could
contribute significantly to the length of the life cycle of
these plumes.

% 4.2
\subsection{\small Waves and turbulence in plumes and inter-plume regions}
\label{Waves}

Waves (usually Alfv\'en waves) are a very promising mechanism for
transporting the energy from the solar surface
into the corona, where
they are partially reflected back down towards the Sun and dissipated by
turbulent processes (see, e.g., Velli 1993; Matthaeus et al. 1999;
Cranmer et al. 2007).
A description of the concept of wave and turbulence-driven SW models
can be found in Cranmer (2009) together with a list of recent reviews on in situ
and remote sensing observations of waves.
On the specific topic of observation
of waves in plumes, we can refer to Banerjee et al. (2009b).

The detection of waves in the outer solar atmosphere is made possible by
analysing the effects these waves have on the plasma. The presence or signature
of compressional waves may be seen in the form of variations or oscillations
in line radiance, due to change in plasma density, and also in the
LOS velocities, due to plasma motions (when they have a
significant component directed towards the observer).
On the other hand, transverse waves give rise to only LOS effects when they
propagate substantially over the plane of sky. Moreover, the latter give no
radiance signature in the theoretical limit of incompressible Alfv\'en waves.
Temporally and spatially resolved motions result in shifts of
the observed profiles, whereas unresolved motions result in broadening of the
spectral lines. These effects can, in principle, be measured from observations
of spectral lines profiles, but the unresolved motions
give rise to an ambiguity in separating thermal and
turbulence effects, which plagues interpretation of some observations.

Quasi-periodic brightness variations in plumes have been observed
with EIT by DeForest et
al. (1997). On 7 March 1996, wave trains in several plumes were
propagating outwards with periods between 10~min and 15~min and speeds between
75~km\,s$^{-1}$ and 150~km\,s$^{-1}$ in the height range from
$h$ = 0.01~$R_\odot$  to 0.2~$R_\odot$. They have been
identified as compressional waves (Ofman et al. 1997, 1999;
DeForest and Gurman 1998).
Plume oscillations with periods of 10~min to 25~min have also been
detected in the O\,{\sc v} 62.9~nm (0.24~MK) line by CDS
(Banerjee et al. 2000b), and with
$\approx 10$~min period in the WL channel of UVCS between
$h$ = 0.9~$R_\odot$ to 1.1~$R_\odot$. Their group velocity was
$\approx 200$~km\,s$^{-1}$ (Ofman et al. 2000a).
Ofman et al. (1999, 2000b) developed a visco-resistive 2.5D
magnetohydrodynamic (MHD) model of plumes with propagating
slow magnetosonic waves, and studied
the effects of wave trapping in high density plumes, non-radial plume expansion,
and solar wind outflow in plumes.
The authors suggest that compressional waves are
propagating upwards from the Sun, and more specifically,
identified the oscillation as slow magneto-sonic waves in plumes with diameters
of $\approx 30$~Mm.
On the other hand, these time scales are commensurate with the
slow photospheric motions thought to drive Alfv\'enic (incompressible)
fluctuations that propagate upwards and are implicated in heating at coronal
altitudes (see, e.g., Dmitruk et al. 2002; Verdini et al. 2010).
Consequently, it is not out of the question that plume formation and dynamics
is in some way related to this broader issue. Very long-period activity
($\approx 170$~min) in a PCH has been reported by Popescu et al. (2005).
In a review article, Ofman (2005) concluded that the energy flux in slow-mode
waves is too small for all of the coronal heating and that other modes must be
considered in addition.

Above BP groups, torsional Alfv\'enic perturbations have been detected through
non-thermal broadenings of the H\,$\alpha$ line
profile (Jess et al. 2009). The authors conclude that the energy flux of these
waves is sufficient to heat the corona.
Indirect evidence for Alfv\'en waves
has been found in CHs by Banerjee et al. (1998, 2009a) as well as by
Dolla and Solomon (2008) from
measurements of line broadenings in spectra obtained with very long exposure
times. The propagation and dissipation of Alfv\'en waves in plumes were
discussed by Ofman and Davila (1995) using a 2.5D MHD model. The
injection of Alfv\'en waves and the formation of a jet was recently studied
by Pinto et al. (2010) using such a model.

\begin{figure}[t]
% Figure 13
\centering
\includegraphics[width=10cm]{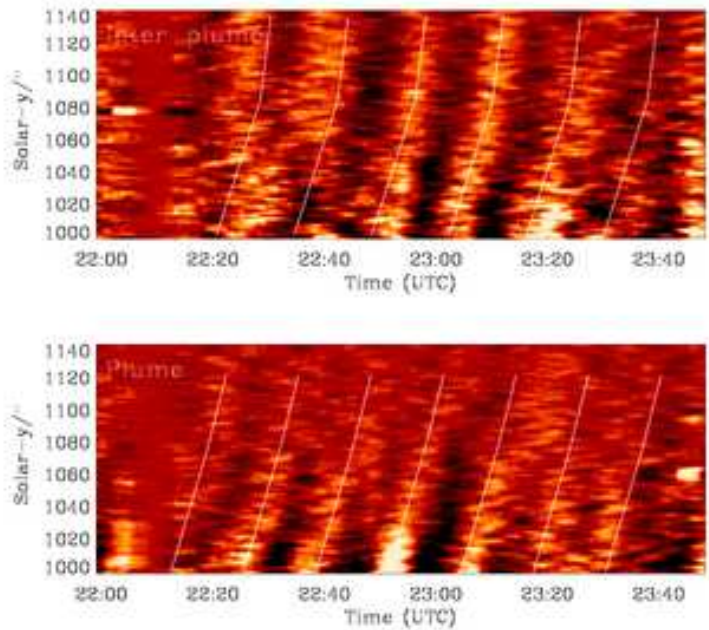} %{xt_luca.ps}
\caption{\label{Figwaves} \small
Maps of radiance along the slot (solar-$y$) versus time
obtained by EIS in Fe\,{\sc xii} on 13 November 2007 by averaging
over 5\arcsec~in the solar-$x$ direction at the selected position.
The maps were smoothed over $\approx 3$~min and
the background trend has been subtracted from each solar-$y$
pixel along time. The height range shown covers the near off-limb and far
off-limb regions of the PCH in an IPR (top panel)
and a plume (bottom panel). The slanted lines represent the disturbances
propagating outward with increasing speed.
It can be seen that in some places the white lines do not
coincide with the enhanced lanes but are nevertheless parallel to them.
This suggests that even if the periodicity changes within a certain range, the
propagation speeds are fairly uniform (after Gupta et al. 2010).}
\end{figure}

Recently, Gupta et al. (2010) detected the presence of propagating waves in an
IPR with a 15~min to 20~min periodicity\,---\,obtained from a wavelet
analysis\,---\,and a propagation speed
increasing from (130 $\pm$ 14)~km\,s$^{-1}$ just
above the limb to (330 $\pm$ 140)~km\,s$^{-1}$ around 160\arcsec~above the limb.
The distant-time map of the Fe\,{\sc xii} radiance over nearly 2~h is shown
in the upper panel of Fig.~\ref{Figwaves}.
Although the waves are best seen in radiance, significant power at those
periodicities is also detected in both Doppler width and shift.
In the adjacent plume region (lower panel), propagating radiance
disturbances also appear to be present (no spectral profiles are available)
with the same range of
periodicity, but with propagation speeds in the range of
(135 $\pm$ 18)~km\,s$^{-1}$ to (165 $\pm$ 43)~km\,s$^{-1}$. The plume
observations might, however, be affected by the IPRs along the LOS, because of
the low electron temperature in plumes, the high formation temperature of
the Fe\,{\sc xii} line (cf., Figs.~\ref{Figtem} and \ref{Figcon}) and the
more favourable conditions for such emissions in IPRs.
Based on the acceleration to supersonic speeds and the signature in Doppler
width and shift, the authors suggest that in IPRs the waves are likely
either Alfv\'enic or fast magneto-acoustic, whereas they are slow
magneto-acoustic in plumes.

An important feature of turbulence models (Dmitruk et al. 2001)
is the non-linear pumping of non-propagating "zero frequency" structures.
These would appear in observations as strong transverse gradients. When
present these "quasi-2D" fluctuations can catalyse a
powerful cascade perpendicular to the large-scale magnetic field that may drive
strong turbulent heating at fine transverse scales
(e.g., Verdini et al. 2010). For this reason in considering turbulence models,
it is essential to examine fluctuations that may not be described by any
linear wave mode (see, e.g., Dmitruk and Matthaeus 2009). It is possible
that plumes, with their characteristic transverse structure, may
participate in these low-frequency dynamical couplings, and thus could
play a direct r$\hat{\rm o}$le in coronal turbulence.

%

% 4.3
\subsection{\small Outflows in plumes, inter-plume regions and coronal holes}
\label{Outflow}

\begin{table}[!t]
% Table~2
\noindent
\caption{Typical outflow speeds in CHs, plumes and IPRs at
some representative heights}
\vspace{0.3cm}
\begin{small}
\begin{tabular*}{14.2cm}{ccccccc}
\hline
Height$^{\rm a}$
& \multicolumn{3}{c}{Outflow speed, $u_{\rm out}/({\rm km\,s}^{-1}$)}
& Method & Date, & Reference\\
$h/R_\odot$ & CH$^{\rm b}$ & PL & IPR & & period &   \\
\hline
\hline
0.6 & $45^{+55}_{-20}$ & $36^{+25}_{-20}$ & & WL &
12 Apr 1993  & Habbal \\
1.2 & $130^{+50}_{-60}$ & $70^{+70}_{-10}$ &
& Spartan and & & et al. 1995 \\
2.2 & 220 & 130 & & Mauna Loa & & \\
\hline
0.5 & & $\approx 50$ & & O\,{\sc vi} 103.2~nm, & Nov 1996 & Corti  \\
1.0 & & 110 & & 103.8~nm ratio & & et al. 1997 \\
\hline
0.5 & $11^{+38}_{-11}$ & & & Model$^{\rm c}$
& Nov 1996/ & Cranmer \\
1.0 & $179^{+78}_{-76}$ ($150^{+49}_{-55}$) & &
& B1: O$^{5+}$ & Apr 1997 & et al. 1999 \\
2.0 & $402^{+44}_{-68}$ ($219^{+30}_{-23}$) & &
& (A1: H$^{0}$)$^{\rm d}$ & &  \\
\hline
0.50 & $45{\scriptstyle \pm 10}$ & & & O\,{\sc vi} 103.2~nm,
& 21 May 1996 & Antonucci \\
2.10 & $360{\scriptstyle \pm 40}$ & & & 103.8~nm ratio
& SOHO roll & et al. 2000 \\
\hline
0.72 & & 0 to & 105 to & O\,{\sc vi} ratio$^{\rm e}$ &
Apr 1996 & Giordano\\
 & & 65 & 150 & & & et al. 2000b \\
\hline
0.05 & & & $67^{+16}_{-14}$ & O\,{\sc vi} ratio & 26 Feb 1998
& Patsourakos \\
&&&&& eclipse & and Vial 2000 \\
\hline
0.4 & $62{\scriptstyle \pm 5}$ ($60{\scriptstyle \pm 5}$) & &
& Semi-empirical & Aug 1996 & Zangrilli \\
1.0 & $234{\scriptstyle \pm 20}$ ($154{\scriptstyle \pm 5}$) & &
& models: & & et al. 2002 \\
1.5 & $370^{+60}_{-90}$ ($232{\scriptstyle \pm 30}$) & &
& O$^{5+}$ (H$^0$)$^{\rm d}$ & & \\
\hline
0.05 & & $\approx 90^{\rm f}$ & 23 & O\,{\sc vi} ratio &
21 May 1996 & Gabriel \\
0.20 & & $\approx 80$ & 30 & & & et al. 2003 \\
0.36 & & $\approx 85$ & 48 & & \\
\hline
0.10 & & static & $< 25$ &  O\,{\sc vi} ratio$^{\rm e}$
& 3 Jun 1996 & Teriaca \\
0.50 & & or & $49{\scriptstyle \pm 28}$ & & & et al. 2003 \\
0.75 & & slow & $84{\scriptstyle \pm 21}$ & & &  \\
1.00 & & outflow & $164{\scriptstyle \pm 38}$ & & \\
\hline
2.1 & 359 to & & & O\,{\sc vi} ratio & 21 May 1996 & Antonucci \\
 & 500 & & & $u_{\rm out}$, $n_{\rm e}$ & & et al. 2004 \\
\hline
0.05 & & 60 & 30 & O\,{\sc vi} ratio & 21 May 1996 & Gabriel \\
0.60 & & 78 & 83 & & & et al. 2005 \\
1.00 & & 115 & 150 & & \\
1.40 & & $130{\scriptstyle \pm 20}$ & $180{\scriptstyle \pm 20}$ & & \\
\hline
0.92 & & $124{\scriptstyle \pm 5}$ & & O\,{\sc vi} ratio &
29 Mar 2006 & Abbo \\
1.07 & & & $208^{+41}_{-27}$ & & eclipse & et al. 2008 \\
\hline

\end{tabular*}

\vspace{0.3cm}
$^{\rm a}$ above the photosphere \\
$^{\rm b}$ not separated into plume (PL) and IPR \\
$^{\rm c}$ parallel motions in thermal equilibrium with electrons \\
$^{\rm d}$ hydrogen speeds in parentheses \\
$^{\rm e}$ depends on isotropy assumptions \\
$^{\rm f}$ with a plume filling factor of $F = 0.5$
\end{small}
\label{Tabout}
\end{table}
\begin{figure}[!t]
% Figure 14
\begin{minipage}[b]{14.2cm}
\begin{minipage}[b]{10.5cm}
\includegraphics[width=10.5cm]{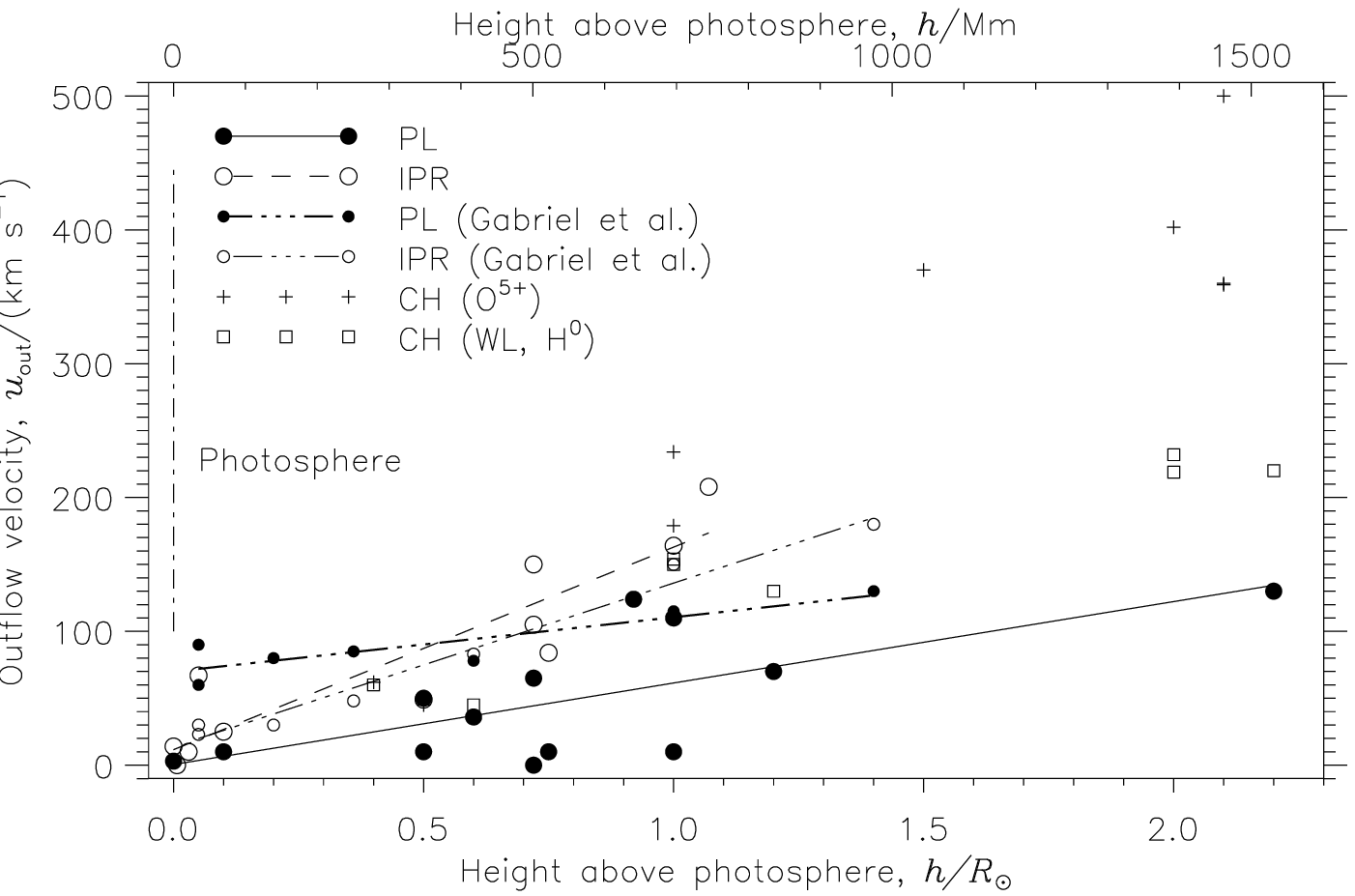} %{OUTFLOW.EPS}
\end{minipage}\hfill
\parbox[b]{3.2cm}{\caption{\label{Figout} \small
Plasma outflow velocities in plumes, IPRs  and CHs
from data in Table~\ref{Tabout} and references cited in the text.
The values obtained by Gabriel et al. (2003, 2005) are plotted with smaller
symbols compared to those of other authors. Separate linear fits are shown for
four groups of data in different line styles.
The speeds of O$^{5+}$ in CHs are consistently higher
than those of H$^0$.}}
\end{minipage}
\end{figure}
Coronal plumes together with other dynamic structures in the solar atmosphere,
such as spicules, macrospicules and
chromospheric jets, are potential sources of the SW. The contribution
of plumes to the fast SW has been disputed in the literature. Some studies
indicate that plumes are a plausible source of the fast SW streams
(Gabriel et al. 2003, 2005), and may even contribute
one half of the fast SW. Other investigations led to different results
(e.g., Wang 1994; Habbal et al. 1995; Wilhelm et al. 1998;
Hassler et al. 1999; Giordano et al. 2000a, b; Teriaca et al. 2003).

Doppler dimming techniques\,---\,developed by Rompolt (1967)
for moving prominences using the H\,$\alpha$
line\,---\,applied to the corona (cf., Kohl and Withbroe 1982; Noci et al. 1987)
have supplied most of the outflow speeds summarized in Table~\ref{Tabout}
using UVCS and SUMER observations of
the O\,{\sc vi} 103.2~nm, 103.8~nm lines.
In this and the following tables, typical results obtained for the
relevant quantities (outflow speed, $u_{\rm out}$; electron density,
$n_{\rm e}$; electron temperature, $T_{\rm e}$;
Doppler velocity, $V_{\rm 1/e}$) by various researchers using different methods
are compiled in order to provide an overview. Many numbers are taken from
diagrams in the referenced papers and are consequently given in rounded values.
If uncertainty margins are available in the original articles in appropriate
formats (either in tables or diagrams),
they are quoted in the tables with the qualification that
their significance varies for different sources.
Our aim is to present a large number
of observations and identify any systematic behaviour.
Since we are comparing solar features over a period of more than one
solar cycle the variations are, in all likelihood, in many cases
larger than any measurement uncertainties.
Nevertheless it should be pointed out
that the stability of the CH conditions during the activity minima is
remarkable.

From Table~\ref{Tabout} and
Fig.~\ref{Figout},
it is clear that the outflow speed in IPRs increases rapidly in the low
corona to high values of $u_{\rm out} \approx 200~{\rm km\,s}^{-1}$ within a
distance of 1~$R_\odot$.
The cases for which only undifferentiated CH outflow speeds have been
published seem to be related more to IPRs than to plumes.
Below $h \approx 0.5~R_\odot$,
faster and slower outflow speeds in plumes than in IPRs have been
deduced in different investigations.
These conflicting findings were one of the motivations for this study and,
consequently, they have to be addressed in some detail. Casalbuoni
et al. (1999) concluded in a study of
pressure-balanced structures (PBS) that
the temperature difference between plumes and IPR has a decisive effect on the
outflow speed. Since there is strong evidence that
plumes are cooler than IPRs both in the electron temperature, $T_{\rm e}$, and
the effective ion temperature, $T_{\rm eff}$, above $h \approx 0.03~R_\odot$
(cf., Table~\ref{Tabtem} and Sect.~\ref{Effective}), it follows
that plumes must be slower.
However, it has also been shown that a plume cannot be maintained if it is
cooler than the IPR at all heights. An additional heat input at the base of a
plume is required for a significant density enhancement
(Del Zanna et al. 1997). Such a scenario
is consistent with the association of BPs and plumes (cf., Sect.~\ref{BP})
and the plume model of Wang (1994) (cf., Sect.~\ref{Models}). In this context,
it must be noted that the electron density height profile assumed in the
evaluation of the outflow speeds plays an important r$\hat{\rm o}$le
and that very different
densities had been assumed in various studies.
Moreover, we point out that the ion temperature in the direction
perpendicular to the LOS has to be known for a determination of the outflow
speed with the help of the Dopper dimming technique. Although there is strong
evidence for temperature anisotropies of heavy ions in IPRs and,
possibly, in plumes
(Cranmer et al. 1999; Giordano et al. 2000b; Teriaca et al. 2003), different
assumptions on the degree of anisotropy led to different results.

At very low altitudes ($h \le 0.03~R_\odot$), LOS Doppler velocities of
$\approx 10$~km\,s$^{-1}$
(corresponding to radial speeds of $\approx 14$~km\,s$^{-1}$ at the observation
site) in the Ne\,{\sc viii} 77.0~nm line are only seen in
darker regions of a CH. A relatively strong plume that could be identified on
the disk showed a speed of less than $\approx 3$~km\,s$^{-1}$ (cf.,
Sect.~\ref{Magnetic}; Hassler et al. 1999; Xia et al. 2003).
Rather high radiance values in
the TR line Si\,{\sc ii} 153.3~nm (27\,000~K)
near the plume footpoint had been found in the same
data set, indicating enhanced heating at the base of the plume.
Tu et al. (2005) found no outflow of C$^{3+}$ ions at a height of 5~Mm,
but a LOS velocity of $\approx 10$~km\,s$^{-1}$ in funnels of
the same CH at 20~Mm (for Ne$^{7+}$).
However, no significant outflow could be detected in a
magnetic plume structure (cf., Sect.~\ref{Magnetic}).
It thus appears as if, indeed, the IPRs have
larger outflow speeds along a height profile and, together with the small
filling factor of plumes in CHs, provide the main contribution to the fast SW
streams as suggested by Wang (1994).
The funnels harbouring the outflows seen in Ne\,{\sc viii}
are most likely rooted
in flux concentrations described by Tsuneta et al. (2008b). The picture
of expanding coronal funnels is supported by the findings of Tian et al. (2010)
that increasingly larger patches of blue-shifted line
profiles (outflows) are observed in hotter spectral lines
with EIS in the on-disk part of a PCH.

Raouafi et al. (2007b) studied the plasma dynamics (outflow speed and
turbulence) inside coronal polar plumes and compared line profiles
(mainly of O\,{\sc vi}) observed by UVCS at the minimum between solar
cycles 22 and 23 with model calculations.
Maxwellian velocity distributions with different widths are assumed for both
plumes and IPRs, and different combinations of the outflow velocities,
$u_{\rm out}$,
and most-probable speeds, $V_{\rm 1/e}$ (cf., Sect.~\ref{Effective})
are considered. The observed profiles are reproduced best
by low outflow speeds close to the Sun in plumes that increased
with height to reach IPR values above $h \approx 3~R_\odot$.
The most-probable speeds in plumes and IPRs assumed are
included in Table~\ref{TabDop}.

In equatorial coronal holes (ECH), with characteristics very similar to PCHs,
outflows along open field lines can be detected in spectral lines with
formation temperatures above 0.1~MK
(cf., Sect.~\ref{Electron}). An average outflow
speed of $u_{\rm out} \approx 5~{\rm km\,s}^{-1}$ was measured
for Ne$^{7+}$ ions
and of $\approx 10~{\rm km\,s}^{-1}$ for Mg$^{8+}$ (Wilhelm et al. 2002a;
Xia et al. 2004; Wiegelmann et al. 2005).
Woo (2007) summarized outflow observations as filamentary structures
on open field lines within the so-called closed corona.

On balance, there seems to be some evidence that IPR outflow speeds become
significantly greater than outflows in plumes at increasing altitude in the
lower corona. In this context, Sheeley et al. (1997) made an interesting
remark on the direction of time that can always be identified in coronal
streamers, but not in polar coronal plumes\,---\,implying that the outflow
signatures in plumes are less pronounced.

% 5
\section{\small Plasma conditions in coronal holes}
\label{Plasma}

\begin{figure}[!t]
% Figure 15
\begin{minipage}[b]{14.2cm}
\begin{minipage}[b]{10cm}
\includegraphics[width=10cm]{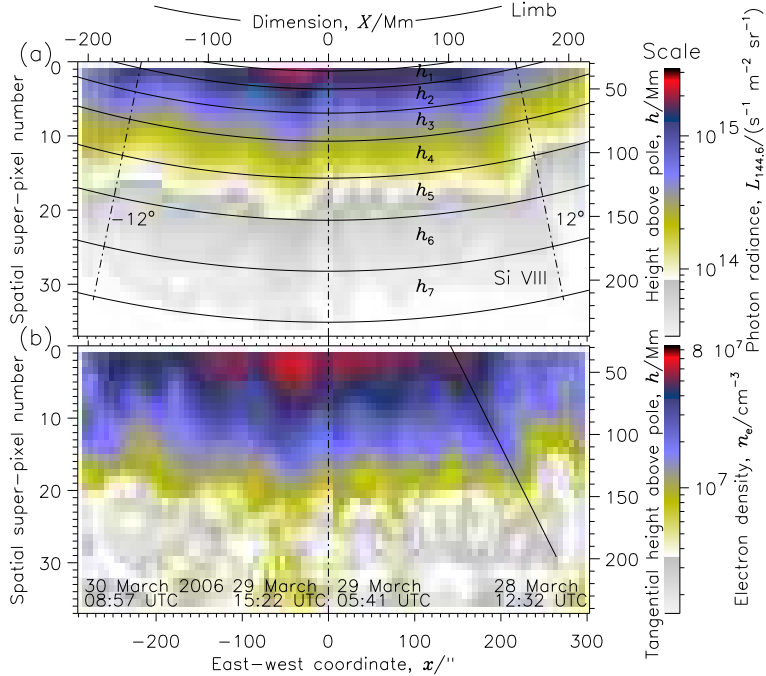} %{AAREVSIDEN.EPS}
\end{minipage}\hfill
\parbox[b]{3.7cm}{\caption{\label{Figecl} \small
(a) Radiance of the Si\,{\sc viii} 144.6~nm line in the southern
CH before and after the total eclipse on 29 March 2006. During the actual
eclipse, the scan was interrupted in favour of high-cadence O\,{\sc vi}
observations, an example of which is shown in
Fig.~\ref{Figjet}.
The concentric
height ranges, $h_1$ to $h_7$, outline the data selection in
Fig.~\ref{FigS06}b.
Radius vectors at $\pm\,12\deg$~are indicated. (b) Electron
density determined from the LOS Si\,{\sc viii} line ratio
$L_{144.6}/L_{144.0}$. The projection of a plume near $x = 200\arcsec$~is
shown for a comparison with the radius vector in panel (a).}}
\end{minipage}
\end{figure}

The knowledge of the plasma conditions in plumes and their environment in CHs
is critical for an understanding of the plume physics. We asked questions as
follows:
-- Can \emph{standard} height profiles of the electron density
in plumes and IPRs be defined considering that the measurements obtained
with various methods agree remarkably well within the general variability
of solar features?
-- Specifically, what is the plume/IPR density ratio and
its potential variation with height?
-- What are the plume and IPR electron and ion temperatures
as a function of height?
-- Is there a significant anisotropy of the ion temperatures
in plumes and IPRs?
-- What can be said about the elemental abundance in plumes
and IPRs, and, specifically, about the first-ionization potential (FIP) effect?
-- What is the expected FIP and ionization state
signature of plumes in the SW, given a
hypothesis for their source?
-- Is it agreed that there are different plasma regimes
present along the LOS in CH observations\,---\,plumes and IPRs?

A brief discussion of the resulting plasma pressures is included in
Sect.~\ref{Wind}.

% 5.1
\subsection{\small Electron densities in plumes and inter-plume regions}
\label{Densities}

Above the solar limb, coronal plumes seen in WL appear brighter than the
surrounding medium which led many authors
to the conclusion that they are denser than the background corona
(called here IPR)
(van de Hulst 1950b; Saito 1965a; Koutchmy 1977;
Ahmad and Withbroe 1977; Fisher and Guhathakurta 1995).
The density measurements in WL utilize the fact that Thomson scattering of
electrons produces polarized light, whereas the much stronger F-coronal radiance
from dust particles is unpolarized. The polarized brightness
\begin{equation}
pB = \sqrt{Q^2 + U^2} \quad ,
\label{Eqpol}
\end{equation}
with $Q$ and $U$ the relevant Stokes parameters, then has to be related to the
electron density along the LOS taking into account the dependence of $pB$ upon
the distance from the plane of the sky (cf., Koutchmy and Bocchialini 1998).

VUV observations rely on atomic data for a determination of the electron
density from line-ratio measurements. In the polar corona the
nitrogen-like ion Si$^{7+}$ and its magnetic dipole transitions
$2{\rm s}^2 2{\rm p}^3\,^4{\rm S}_{3/2} - 2{\rm s}^2 2{\rm p}^3\,^2{\rm
D}_{3/2}$ and $2{\rm s}^2 2{\rm p}^3\,^4{\rm S}_{3/2} - 2{\rm s}^2 2{\rm
p}^3\,^2{\rm D}_{5/2}$ with the corresponding emission lines Si\,{\sc viii}
144.6~nm and 144.0~nm
provide a convenient means of deducing $n_{\rm e}$ through
the radiance ratio $R_{\rm Si} = L_{144.6}/L_{144.0}$. It is density sensitive,
because de-excitation of the D$_{5/2}$ level occurs not only radiatively, but
also collisionally (see Laming et al. 1997; Doschek et al. 1997 for a
conversion procedure). Compared with this procedure, the
electron densities determined from $R_{\rm Si}$ with the help of the CHIANTI
atomic data base yield slightly higher values (Banerjee et al. 1998).
Warren and Hassler (1999) discussed, in addition, other density-sensitive line
ratios, and derived CH densities that are in good agreement
with the Si\,{\sc viii} values.
In an ECH, Del Zanna and Bromage (1999) measured coronal electron densities
of $n_{\rm e} \approx 3 \times 10^8$~cm$^{-2}$ (approximately a factor of two
lower than in the adjoining QS regions).
A selection of typical electron density measurements, obtained with the help
of spacecraft and eclipse observations, is compiled in Table~\ref{Tabden}
for some representative heliocentric distances.

For an isothermal plasma, optically thin for a certain emission line, the
LOS-integrated electron density (along the $z$ direction)
can be deduced from
$\int n_{\rm e}^2\,{\rm d}z$, the emission measure (EM)
(see, e.g., Raymond and Doyle 1981),
which in turn can be obtained from radiance observation,
if the electron temperature and the element abundance are known (cf.
Sects.~\ref{Electron} and \ref{Abundances}).
It is important to note that the Thomson scattering depends linearly on the
electron density, whereas the emission process is a function of
$n_{\rm e}^2$. So that WL observations yield the
mean value of the electron density, $\langle n_{\rm e} \rangle$,
whereas spectroscopic line ratios yield $\sqrt{\langle n_{\rm e}^2 \rangle}$.
These will not be the same where there are
inhomogeneities in the plasma, and there is some evidence for this. Such
inhomegeneities could have different scales, for example beam plumes
and IPRs, network plumes, or even much finer structures
due to turbulence.
\begin{table}
% Table~3
\noindent
\caption{Typical electron densities in plumes and
IPRs at some heliocentric distances}
\vspace{0.3cm}
\begin{small}
\begin{tabular*}{14.2cm}{cccccc}
\hline
Distance$^{\rm a}$ &
\multicolumn{2}{c}{Density, $n_{\rm e}/(10^7~{\rm cm}^{-3}$)} &
Method & Date, & Reference\\
$R/R_\odot$ & PL & IPR$^{\rm b}$ & &period &   \\
\hline
\hline
1.20 & 20 & 4 & WL isophots & 1900 eclipse & van de Hulst 1950b \\
\hline
1.18 & 10 & 2.8 & Eclipse, & 1962 eclipse & Saito 1965a, b \\
1.33 & 4 & 1.1 & WL radiance && \\
1.67 & 1 & $\approx 0.2$ &&& \\
\hline
1.10 & $\approx 10^{\rm c}$ & 0$^{\rm d}$ & Eclipses, & 1962, 1963, &
Newkirk and \\
&&& WL radiance & 1965  & Harvey 1968 \\
\hline
2.00 & & 0.052 & WL & Jul  & Munro and \\
2.50 & & 0.013 & coronagraph & 1973 & Jackson 1977 \\
5.00 & & 0.0013 & Skylab$^{\rm e}$ && \\
\hline
1.60 & $0.21{\scriptstyle \pm 0.07}$ & $0.08{\scriptstyle \pm 0.03}$ &
Spartan & 11/12 Apr & Fisher and \\
2.00 & $0.05{\scriptstyle \pm 0.008}$ & $0.02{\scriptstyle \pm 0.003}$ &
201-01 & 1993 & Guhathakurta \\
4.00 & $0.003{\scriptstyle \pm 0.0005}$ & $0.001{\scriptstyle \pm 0.0002}$ &
WLC ($pB$) &  & 1995  \\
\hline
1.5 & $0.32^{+0.09}_{-0.08}$ & & O\,{\sc vi} 103.2~nm, & Nov 1996 & Corti \\
2.0 & 0.042 & &103.8~nm & & et al. 1997 \\
2.3 & 0.017 & & (EM)& & \\
\hline
1.02 & & $8.0{\scriptstyle \pm 0.5}$ & Si\,{\sc viii} 144 nm, &
4 Nov 1996 & Dosckek et al.\\
1.10 & & $3.0{\scriptstyle \pm 0.5}$ & 144.6 nm & & 1997 \\
1.30 & & $0.4^{+0.4}_{-0.3}$ & line ratio & & \\
\hline
1.10 & $7{\scriptstyle \pm 1}$ & 1.0  & Eclipses (WL)  &  &
Koutchmy and \\
1.50 & $0.2{\scriptstyle \pm 0.06}$ & 0.1 & star calibr. & &
Bocchialini 1998\\
\hline
1.03 & $13{\scriptstyle \pm 3}$  & $8{\scriptstyle \pm 4}$ & Si\,{\sc viii}  &
Nov 1996/  & Wilhelm  \\
1.08 & $6{\scriptstyle \pm 2}$ & $5{\scriptstyle \pm 2}$ & line ratio &
Jan 1997 & et al. 1998\\
1.30 & $2.0{\scriptstyle \pm 0.5}$ & $0.73{\scriptstyle \pm 0.3}$ & & & \\
\hline
1.03 & & 11 & Si\,{\sc viii} & May, Nov, & Banerjee \\
1.26 & & 1.6 & line ratio & Dec 1996 &  et al. 1998\\
\hline
1.05 & 5.0 & 3.0 & Si\,{\sc viii} & 3 Sep 1997 & Wilhelm and \\
1.10 & 3.4 & 1.7 & line ratio & SOHO roll & Bodmer 1998 \\
\hline
1.05 & & 15 & SUMER & Nov/Dec &  Doyle \\
1.30 & & 1.5 & (Si\,{\sc viii}) & 1996 &  et al. 1999\\
2.00 & & 0.02 & UVCS (WL)& 1996/1997 &  \\
8.00 & & 0.0004 & LASCO (WL)& 1996 &  \\
\hline
1.00 & $20^{+5}_{-4}$  & $11{\scriptstyle \pm 4}$ & Si\,{\sc ix} 34.2 nm,
& Aug/Sep & Fludra  et al.\\
1.10 & $8{\scriptstyle \pm 2}$ & &  35.0 nm ratio$^{\rm f}$& 1996 &  1999\\
\hline
1.00 & $45^{+55}_{-20}$  & $16{\scriptstyle \pm 6}$ &
Si\,{\sc ix}$^{\rm g}$ & 25 Oct 1996 &
Young et al. \\
1.10 & $40^{+140}_{-40}$  & $7^{+3}_{-2}$  & line ratio & & 1999 \\
\hline
1.015 & $22{\scriptstyle \pm 4}$ & & Ne\,{\sc vii } line & 27 Aug 1996
& Warren and\\
& & & ratio && Hassler 1999\\
\hline
\end{tabular*}
\label{Tabden}

Continued on next page\\
------------------

$^{\rm a}$ values at the base of the corona are indicated by $R = 1~R_\odot$\\
$^{\rm b}$ below $R < 1.05~R_\odot$ some plume projections seem to
merge and could hide IPRs \\
$^{\rm c}$ density at centre of cylindrical plume \\
$^{\rm d}$ no electrons outside plumes assumed \\
$^{\rm e}$ CH values not separated into PL and IPR\\
$^{\rm f}$ Pl values from spatial averages with IPR contributions\\
$^{\rm g}$ CDS observations of very strong plume with
jet characteristics; Si\,{\sc ix} (1.15~MK)

\end{small}
\end{table}

\addtocounter{table}{-1}

\begin{table}[!t]
%Table 3 continued
\noindent
\caption{continued}
\vspace{0.3cm}
\begin{small}
\begin{tabular*}{14.2cm}{cccccc}
\hline
Distance &
\multicolumn{2}{c}{Density, $n_{\rm e}/(10^7~{\rm cm}^{-3}$)} &
Method & Date, & Reference\\
$R/R_\odot$ & PL & IPR & &period &   \\
\hline
\hline
1.102 & 13.2 & 2.87 & Eclipse & 26 Feb 1998 & Lites et al. 1999 \\
1.60 & 0.342 & 0.0883 & (WL) & &  \\
2.20 & & 0.0159 & & & \\
\hline
1.03 & 20 to 50 & & Eclipses & 1994 to 1998 & Hiei et al. 2000  \\
1.20 & 1 to 7 & & (WL) &  & \\
\hline
1.05 & 5.2 & 1.9 & Si\,{\sc viii} ratio & 3 Sep 1997 &
Dwivedi et al. 2000  \\
1.05 & 6.4 & 3.0 & Mg\,{\sc viii} ratio$^{\rm g}$ & SOHO roll & \\
\hline
1.4 & & $3.09{\scriptstyle \pm 0.1}$ & O\,{\sc vi} 103.2 nm, &
Aug 1996 & Zangrilli \\
2.0 & & $0.40{\scriptstyle \pm 0.03}$ & 103.8 nm, & & et al. 2002$^{\rm e}$ \\
2.5 & & $0.14^{+0.04}_{-0.03}$  & H\,{\sc i} Ly\,$\alpha$ & & \\
\hline
1.00 & $120{\scriptstyle \pm 20}$ & $50{\scriptstyle \pm 20}$ &
Line ratios & Oct 1997 & Del Zanna et al. 2003\\
1.00 & $\approx 10$ &$\approx 5$ & & Aug 1996 & \\
\hline
1.70 & & $0.08{\scriptstyle \pm 0.03}$ & O\,{\sc vi} 103.2 nm, & 21 May 1996 &
Antonucci \\
3.10 & & $0.003{\scriptstyle \pm 0.001}$ & 103.8 nm & &
et al. 2004$^{\rm e}$ \\
\hline
1.07 & 6.9 & 1.3 & Si\,{\sc viii} & 24 May 2005
& Wilhelm 2006 \\
1.11 & 4.1 & 1.1 & line ratio & &  \\
1.19 & 3.4 & 0.78 &  &  &  \\
\hline
1.03 & & $13^{+3}_{-2}$  & Si\,{\sc viii} & 3 Nov 1996 & Landi 2008 \\
1.12 & & $2.8{\scriptstyle \pm 0.3}$ & line ratio & &  \\
\hline
1.2 & $1.5{\scriptstyle \pm 0.3}$ & $1.3{\scriptstyle \pm 0.3}$ &
EKPol & 29 Mar 2006 & Abbo et al. 2008 \\
1.5 & $0.19{\scriptstyle \pm 0.04}$ & $0.18{\scriptstyle \pm 0.04}$ &
polarimeter & eclipse & \\
2.0 & $0.022{\scriptstyle \pm 0.004}$ & $0.021{\scriptstyle \pm 0.004}$ &
$pB$ & & \\
\hline
1.03 & 22 & 19 & Fe\,{\sc xii} 18.7~nm, & 10 Oct 2007 &
Banerjee et al. 2009a \\
1.15 & 16 & 8.7 & 19.5~nm ratio & & \\
\hline
1.06 ($h_1$) & 4.5 (4.5) & 0.68 (2.34) & Si\,{\sc viii} ratio$^{\rm h,i}$ &
28/29 Mar & this work\\
1.14 ($h_4$) & 2.7 (3.5) & 0.52 (1.09) & tangential & 2006 & \\
1.24 ($h_6$) & 1.3 (2.3) & 0.13 (0.17) & (concentric) & & \\
\hline
1.04 ($h_1$) & 8.0 (5.1) & 1.1 (3.9) & Si\,{\sc viii} ratio$^{\rm j}$ &
7/8 Apr & this work\\
1.13 ($h_4$) & 4.4 (6.0) & 0.52 (1.2) & tan. (con.) & 2007 & \\
\hline
\end{tabular*}

$^{\rm e}$ see previous page\\
$^{\rm g}$ Mg\,{\sc viii} (0.79~MK) \\
$^{\rm h}$ Ratios:
$\rho^{\rm tan}_{\rm PL}(h_1,h_4,h_6)$ = (5.93, 4.05, 2.43);
$\rho^{\rm tan}_{\rm IPR}(h_1,h_4,h_6)$ = (1.77, 1.50, 1.00); \newline
$\rho^{\rm con}_{\rm PL}(h_1,h_4,h_6)$ = (4.54, 3.46, 2.31);
$\rho^{\rm con}_{\rm IPR}(h_1,h_4,h_6)$ = (2.34, 1.09, 1.69)\\
$^{\rm i}$ margins of correlation coefficients ($3\,\sigma$) given in
Fig.~\ref{FigS06}\\
$^{\rm j}$ Ratios:
$\rho^{\rm tan}_{\rm PL}(h_1,h_4)$ = (8.90, 5.79);
$\rho^{\rm tan}_{\rm IPR}(h_1,h_4)$ = (2.18, 1.56); \newline
$\rho^{\rm con}_{\rm PL}(h_1,h_4)$ = (6.50, 7.30);
$\rho^{\rm con}_{\rm IPR}(h_1,h_4)$ = (5.24, 2.41)
\end{small}
\end{table}
A radiance map of the southern low corona in
the Si\,{\sc viii} 144.6~nm line during the eclipse campaign
2006 is shown in
Fig.~\ref{Figecl}a
together with the electron densities in
panel (b), derived from the $L_{144.6}/L_{144.0}$ photon radiance ratio
observed in 96 raster steps of the SUMER slit from W to E.
The height resolution is limited by the count
statistics to a super-pixel of eight detector pixels. Several plume
signatures with enhanced density can be identified. The line-ratio
method\,---\,as applied here so far\,---\,obviously suffers from LOS effects
if there are density
(and temperature) variations along the integration path, complications discussed
by Habbal et al. (1993).
The polarization brightness, on the other hand, depends more on the conditions
near the plane of the sky (Munro and Jackson 1977; Koutchmy and
Bocchialini 1998), so that plumes and IPRs can probably be separated more
effectively.
From an analysis of the polarized K-corona measurements obtained with
the EKPol polarimeter, electron density profiles
have been derived in plume and IPR structures for the total eclipse
on 29 March 2006 (Abbo et al. 2008; see Table~\ref{Tabden}
for representative values).

\begin{figure}[!t]
% Figure 16
\begin{minipage}[b]{14.2cm}
\begin{minipage}[b]{10.5cm}
\includegraphics[width=10.5cm]{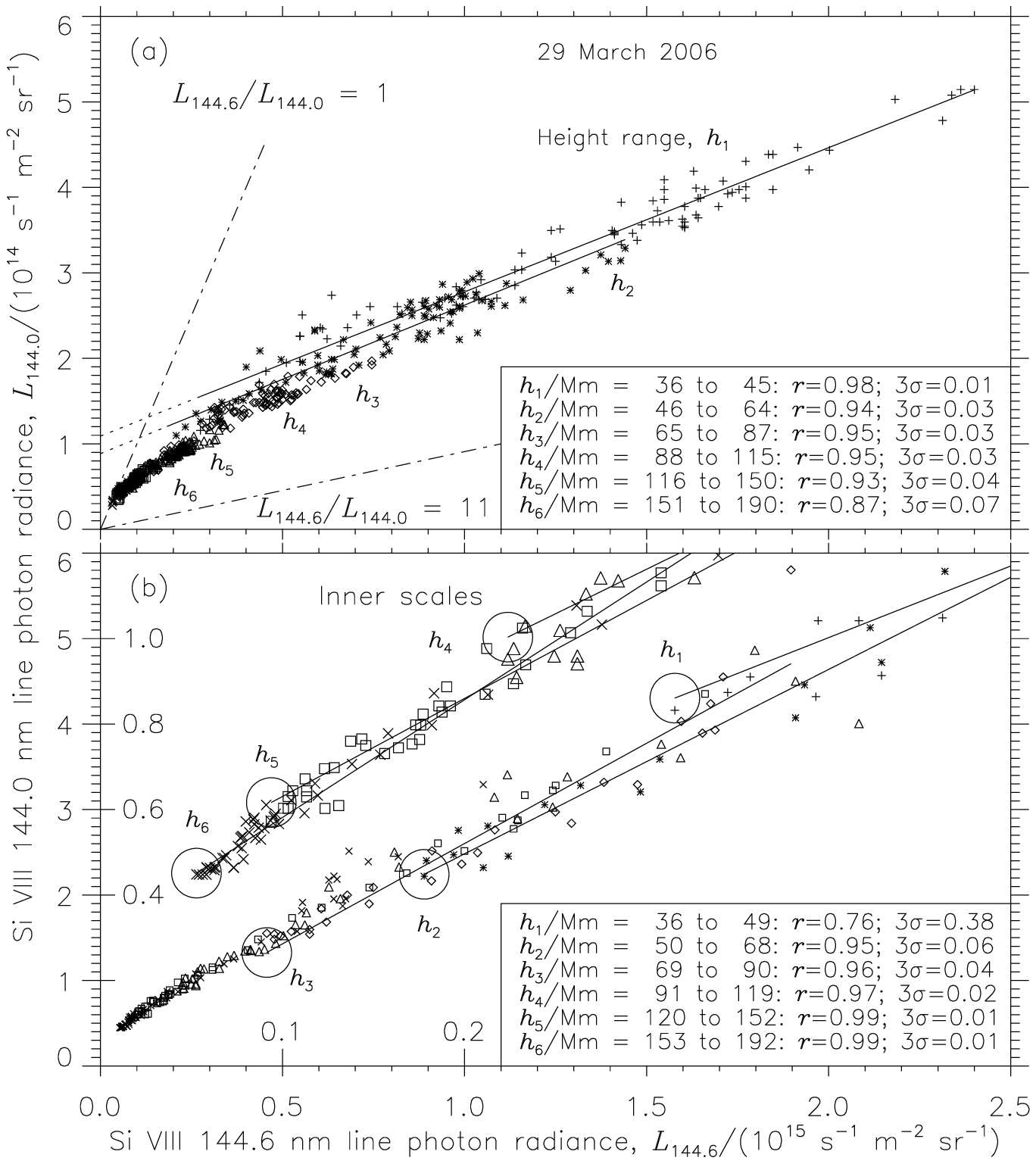} %{AAREV_06S.EPS}
\end{minipage}\hfill
\parbox[b]{3.2cm}{\caption{\label{FigS06} \small
Radiance of the Si\,{\sc viii} 144.0~nm line
as a function of the 144.6~nm radiance
(a) in tangential height ranges and (b) in concentric rings
(see Fig.~\ref{Figecl}).
The correlation coefficients, $r$, the
confidence levels and the height
ranges, $h_1$ to $h_6$, are given. Some of the linear fits are shown.
To highlight the crowded low-radiance portion of the diagram, it
is repeated in (b) increased by a factor of five
in both axes and with larger symbols.
The line labelled $L_{144.6}/L_{144.0} = 1$ shows the asymptotic value of the
ratio reached at $n_{\rm e} \approx 1 \times 10^6~{\rm cm}^{-3}$,
whereas  $L_{144.6}/L_{144.0} = 11$ corresponds to an electron density of
$n_{\rm e} = 1 \times 10^8~{\rm cm}^{-3}$.}}
\end{minipage}
\end{figure}

In an attempt to improve the
separation of plume and IPR plasma regimes with the help
of line-ratio observations, the radiances $L_{144.6}$ and $L_{144.0}$
measured with SUMER on 29 March 2006 are plotted in
Fig.~\ref{FigS06}(a) and (b)
averaged over certain height intervals per W-E step.
In the upper panel, linear fits are calculated from data points obtained along
tangential height ranges following a procedure developed earlier
for 2005 observations (Wilhelm 2006). In the lower panel, the same
measurements are organized according to concentric height ranges defined in
Fig.~\ref{Figecl}.
The highest range, $h_7$, has been omitted, because the data were too noisy.
In both plots, the extrapolations of the linear fits
do not pass through the origins of the diagrams as
indicated for the lowest heights, $h_1$ and $h_2$ in panel (a).
The only explanation for this behaviour
appears to be that more than one density regime is encountered at most of the
LOS directions. Under the assumption that the lowest radiances observed for each
height range\,---\, encircled in panel (b)\,---\,represent pure IPR plasma
(or at least the best estimate we can get),
the ratio of these radiances then provides the density of the IPR
plasma.\footnote{This
method has some similarities with the radial background subtraction employed by
DeForest and Gurman (1998).} The highly significant
regression lines allow us to conclude that there must be
(at a given height) another plasma with a rather
well-defined higher density as determined from the slopes.
A varying amount of this ``plume'' plasma is sampled at different
LOS positions. The variations caused by the reduced IPR contribution are not
significant, in particular, as the longest path length through plume material
is small relative to that of the low-density plasma
(cf., Sect.~\ref{Morphology}).

The observations taken in April 2007 led to diagrams very similar to those
shown in Fig.~\ref{FigS06}, and some of the 2007 results are included in
Table~\ref{Tabden}. Considering that atomic physics data could have
higher uncertainties than the measurements, the actual line ratios,
$\rho = L(144.6~{\rm nm})/L(144.0~{\rm nm})$,
observed in 2006 and 2007 are given in table footnotes
for a future re-evaluation should new conversion functions
become available.

\begin{figure}[!t]
% Figure 17
\includegraphics[width=\textwidth]{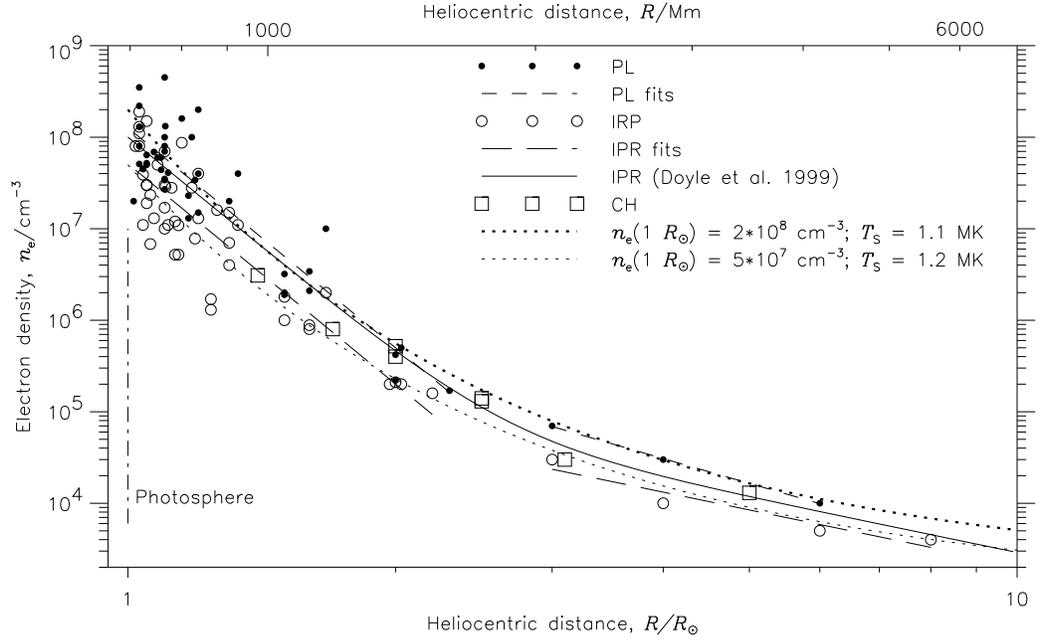} %{ELECDENSITY.EPS}
\caption{\label{Figden} \small Electron density measurements inside plumes
and IPRs plotted from data in Table~\ref{Tabden}, except for the values
at $R = 1~R_\odot$. Undifferentiated CH values are also shown.
The short and long dashed lines show power-law fits to
the values smaller and greater than $R = 2.5~R_\odot$ for plumes and IPRs,
respectively.
The solid curve is the density profile according to
Eq.~\ref{Eqden}. Density profiles of plumes
and IPRs under hydrostatic conditions are shown as dotted lines.
The initial conditions have been adjusted to produce best visual
fits to the profiles.}
\end{figure}

Electron densities range up to about $10^9~{\rm cm}^{-3}$ in plumes
(Young et al. 1999; Del~Zanna et al. 2003), with little
decrease over the first 70~Mm in height.
The IPR density at the base of the
corona is $\approx 10^8$ cm$^{-3}$ and falls sharply with height.
This can be seen from
Fig.~\ref{Figden}, where the data are plotted
versus heliocentric distance.
Results for $R = 1~R_\odot$ have been omitted, because
they are probably not directly related to the plume densities, but to the base
heating or an associated BP.\footnote{An example of such a situation
can be found in Fig.~3 of
Banerjee et al. (2009a).}
The data are plotted together with an empirical fit
to the electron densities
\begin{equation}
n_{\rm e} = \left[\frac{1 \times 10^8}{(R/R_\odot)^8} +
\frac{2.5 \times 10^3}{(R/R_\odot)^4} +
\frac{2.9 \times 10^5}{(R/R_\odot)^2}\right]\,{\rm cm}^{-3} \quad ,
\label{Eqden}
\end{equation}
derived by Doyle et al. (1999) for IPR conditions. Also shown are hydrostatic
density curves adjusted to the plume and IPR data points by selecting
appropriate base densities, $n_{\rm e,0} = n_{\rm e}(R_\odot)$, and
hydrostatic temperatures, $T_{\rm S}$:
\begin{eqnarray}
n_{\rm e}^{\rm PL,IPR}(R) =
n_{\rm e,0}^{\rm PL,IPR}\,\exp \left[\frac{G_{\rm N}\,M_\odot\,m_{\rm p}\,\mu}
{k_{\rm B}\,T_{\rm S}^{\rm
PL,IPR}}\,\left(\frac{1}{R} - \frac{1}{R_\odot} \right) \right] \quad ,
\label{Equilibrium}
\end{eqnarray}
where $G_{\rm N}$ is the gravitational constant, $M_\odot$
the mass of the Sun, $m_{\rm p}$ the proton mass,
$k_{\rm B}$ the Boltzmann constant and $\mu = 0.56$ for a fractional
helium abundance of $n_\alpha/n_{\rm p} = 5~\%$ (cf., Bame et al. 1977).
However, with the high outflow speeds that have been observed in IPRs, the
hydrostatic model does, in all likelihood, not provide a viable option for
that case.

\begin{figure}[!t]
% Figure 18
\begin{minipage}[b]{14.2cm}
\begin{minipage}[b]{10.5cm}
\includegraphics[width=10.5cm]{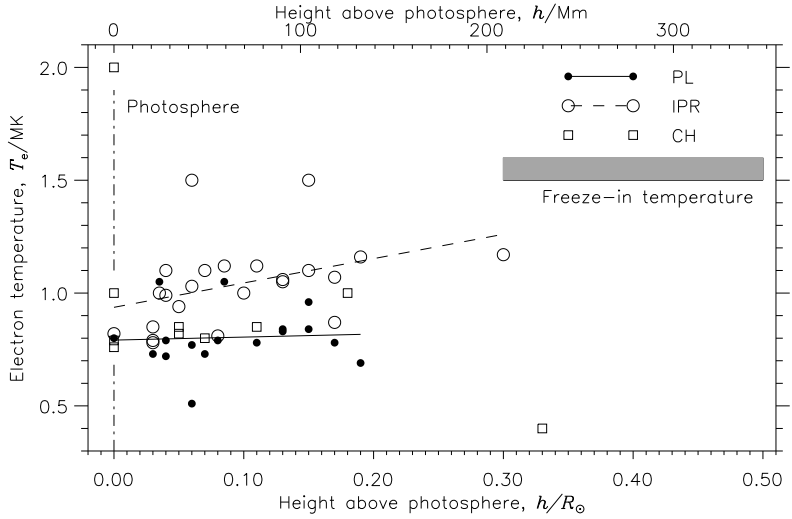} %{EL_TEM.EPS}
\end{minipage}\hfill
\parbox[b]{3.2cm}{\caption{\label{Figtem} \small
The electron temperatures summarized in Table~\ref{Tabtem} from line-ratio and
DEM studies are displayed together with the temperature range consistent with
charge-state measurements in the fast SW.
The CH temperature of 2~MK at $h = 0$ is probably related to a BP.
A linear fit of the plume data is plotted
as solid line and that for the IPR values as dashed line.}}
\end{minipage}
\end{figure}
\begin{figure}[!t]
% Figure 19
\includegraphics[width=\textwidth]{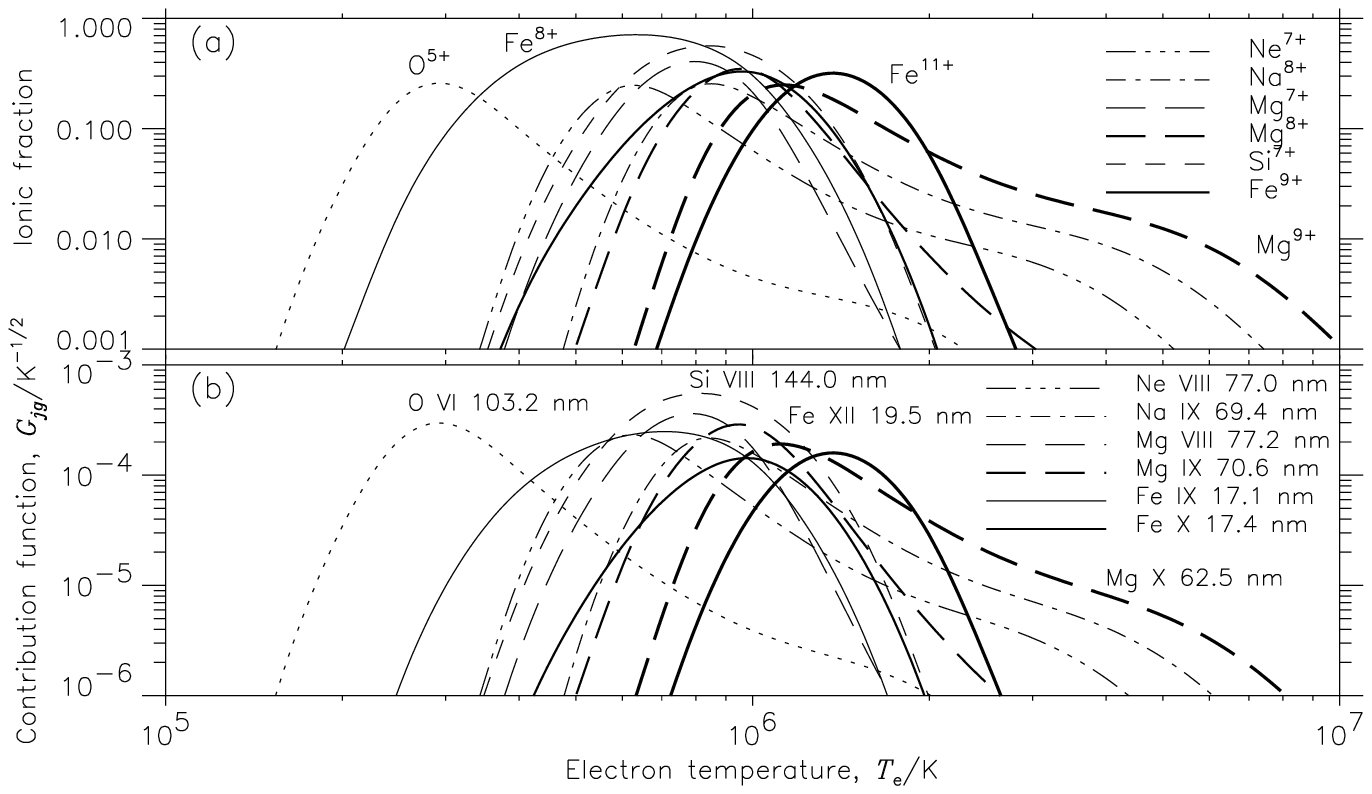} %{AAREV_CONTRIB.EPS}
\caption{\label{Figcon} \small (a) Ionic fractions of some ions
(treated in this review) in a plasma in ionization
equilibrium as a function of the electron temperature
(data from Mazzotta et al. 1998);
(b) contribution functions (see Eq.~\ref{Eqcon}) of
spectral lines from these ions in optically thin plasmas.
Note the long tails to high
temperatures of the lithium-like ions O$^{5+}$, Ne$^{7+}$, Na$^{8+}$ and
Mg$^{9+}$.}
\end{figure}

Considering that the various data points and the fits have been obtained with
the help of different methods and observations of many plumes and IPRs
spread over several
years in two sunspot minimum periods, it must be concluded that the excellent
agreement can only be explained, if CH conditions in general
are remarkably stable.
It therefore appears to be reasonable to assume standard plume and
IPR densities close to the graphs in
Fig.~\ref{Figden}
and examine, in particular,
the density ratio between these plasma regimes.
If we take the power-law fits at face value, we find a plume-IPR
density ratio of three below $R = 2~R_\odot$ with
$n_{\rm e} \propto (R_\odot/R)^8$
both in plumes and IPRs. Above $R = 3~R_\odot$,
it is
$n^{\rm PL}_{\rm e} \propto (R_\odot/R)^3$,
whereas the IPR density
$n^{\rm IPR}_{\rm e}$ is proportional to
$(R_\odot/R)^2$.
In the low corona,
the densities of plumes
and IPRs thus decline at the same rate. Higher up, the plume densities seem to
decrease faster than those of IPRs and the density ratio becomes smaller.
This ratio is four to seven in
the low corona from WL eclipse observations, but, in general, the value is near
two if obtained from LOS line-ratio studies, such as that in
Fig.~\ref{Figecl}. This discrepancy could be resolved
by the above consideration of density variations along the
LOS caused by two plasma regimes in
Fig.~\ref{FigS06},
leading to density ratios in agreement with WL observations.

% 5.2
\subsection{\small Plasma temperatures and non-thermal motions}
\label{Temperatures}

% 5.2.1
\subsubsection{\small Electron temperature}
\label{Electron}

Early electron-temperature determinations of CH plasmas have been
summarized by Habbal et al. (1993). Data obtained in wavelength ranges
from X-rays
to WL as well as charge-state measurements have been considered together
with relevant evaluation methods. The conclusion was that large uncertainties
have to be accepted and that $T_{\rm e}$ increases in CHs between $h > 0$ and
$h = 0.6~R_\odot$ from 0.8~MK to $\approx 1.3$~MK with some indication of higher
temperatures on the disk.

Typical temperatures found in CHs, IPRs and plumes (using methods described
below) during the last two decades are summarized in Table~\ref{Tabtem} and
Fig.~\ref{Figtem}.
It appears as if the temperature increase with height in CHs is mainly
related to IPRs, but is not typical for plumes.
\begin{table}[!th]
% Table~4
\noindent
\caption{Typical electron temperatures in CHs, plumes and IPRs at some heights}
\vspace{0.3cm}
\begin{small}
\begin{tabular*}{14.2cm}{ccccccc}
\hline
Height,  & \multicolumn{3}{c}{Temperature, $T_{\rm e}$/MK} &
Method & Date & Reference\\
$h/R_\odot$ & CH$^{\rm a}$ & PL & IPR$^{\rm b}$ & & &  \\
\hline
\hline
0.05 & 0.85 & & &  Mg\,{\sc x},  & 13 Dec 1973
& Habbal\\
&&&& Ne\,{\sc vii}$^{\rm c}$ & & et al. 1993\\
\hline
0.00 & 1.00 & & & DEM  & 25 Feb 1996 & Mason \\
&&&&(on disk) & & et al. 1997\\
\hline
0.07 & $0.8^{+0.07}_{-0.1}$  & & & O\,{\sc vi} 17.3~nm, & 21 May 1996 & David \\
0.18 & $1.0^{+0.3}_{-0.2}$  & & & 103.2~nm & SOHO roll & et al. 1998  \\
0.33 & $0.4^{+0.6}_{-0.3}$  & & & line ratio & & \\
\hline
0.03 & & 0.73 & 0.78 & Mg\,{\sc ix} & 24 Jan 1997 & Wilhelm \\
0.08 & & 0.79 & 0.81 & (70.6,75.0)\,nm & 15 Nov 1996 & et al. 1998  \\
0.17 & & 0.78 & 0.87 & line ratio & & \\
\hline
0.03 & & & 0.85 & Si\,{\sc viii}/Si\,{\sc vii}  &
Jul/Dec & Doschek \\
0.05 & & & 0.94 & line ratio & 1996 &  et al. 1998 \\
\hline
0.00 & 0.76 & & $\approx 0.82$ & Mg\,{\sc ix} 36.8 nm & Aug/ &
Fludra \\
0.05 & $0.82{\scriptstyle \pm 0.2}$ & & & Mg\,{\sc x} 62.5 nm & Sep 1996 &
et al. 1999 \\
0.11 & 0.85 & & & line ratio & & \\
\hline
0.00 & 2.00 & & & BP at base, & 25 Oct 1996 & Young \\
0.035 & & $1.05{\scriptstyle \pm 0.05}$ & 1.00 & discretized & &
et al. 1999 \\
0.085 & & $1.05{\scriptstyle \pm 0.05}$ & 1.12 & DEM & & \\
\hline
0.00 & & $0.80{\scriptstyle \pm 0.05}$ & & DEM  &
23 Aug 1996 & Del Zanna and \\
&&&&  (on disk) & & Bromage 1999 \\
\hline
0.00 & 0.79 & & & DEM & 10/11 Oct  & Del Zanna \\
&&&(on disk) &  & 1997 & et al. 2003\\
\hline
0.07 & & 0.73 & 1.10 & Mg\,{\sc ix} & 24 May 2005 & Wilhelm  \\
0.11 & & 0.78 & 1.12 & 70.6 nm, & & 2006 \\
0.19 & & 0.69 & 1.17 & 75.0 nm & & \\
\hline
0.03 & & & $0.79^{+0.1}_{-0.08}$  & EM & 3 Nov 1996 & Landi \\
0.10 & & & $1.00^{+0.1}_{-0.09}$  & DEM    & &  2008 \\
0.17 & & & $1.07^{+0.13}_{-0.1}$  &     &     & \\
\hline
0.3 & & & 1.16 & Mg\,{\sc ix} ratio & 1996/1997 & Del Zanna \\
&&&&&& et al. 2008 \\
\hline
0.06 ($h_1$) & & 0.77 (0.51) & 1.50 (1.03) & Mg\,{\sc ix} ratio &
28/29 Mar & this work \\
0.15 ($h_4$) & & 0.96 (0.84) & 1.50 (1.10) & tan. (con.)$^{\rm d,e}$
& 2006 &  \\
\hline
0.04 ($h_1$) & & 0.79 (0.72) & 1.10 (0.99) & Mg\,{\sc ix} ratio$^{\rm f}$ &
7/8 Apr & this work \\
0.13 ($h_4$) & & 0.84 (0.83) & 1.06 (1.05) & tan. (con.) & 2007 & \\
\hline
\end{tabular*}

\vspace{0.3cm}
$^{\rm a}$ possibly affected by plume plasma \\
$^{\rm b}$ below $h < 0.05~R_\odot$ some plume projections seem to
merge and could hide IPRs\\
$^{\rm c}$ Ne\,{\sc vii} (0.51~MK) \\
$^{\rm d}$ Ratios: tangential --
$\rho^{\rm tan}_{\rm PL}(h_1,h_4)$ = (6.52, 5.67);
$\rho^{\rm tan}_{\rm IPR}(h_1,h_4)$ = (4.53, 4.54); \newline
concentric --
$\rho^{\rm con}_{\rm PL}(h_1,h_4)$ = (9.04, 6.14);
$\rho^{\rm con}_{\rm IPR}(h_1,h_4)$ = (5.45, 5.16)\\
$^{\rm e}$ margins of correlation coefficients ($3\,\sigma$) given in
Fig.~\ref{FigM06}\\
$^{\rm f}$ Ratios:
$\rho^{\rm tan}_{\rm PL}(h_1,h_4)$ = (6.61, 6.15);
$\rho^{\rm tan}_{\rm IPR}(h_1,h_4)$ = (5.15, 5.29); \newline
$\rho^{\rm con}_{\rm PL}(h_1,h_4)$ = (6.99, 6.38);
$\rho^{\rm con}_{\rm IPR}(h_1,h_4)$ = (5.60, 5.33)
\end{small}
\label{Tabtem}
\end{table}
The electron temperature in the solar corona can be measured using
several different methods, most of them depend, at this stage, on the
observations of electromagnetic radiation from atoms and ions:
Under the assumption of ionization equilibrium, the ionic fractions of the
various species in a plasma
as a function of the electron temperature
can be determined (cf., e.g., Mazzotta
et al. 1998). The excitation and de-activation processes of the species also
depend on the electron temperature. For allowed transitions, spontaneous photon
emission, in general, is the dominant de-activation process. It is controlled by
the contribution function
\begin{eqnarray}
G_{jg}(T_{\rm e}) =
\frac{n_g}{n_{\rm X}}\,\frac{1}
{\sqrt{T_{\rm e}}}\,\exp\left(\frac{-\,{\mathrm\Delta\epsilon_{gj}}}
{k_{\rm B}\,T_{\rm e}}\right) \quad ,
\label{Eqcon}
\end{eqnarray}
where $n_{\rm g}/n_{\rm X}$ is the ionic fraction of the element~X,
if most of the ions are in the ground state and
$\mathrm\Delta\epsilon_{gj}$ the energy difference between
the states $g$ and $j$
(cf., Pottasch 1963; Gabriel and Jordan 1973; Mariska 1992;
Wilhelm et al. 2004).

The ionic fractions and contribution functions of some ions and their emission
lines of importance in this context are plotted in Fig.~\ref{Figcon}.
Most of the contribution functions have a pronounced maximum at an electron
temperature that is called the formation temperature, $T_{\rm F}$, of a spectral
line.
\begin{figure}[!t]
% Figure 20
\begin{minipage}[b]{14.2cm}
\begin{minipage}[b]{10.5cm}
\includegraphics[width=10.5cm]{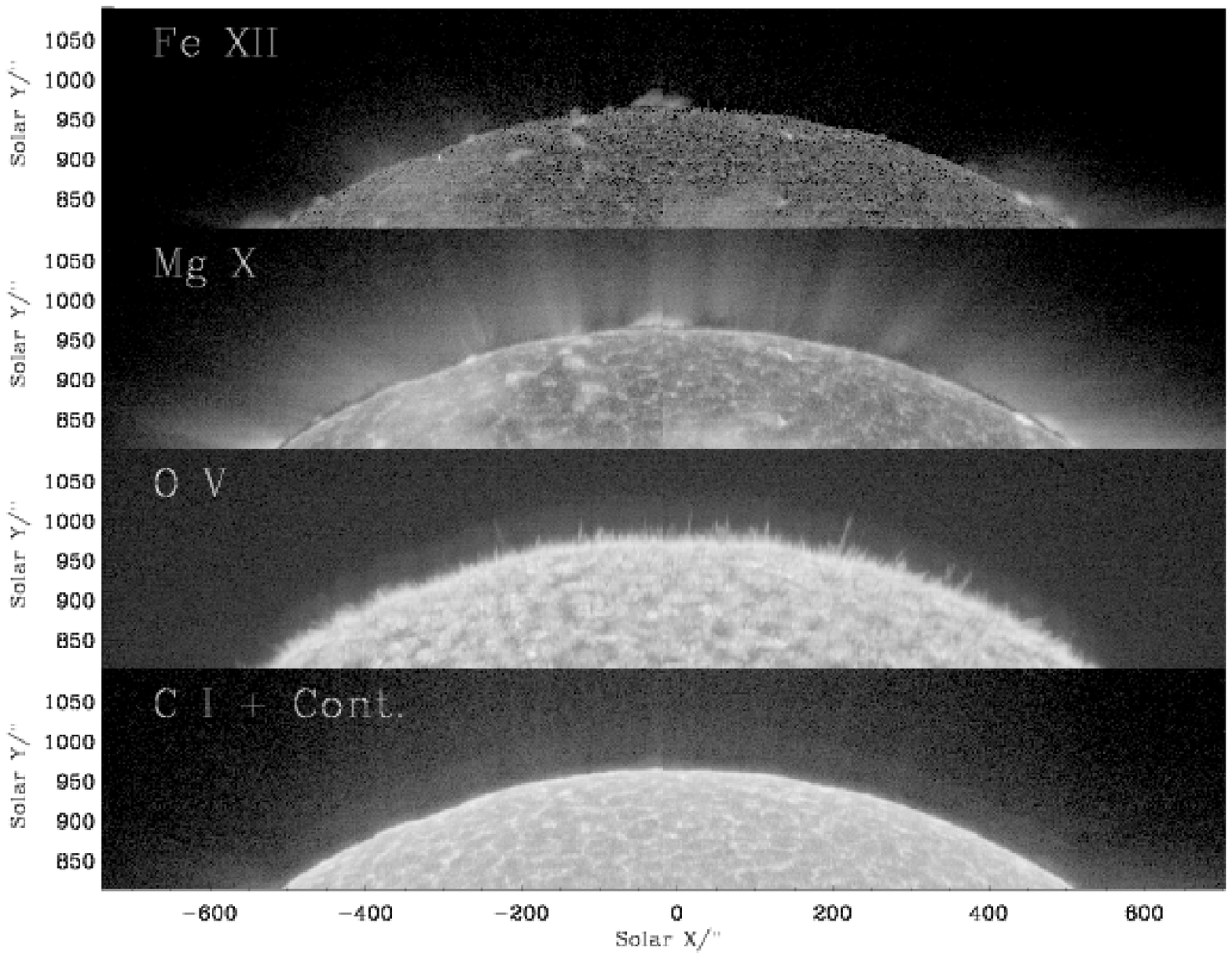} %{ALL_0996_NCH_KLAUS_BW.EPS}
\end{minipage}\hfill
\parbox[b]{3.2cm}{\caption{\label{FigVUV} \small
N polar region observed on 31~August / 1~September 1996 in four
spectral lines:
C\,{\sc i} ($T_{\rm F} \le 30\,000$~K; showing the chromospheric network in the
bottom panel),
O\,{\sc v} with network, spicules  and macrospicules
at the limb, no plumes, Mg\,{\sc x} showing CH, BPs
and plumes,
Fe\,{\sc xii} with CH and BPs and no plumes. See Feldman et al. (2003)
for further monochromatic CH images.}}
\end{minipage}
\end{figure}
The N polar region of the Sun is shown in
Fig.~\ref{FigVUV}
as four monochromatic images simultaneously obtained
in a raster scan with SUMER.
In the C\,{\sc i} image (shown without background subtraction) the
chromospheric network is prominent. It can also be seen in the
O\,{\sc v} image together with many short, irregular spikes of spicule and
macrospicule activity. Only the Mg\,{\sc x} image exhibits coronal plumes.
The knowledge of
$T_{\rm F}$ of the spectral lines allows us to conclude that plumes
are hotter than 0.24~MK and cooler than 1.38~MK.
Note in this context that the
plume structures seen in EIT 19.5~nm images are not related to Fe\,{\sc xii}
emission, but are the result of two Fe\,{\sc viii} lines at 19.47~nm and
19.6~nm
and other cooler iron lines (cf., Del Zanna and Bromage 1999;
Del Zanna et al. 2003).
By observing two or more spectral lines with different contribution
functions and formation temperatures, more information on the electron
temperature can be gained. A radiance ratio of the EIT bands 19.5~nm
(Fe\,{\sc xii}) and 17.1~nm (Fe\,{\sc ix}, Fe\,{\sc x}), for instance,
clearly shows a CH
temperature near 1~MK (Moses et al. 1997; Wilhelm et al. 2002b).
\begin{figure}[!t]
% Figure 21
\begin{minipage}[b]{14.2cm}
\begin{minipage}[b]{10.5cm}
\includegraphics[width=10.5cm]{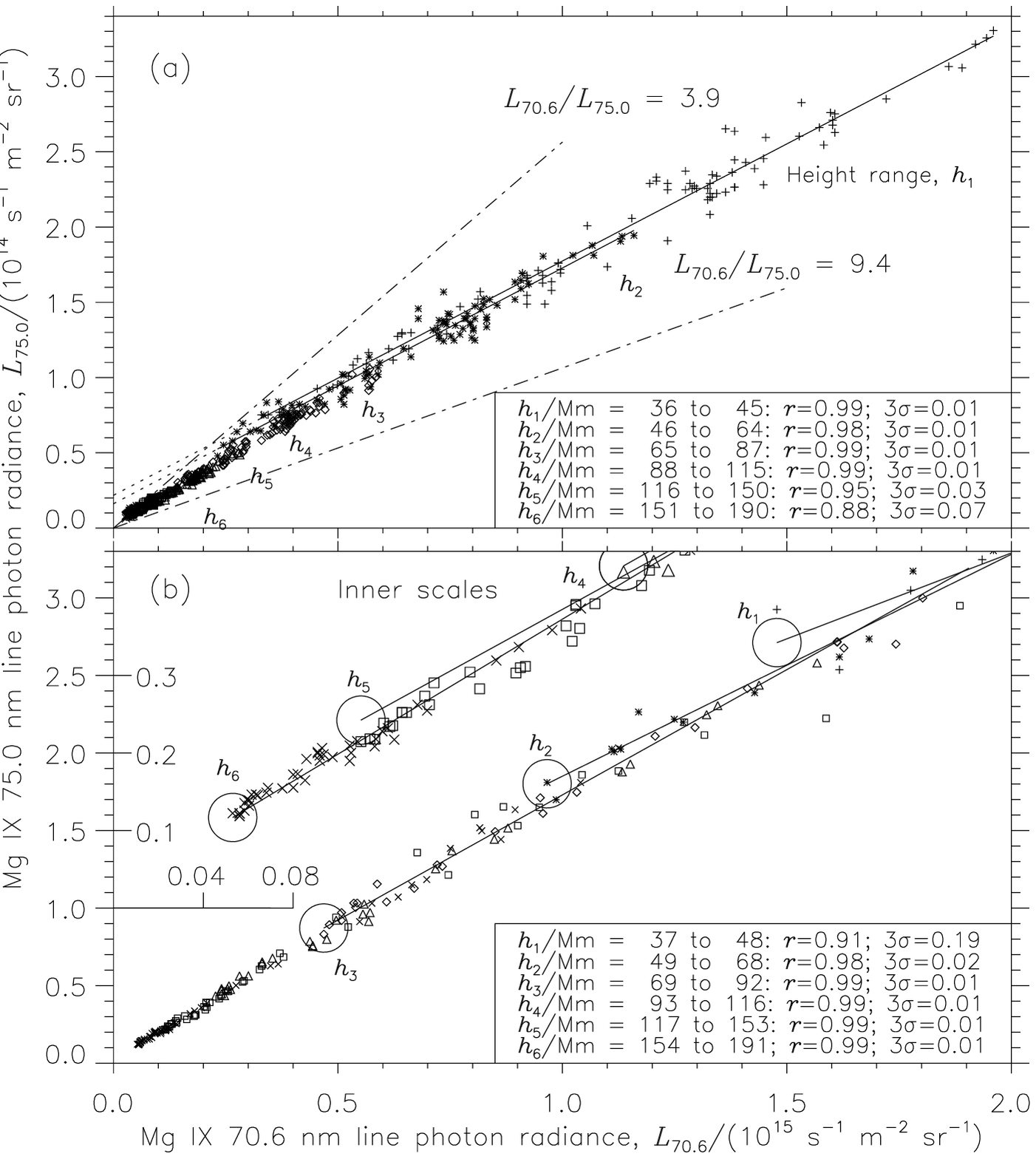} %{AAREV_06M.EPS}
\end{minipage}\hfill
\parbox[b]{3.2cm}{\caption{\label{FigM06} \small
Radiance of the Mg\,{\sc ix} 75.0~nm line as a function of
the 70.6~nm radiance on 29 March 2006.
The observations have been simultaneously obtained with
the Si\,{\sc viii} data in Fig.~\ref{FigS06},
and are arranged in tangential heights in panel (a) and
concentric ones in (b).
The correlation results are given in the inset.
The broken lines labelled $L_{70.6}/L_{75.0} = 9.4$ and 3.9
correspond to $T_{\rm e} \approx 0.5$~MK and 2~MK, respectively.
The lowest radiance values are encircled and taken as the best estimates for IPR
conditions. The low-radiance portion is repeated in panel (b)
enlarged by a factor of five
with larger symbols and a shifted scale of the ordinate.}}
\end{minipage}
\end{figure}
\begin{figure}[!b]
% Figure 22
\begin{minipage}[b]{14.2cm}
\begin{minipage}[b]{10.7cm}
\includegraphics[width=10.7cm]{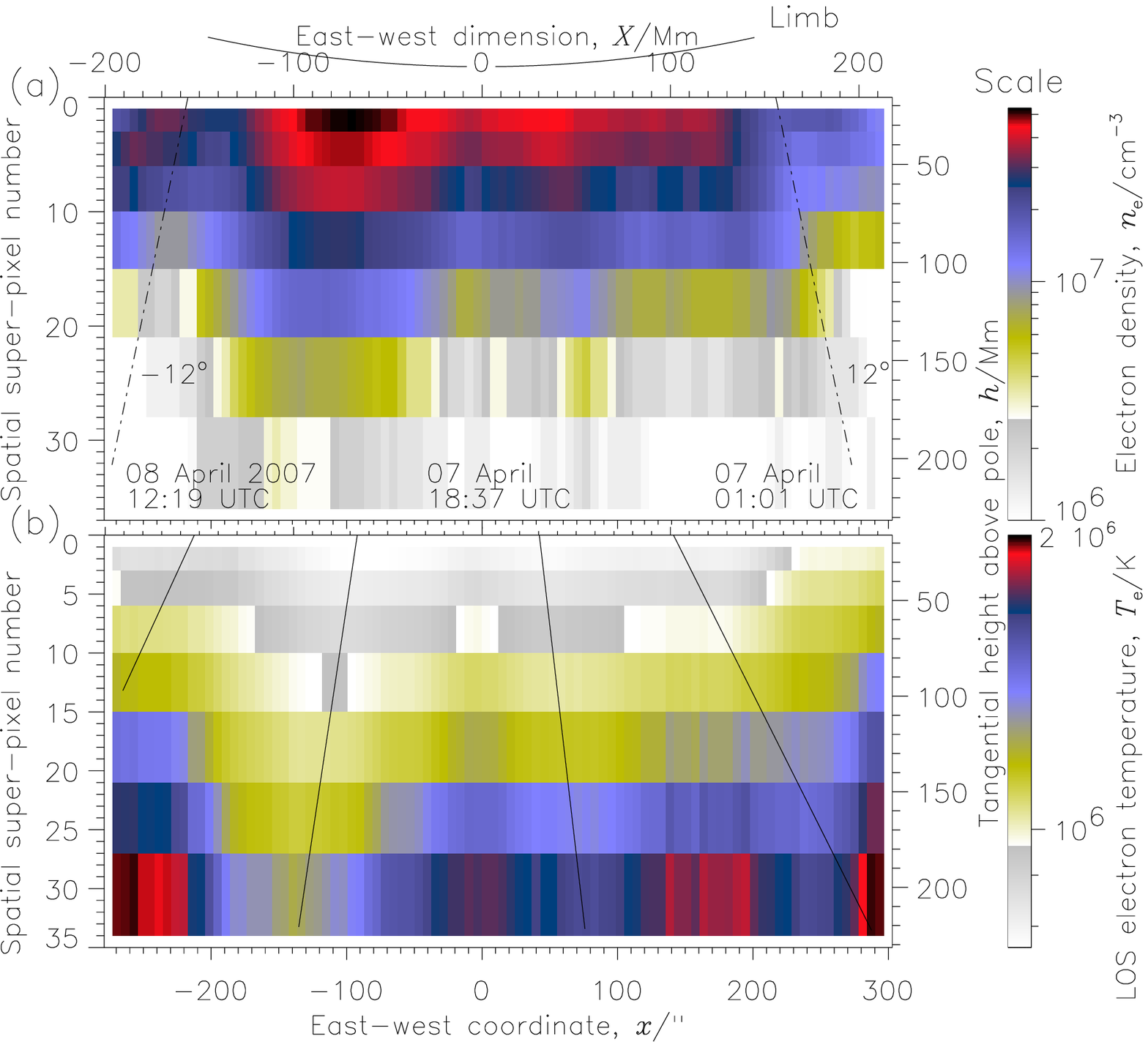} %{PLUMES_F8_2007.EPS}
\end{minipage}\hfill
\parbox[b]{3.2cm}{\caption{\label{FigM07} \small
(a) Electron density map of the southern CH in April 2007
obtained from the Si\,{\sc viii} line ratios.
(b) Electron temperatures simultaneously
measured in the same FOV with the Mg\,{\sc ix} line-ratio method. The weak
line at 75.9~nm required combining detector pixels to super-pixels and
averaging over relatively large height ranges. Some
plume projections are indicated that could be identified in
Fig.~\ref{FigNeNa},
where additional data are displayed at higher spatial
resolution (from Wilhelm et al. 2010).}}
\end{minipage}
\end{figure}

One of the methods to check whether the condition of thermal
equilibrium is fulfilled is to compare the effective ion temperatures
(cf., Sect.~\ref{Effective}) with the electron temperature estimated from
ionization equilibrium assumptions and to separate the ion thermal velocity
(cf., Sect.~\ref{motion}) from the observed line width by using two emission
lines of different atomic species (e.g., Seely et al. 1997; Imada et al. 2009).

Without strong temperature gradients along the LOS, an EM
analysis utilizing many lines and their contribution functions provides a
reliable electron temperature determination (cf., e.g., Landi 2008). If there
are temperature gradients, a differential emission measure (DEM)
procedure, where only the maxima of the contribution functions are considered,
leads to an electron temperature estimate (cf., e.g., Del Zanna et al. 2003).

Finally, line-ratio investigations have to be mentioned
(Flower and Nussbaumer 1975).
The radiance ratio of two lines (preferably) from the same ion is
temperature dependent, if the excitation energies are very different, e.g.,
O\,{\sc vi} 17.3~nm, 103.2~nm, or, if one line is excited from
a metastable level, e.g., in Mg$^{8+}$.
CH temperature
measurements using the O\,{\sc vi} ratio have been reported by David et al.
(1998), and plume as well as IPR results from the Mg\,{\sc ix} ratio
have been obtained by Wilhelm et al. (1998).
In observing temperature-sensitive line ratios, complications similar to
those for density-sensitive pairs arise (see Sect.~\ref{Densities}), if
the electron temperature along the LOS is not uniform. A plot of the radiances
of the Mg\,{\sc ix} 70.6~nm, 75.0~nm lines in
Fig.~\ref{FigM06}
therefore exhibits the same
characteristic features of the regression lines with a positive offset on the
ordinate. The diagram allows us to
estimate the Mg\,{\sc ix} ratios in plumes and IPRs in the various height
ranges with a method in analogy to the density determination.
However, the small variation of the slopes as a function of the relevant
temperatures from 0.5~MK to 2~MK leads to relatively large uncertainties.
The Mg\,{\sc ix} data
were recorded by SUMER in March 2006 simultaneously with the Si\,{\sc viii}
observations in
Fig.~\ref{FigS06}.
The conversion of the measured ratios into temperatures requires atomic physics
data. They were taken from Zhang et al. (1990) for the O$^{5+}$ ion and from
Keenan et al. (1984) for Mg$^{8+}$.

Recently, a
re-calculation for the Be-like Mg$^{8+}$ ion has been performed by Del Zanna et
al. (2008) according to which
the plume temperatures from the Mg\,{\sc ix} ratio would increase by
approximately 0.1~MK and the IPR values by $\approx 0.4$~MK. From the table, in
particular, if the revised Mg$^{8+}$ data would be taken into account,
it is clear
that the electron temperature in plumes is lower than in IPRs except at very low
heights near the base of a plume. These findings are in excellent agreement with
theoretical considerations (cf., Wang 1994; see Sect.~\ref{Models}).
They are also consistent with the so-called freeze-in temperatures determined
from charge-state measurements of SW ions, namely
$1.5~{\rm MK} \le T_{\rm e} \le 1.6~{\rm MK}$ in the height range
$0.3~R_\odot \le h \le 0.5~R_\odot$ (Ko et al. 1997; cf.,
Fig.~\ref{Figtem}).

The results of later campaigns in 2007 and 2008 have not
yet been analysed in detail,
but judged from first assessments they are in agreement with the 2006 and
also with the 2005 results. Some preliminary values for 2007 are included in
Table~\ref{Tabtem}. The LOS Si\,{\sc viii} and
Mg\,{\sc ix} ratios have been used to determine the electron densities and
temperatures in Fig.~\ref{FigM07} during the Hinode campaign in April 2007.
The diagram clearly shows low temperatures along the plume projections and
higher ones in IPRs.

\begin{figure}[!b]
% Figure 23
\begin{minipage}[b]{14.2cm}
\begin{minipage}[b]{10cm}
\includegraphics[width=10cm]{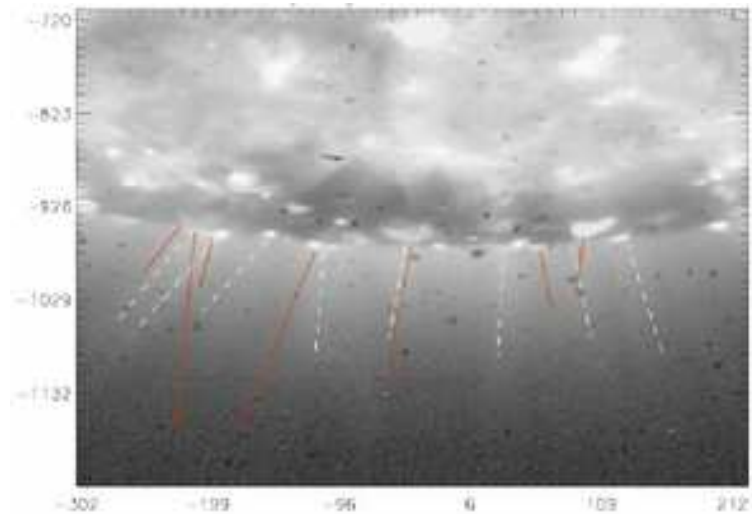} %{Giannina2.eps}
\end{minipage}\hfill
\parbox[b]{3.7cm}{\caption{\label{XRT} \small
XRT image with Al\_poly filter on 1 July 2008. The white
dashed lines are drawn
to guide the eye in plume identification.
Seven jets observed over a 3~h time interval are drawn as solid red lines.
The scales are in seconds of arc from the centre of the solar disk.
An association is apparent between plumes and jets.}}
\end{minipage}
\end{figure}

Fig.~\ref{XRT}
gives a sample Al\_poly image, recorded by XRT on 1 July 2008
in the time interval from 20:54 to 21:11~UTC. Similar images have been obtained
in the Al\_mesh channel, allowing us to build maps of the
Al\_poly-over-Al\_mesh radiance ratios in the FOV.
They can be converted into temperature maps with the help of a
calibration procedure (Golub et al. 2007).
Dashed lines are drawn
to guide the eye in the plume identification (the red
lines indicate jets as explained in Sect.~\ref{Jets}).
A comparison between plumes seen both in the TRACE 17.1~nm bandpass and
the ratio maps of XRT yields a one-to-one correspondence\,---\,at least
for the best visible, high contrast, structures.
This is not surprising, because the temperature response of this TRACE
channel peaks at a temperature just below 1~MK and
the XRT maps show plumes to be
at a temperature close to (but probably lower than) 1~MK.
The IPR plasma is slightly hotter than that of plumes,
at least in the low corona.
These results are consistent with those obtained with
different instrumentation and methods given in Table~\ref{Tabtem}.
We conclude from this preliminary analysis that
Hinode appears to offer valuable data
to advance our knowledge on plume and IPR plasmas, and
the association between plumes and jets (cf., Sect.~\ref{Jets}).

% 5.2.2
\subsubsection{\small Line profiles and effective ion temperatures}
\label{Effective}

First observations of broadenings of the Mg\,{\sc x} 60.9~nm, 62.5~nm lines
corresponding to velocities of
$V_{\rm 1/e} = 45$~km\,s$^{-1}$ at the base of the corona to
$\approx 55$~km\,s$^{-1}$ at $h = 0.2~R_\odot$
have been reported by Hassler et al. (1990) from a sounding rocket flight
on 17 March 1988. The quantity $V_{\rm 1/e}$ is the speed at which
the Doppler effect decreases the spectral radiance of a line
profile to 1/e of the peak\,---\,it can be considered as the most-probable
speed.
The spectrometers on SOHO and Hinode
can now measure the spectral profiles of many
emission lines in the VUV wavelength range on the solar disk and
in the corona out to several solar radii.
The speed, $V_{\rm 1/e}$, called Doppler velocity here,
is related to the Doppler width,
$\mathrm{\Delta} \lambda_{\rm D}$, and
the effective temperature, $T_{\rm eff}$, by
\begin{eqnarray}
\mathrm{\Delta} \lambda_{\rm D} = \frac{\lambda_0}{c_0}\,V_{\rm 1/e}
\label{Eqv1e}
\end{eqnarray}
and
\begin{eqnarray}
V_{\rm 1/e} = \sqrt{\frac{2\,k_{\rm B}\,T_{\rm eff}}{m_{\rm i}}} \quad ,
\label{Eqeff}
\end{eqnarray}
where $m_{\rm i}$ is the ion mass, $c_0$ the speed of light in vacuum and
$\lambda_0$ the rest wavelength of the spectral line. The
Doppler widths\footnote{For a Gaussian distribution, it is
$\mathrm{\Delta} \lambda_{\rm D} = \sigma\,\sqrt{2} = \mathrm{\Delta}
\lambda_{\rm FWHM}/(2\,\sqrt{\ln 2})$,
where $\sigma$ is the standard deviation.}
are of particular interest as they indicate temporally unresolved LOS motions
of the emitters (atoms or ions; see the next section for more details).

\begin{table}
% Table~5
\caption{Widths of spectral lines in CHs, plumes and IPRs as LOS
Doppler velocities, $V_{\rm 1/e}$}
\vspace{0.3cm}
\noindent
\begin{small}
\begin{tabular*}{14.2cm}{ccccccc}
\hline
Height$^{\rm a}$ & \multicolumn{3}{c}{Velocity,
$V_{\rm 1/e}/({\rm km\,s}^{-1})$} & Method & Date, & Reference \\
$h/R_\odot$ & CH & PL & IPR & & Period &   \\
\hline
\hline
0.50 & & $120{\scriptstyle \pm 4}$ &
$150{\scriptstyle \pm 8}$ & O\,{\sc vi} 103.2~nm & 6 Apr 1996 &
Antonucci \\
1.00 & & 220 & 270 & line width & & et al.\\
1.30 & & 240 & 300 & & & 1997\\
\hline
0.02 & 42 & & &  Mg\,{\sc x} & 20 Jul 1996 & Marsch \\
0.11 & 50 & & &  62.5 nm & & et al.\\
0.16 & 57 & & &  line width$^{\rm b}$ & & 1997\\
\hline
0.05 & & $44{\scriptstyle \pm 2}$ & $48{\scriptstyle \pm 2}$ &
O\,{\sc vi} 103.2~nm & 22 May 1996
& Hassler \\
&&&&&& et al. 1997 \\
\hline
0.50 & 90 & & & O\,{\sc vi} & 3 Jun 1996 & Kohl  \\
1.00 & $250{\scriptstyle \pm 25}$ & & & line width &  &  et al. 1998 \\
\hline
0.03 & & & 36 & Si\,{\sc viii} 144.6 nm & 4 Nov 1996 & Banerjee \\
0.12 & & & 46 & line width$^{\rm c}$ & & et al.1998 \\
\hline
0.06 & & 48 & 56 & O\,{\sc vi} 103.8~nm & 13 Mar 1996 & Wilhelm 1998 \\
\hline
0.05 & $39{\scriptstyle \pm 3}$ & & &
Si\,{\sc viii} 144.6 nm & 5 Nov 1996 & Tu \\
0.18 & $63^{+4}_{-5}$  & & & line width$^{\rm d}$ &
3 Oct 1996 & et al. 1998 \\
\hline
0.03 & & 56 (41) & 53 (41) & Mg\,{\sc ix}, & 24 Jan 1997
& Wilhelm \\
0.08 & & 59(44) & 50(45) & (Si\,{\sc viii}) & &  et al. 1998 \\
0.17 & & 60(47) & 62(50) & line widths & & \\
\hline
0.34 & & $60{\scriptstyle \pm 5}$ & $180{\scriptstyle \pm 25}$ &
O\,{\sc vi} 103.2~nm & Sep 1997
& Kohl \\
0.94 & & $60{\scriptstyle \pm 10}$ & $380{\scriptstyle \pm 30}$ &
line width & &  et al. 1999 \\
\hline
0.35 & $186{\scriptstyle \pm 22}$($104{\scriptstyle \pm 6}$)& & &
H\,{\sc i} Ly\,$\alpha$, (Mg\,{\sc x}) & Aug/Sep &
Esser \\
1.00 & $207{\scriptstyle \pm 25}$($200{\scriptstyle \pm 20}$) & & &
line widths$^{\rm e}$ & 1997 &  et al. 1999 \\
\hline
0.5 & 89(190) & & & O\,{\sc vi} 103.2~nm, & Nov 1996/ & Cranmer  \\
1.0 & 310(220) & & & (H\,{\sc i} Ly\,$\alpha$) & Apr 1997 & et al. 1999\\
1.5 & 420(240) && & line widths & &  \\
\hline
0.72 & & $167{\scriptstyle \pm 10}$ & $182{\scriptstyle \pm 15}$ &
O\,{\sc vi} 103.2~nm & 6 Apr 1996 & Giordano \\
     & &     &     & line width$^{\rm e}$ & & et al. 2000b\\
\hline
0.06 & & $54{\scriptstyle \pm 1}$ & $56{\scriptstyle \pm 1}$ &
O\,{\sc vi} 103.2~nm & 3 Jun 1996 & Banerjee \\
0.20 & & $62{\scriptstyle \pm 1}$ & $67{\scriptstyle \pm 2}$ &
line width & & et al. 2000a\\
0.28 & & $69{\scriptstyle \pm 1}$ & $75{\scriptstyle \pm 2}$ & & \\
\hline
0.60 & 110 & & & O\,{\sc vi} 103.2~nm & Feb 2001 & Miralles  \\
1.40 & 340 & & & line width & & et al. 2001 \\
1.85 & 429 & & & & & \\
\hline
0.06 & & 54 & 56 & O\,{\sc vi} 103.2~nm & 3 Jun 1996
& Teriaca \\
0.50 & & 80 & 90 & line width & & et al. 2003 \\
1.00 & & $270{\scriptstyle \pm 30}$ & $300{\scriptstyle \pm 30}$ & & & \\
1.50 & & 360 & 400 & & & \\
0.50 & & $160{\scriptstyle \pm 10}$ & $170{\scriptstyle \pm 10}$ &
H\,{\sc i} Ly\,$\alpha$ & & \\
1.00 & & $190{\scriptstyle \pm 15}$ & $200{\scriptstyle \pm 15}$ &
line width & & \\
\hline
\end{tabular*}
\label{TabDop}
Continued on next page\\
------------------

$^{\rm a}$ above the photosphere \\
$^{\rm b}$ tendency of wider profiles in IPR \\
$^{\rm c}$ $V_{\rm 1/e}$ calculated from non-thermal motion
$\xi^{\rm IPR} = (27, 39)~{\rm km\,s}^{-1}$ at $h_1$ and $h_2$
with $T_{\rm i} = 1$~MK \\
$^{\rm d}$ widths of other emission lines measured as well\\
$^{\rm e}$ $V_{\rm 1/e}$ calculated from $T_{\rm eff}$ \\

\end{small}
\end{table}

\addtocounter{table}{-1}

\begin{table}[!t]
%Table 1 continued
\noindent
\caption{continued}
\vspace{0.3cm}
\begin{small}
\begin{tabular*}{14.2cm}{ccccccc}
\hline
Height$^{\rm a}$ & \multicolumn{3}{c}{Velocity,
$V_{\rm 1/e}/({\rm km\,s}^{-1})$} & Method & Date, & Reference \\
$h/R_\odot$ & CH & PL & IPR & & Period &   \\
\hline
\hline
0.05 & & $\approx 35$ & 59 & Mg\,{\sc ix} 70.6~nm & 24 May 2005 &
Wilhelm 2006 \\
0.08 & & & 66 & line widths$^{\rm f}$ &  &  \\
0.13 & & & 69(48) & (Si\,{\sc viii} 144.6~nm) & & \\
\hline
0.0 & &45 & 45 &  Empirical model$^{\rm g}$ & 1996 & Raouafi et al. 2007b \\
0.5 & &50 & 125 &  O\,{\sc iv} & & \\
1.0 & &65 & 200 & & &\\
1.5 & &110 & 220  & & &\\
\hline
0.01 && 31 (33) & 38 (37)  & Fe\,{\sc xii} 19.5~nm & 10 Oct 2007 &
Banerjee  et al. 2009a\\
0.03 && 33 (38) & 39 (38)  & (Fe\,{\sc xiii} 20.2~nm) & & \\
0.15 && 45 (45) & 47 (49)  & line widths$^{\rm h}$ & & \\
\hline
\end{tabular*}

\vspace{0.3cm}

$^{\rm f}$ results of a comparison of observations with a forward model
calculation \\
$^{\rm g}$ the model yields
$V_{\rm 1/e}^{\rm PL} \approx V_{\rm 1/e}^{\rm IPR}$ for $h > 3~R_\odot$\\
$^{\rm h}$ $V_{\rm 1/e}$ calculated from non-thermal motions with
$T_{\rm i} = T_{\rm F}$

\end{small}
\end{table}

\begin{figure}[!t]
% Figure 24
\includegraphics[width=\textwidth]{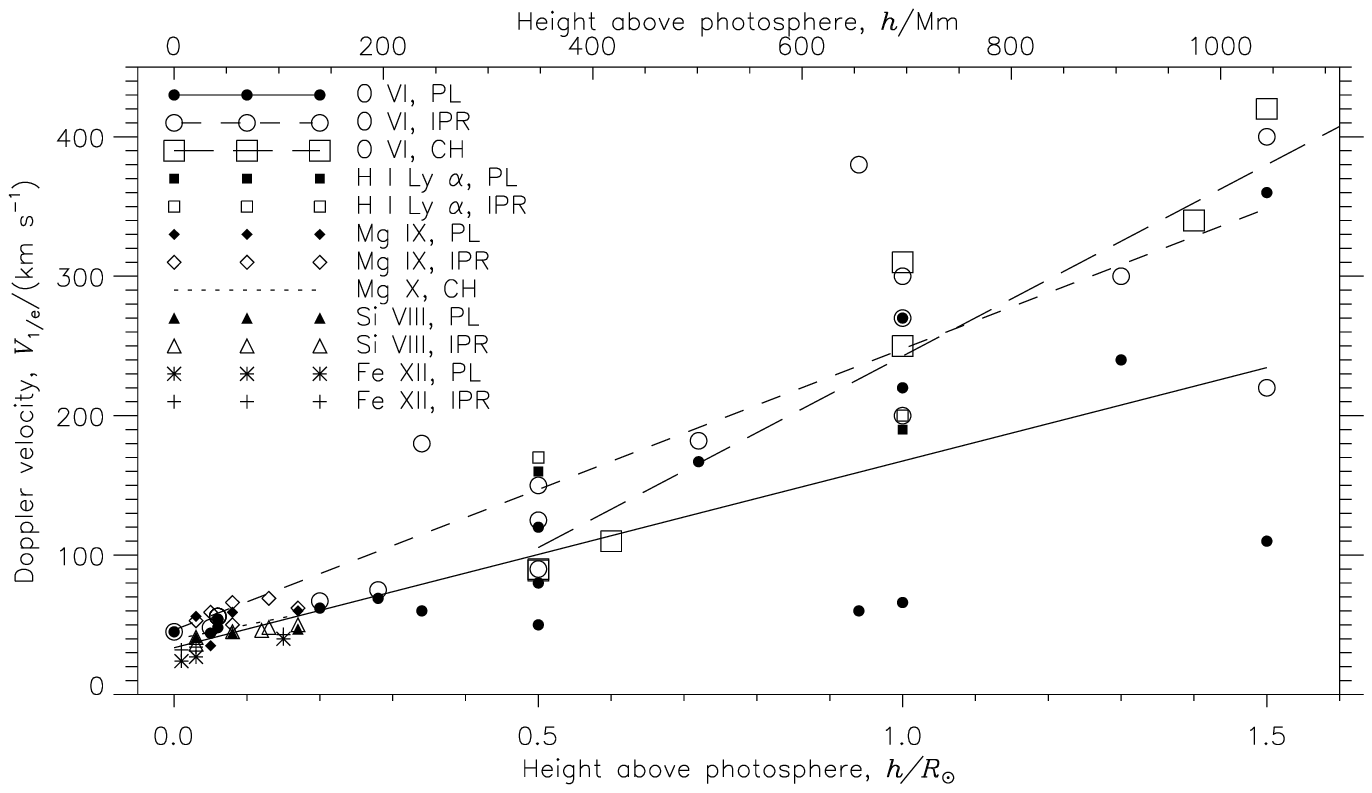} %{VDOPPLER.EPS}
\caption{\label{Figdop} \small Doppler velocities of prominent coronal emission
lines in plumes,
IPRs and CHs in general. Linear regression lines are given for four groups of
data points (O\,{\sc vi}: PL, IPR, CH; Mg\,{\sc x}: CH).
At low altitudes, $V_{\rm 1/e}$ is relatively constant near
$50~{\rm km\,s}^{-1}$, but with a positive slope.}
\end{figure}

As can be seen from the Table~\ref{TabDop} and Fig.~\ref{Figdop},
the line widths are, in general, smaller in
plumes than in IPRs. The relative decrease is $\approx 10$~\% to 20~\%
in many observations, but more pronounced variations also occurred.
The scatter of the data points is rather large, but the trends confirm
wider profiles in IPRs than in plumes.
In accord with this result,
the radiance of many emission lines were anti-correlated with the line
widths during the SOHO roll man{\oe}uvre on 3 September 1997 between
$h$ = 0.05~$R_\odot$ and 0.10~$R_\odot$ (Wilhelm and
Bodmer 1998). The widths increase with height both in plumes and IPRs.
Below $h = 0.1~R_\odot$, it is sometimes difficult to find pure
lane conditions as contamination by plume material might occur.
When only CH data are available, they seem to reflect the IPR widths.
High-spatial-resolution observations of the UV emission of a CH from
$R = 1.45~R_\odot$ to 2.05~$R_\odot$ at solar minimum were performed with
UVCS over an interval of 72~h. In the O\,{\sc vi} 103.2 nm radiance map
reconstructed from spectral data, four plumes can be identified
as bright features (Fig.~1 of Giordano et al. 2000b).
The line width in these plumes were narrower than in the IPRs.

A detailed study of coronal emission line profiles using UVCS observations
indicated some deviations from a Gaussian shape near the peak of many coronal
lines (Kohl et al. 1999). The H\,{\sc i}~Ly\,$\alpha$ line, however,
can be approximated
rather well by a Gaussian profile at all heights, whereas
the O\,{\sc vi} profiles are more complex: at low altitudes
(below $h \approx 0.35~R_\odot$) they
are nearly Gaussians, but more pronounced deviations occur at greater heights.
The O\,{\sc vi} profiles can be represented there by narrow and broad
components. The latter appears to be
associated with low-density IPRs, and the former with denser plumes
along the LOS. They
are mainly observed at the centre of the
line close to the rest wavelength, so that no
significant Doppler shifts are observed for these plume contributions.
The width of the narrow component hardly changes with height.
Beyond $R = 2.5~R_\odot$,
the line profiles are formed only by the broad component, and the narrow
component is no longer observed.

For SUMER observations from
low altitudes, the separation of a narrow plume line profile from a
somewhat wider IPR contribution is not easy. In a forward calculation, the
observed line widths in Mg\,{\sc ix} 70.6~nm could be reproduced under the
assumption of a constant $V_{\rm 1/e} = 35$~km\,s$^{-1}$ over a
height range from
0.05~$R_\odot$ to 0.13~$R_\odot$ and the IPR widths nearly twice as wide
(see Table~\ref{TabDop}).
This result would explain the observations by Doschek et al. (2001) indicating
line-width increases with height for Si\,{\sc viii} 144.6~nm on 3 May 1996 and
no effect for several ions on 3 November 1996, if the SUMER slit was aligned
with an IPR in May and with a plume in November. It should be mentioned here
that Banerjee et al. (2000a) and Xia et al. (2004)
found wider O\,{\sc vi} line widths in plumes (or
near their footpoints) than in darker regions on the disk.
We also have to note that the
$V_{\rm 1/e}$ speeds of O$^{5+}$ ions are much higher than those of
Si$^{7+}$ in IPR (Banerjee et al. 2000a), and that magnesium ions have
higher speeds
than hydrogen (Esser et al. 1999) or Si$^{7+}$ (Wilhelm et al. 1998; Wilhelm
2006). Finally, O$^{5+}$ exhibits about twice the difference in $V_{\rm 1/e}$
between plumes and IPR than hydrogen at $h \approx$ 0.5~$R_\odot$ to
1~$R_\odot$ (Giordano et al. 1997).

% 5.2.3
\subsubsection{\small Ion temperatures and non-thermal motions}
\label{motion}

The Doppler width, $\Delta \lambda_D$, of an emission line is caused by the
thermal motions along the LOS of the atom or ion under study and the non-thermal
contributions, which are either turbulence or temporally unresolved waves.
We can thus write
\begin{eqnarray}
\Delta \lambda_{\rm D} =
\frac{\lambda_0}{c_0}\,\sqrt{\frac{2\,k_{\rm B}\,T_{\rm i}}{m_{\rm i}}
+ \xi^2} \quad ,
\label{Eqxi}
\end{eqnarray}
where $T_{\rm i}$ is the ion
temperature and $\xi$ the non-thermal speed (cf., Mariska 1992).

Under the assumption of $T_{\rm i} =1$~MK for Si$^{7+}$ ions,
Banerjee et al. (1998)
derived non-thermal motions of 27~km\,s$^{-1}$ at 0.03~$R_\odot$ above the
limb in PCHs and 46~km\,s$^{-1}$ at 0.26~$R_\odot$. Both
values were obtained in IPR. An ion temperature of 2~MK did not lead to a
consistent result.
Hassler et al. (1997) and Banerjee et al. (2000a) put forward the
hypothesis that the ion temperatures in plumes and IPRs are not very
different, but that higher non-thermal motions are present in IPRs.

The non-thermal speed is related to the wave amplitude by
$\xi^2 = \langle \delta v^2 \rangle /\zeta$ (with $\zeta \approx 2$
depending on the polarization and LOS), if waves are
responsible for this contribution (cf., e.g., Hassler et al. 1990).
Variations of $\xi$ as function of height above the limb
have been calculated from EIS observations in the lines Fe\,{\sc xii} 19.5~nm
and Fe\,{\sc xiii} 20.2~nm (Banerjee et al. 2009a).
Since the separation of $V_{\rm 1/e}$ into $T_{\rm i}$ and $\xi$ is not a
straightforward procedure\,---\,as discussed in detail by
Dolla and Solomon (2008)\,---\,we include
in Table~\ref{TabDop} only LOS Doppler velocities directly derived
from line-width measurements or calculated from $\xi$ and $T_{\rm i}$ data in
the literature.

% 5.3
\subsection{\small Elemental abundances and first ionization potentials}
\label{Abundances}

The composition of the solar photosphere can, with the exception of the noble
gases, be directly determined from spectroscopic observations
(cf., Grevesse and Sauval 1998).
In the corona,
attempts to measure the helium abundance have been made by Gabriel et al. (1986)
and Laming and Feldman (2001); for argon and neon abundances see Phillips et al.
(2003) and Landi et al. (2007).
Methods of measuring the solar and coronal abundances
of an element~X have been reviewed by von Steiger et al. (2001).
In the corona, the abundance of elements is
varying and the so-called FIP effect seems to play a dominant
r$\hat{\rm o}$le.
Elements with a FIP of $I_{\rm X} < 10$~eV are defined as low-FIP elements and
those with FIP of $I_{\rm X} > 10$~eV as high-FIP elements,
separated by the photon
energy $h\,\nu = 10$~eV of the H\,{\sc i} Ly\,$\alpha$ line.
The low-FIP elements are\,---\,with respect to photospheric
values\,---\,overabundant in the (equatorial) corona
by a factor of about four to five relative to the high-FIP elements
(cf., Widing et al. 2005).
The magnesium/neon abundance ratio in plumes is enhanced
relative to IPRs by a factor of 1.5 (Young et al. 1999)
to 1.7 at $h = 0.05~R_\odot$ and 3.5 at $0.1~R_\odot$ (Wilhelm and Bodmer 1998).
These abundance determinations depend on the electron temperatures
employed, and might have to be re-evaluated  should better temperature data
become available.
According to Feldman et al. (1998), there was no significant FIP effect
observed at $h = 0.03~R_\odot$ above the northern solar limb on
3~November 1996;
these data were taken by SUMER in an IPR of a CH (cf., Landi 2008). If the IPRs
are indeed the source regions of the fast SW, no composition changes would be
expected in the high-speed streams in accordance with observations,
except probably for helium with its long first ionization time (FIT)
(Geiss et al. 1995).

\begin{figure}[!t]
% Figure 25
\begin{minipage}[b]{14.2cm}
\begin{minipage}[b]{10.2cm}
\includegraphics[width=10.2cm]{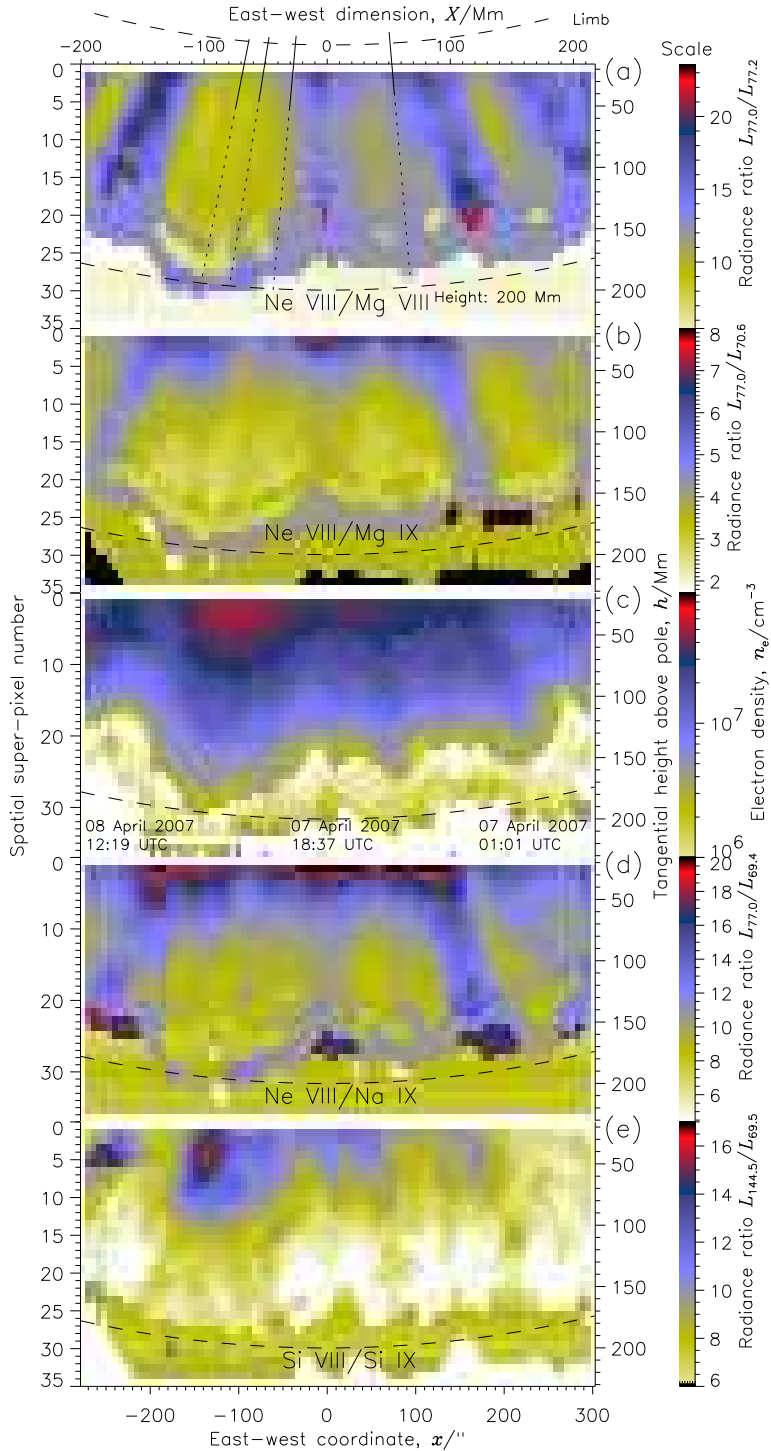} %{AAREV_BANG.EPS}
\end{minipage}\hfill
\parbox[b]{3.7cm}{\caption{\label{FigNeNa} \small
Ratios of photon radiances for several spectral lines as indicated in the
panels (a), (b), (d) and (e).
The maps obtained in one raster scan of 96 steps from right to left with SUMER
show the same region of the southern CH in April 2007. From the limb on
top of the figure, some projections of plume reconstructions from EUVI
stereoscopic observations are indicated continuing into
panel~(a) (cf., Curdt et al. 2008).
The electron densities in panel~(c) are converted from the LOS
Si\,{\sc viii} 144.6~nm, 144.0~nm ratios.
Beyond $\approx 150$~Mm altitude, the maps are unreliable, because of noise
contributions from the weaker lines of the pairs.}}
\end{minipage}
\end{figure}

Observations  of the Ne\,{\sc viii} 77.0~nm and Mg\,{\sc viii} 77.2~nm lines
in April 2007 during the most recent solar minimum
showed a very clear plume/IPR structure
in a map of the radiance ratio $L_{77.0}/L_{77.2}$ (Curdt et al. 2008), which
is repeated in
Fig.~\ref{FigNeNa}(a).
Any doubt that this effect is caused by the high-temperature tail of the
lithium-like Ne\,{\sc viii} contribution function (see
Fig.~\ref{Figcon})
can be dispelled by inspecting
Fig.~\ref{FigNeNa}(b) and (d)
showing radiance ratios of the Ne\,{\sc viii} line
with the Mg\,{\sc ix} and Na\,{\sc ix} 69.4~nm (0.85~MK)
lines.\footnote{Temperature
effects, however, cause the high Ne\,{\sc viii}/Mg\,{\sc ix} and
Ne\,{\sc viii}/Na\,{\sc ix} ratios at low altitudes in panels~(b) and (d), and
the increase of the Ne\,{\sc viii}/Mg\,{\sc viii} ratio in panel~(a) above
120~Mm in the hot IPRs.}
Although these lines have different contribution
functions compared to Mg\,{\sc viii}, the plume/IPR patterns are basically the
same. In particular, as the contribution function of Na\,{\sc ix}
is always larger
than that of Ne\,{\sc viii} above 0.8~MK,
it must be concluded that a FIP effect in plumes
operates with an enhancement of the magnesium abundance by a factor of 1.5 to 2
relative to neon,
and that the relative abundance of the low-FIP element sodium is
enhanced as well.
A verification that there are higher electron temperatures
in IPRs than in plumes follows from
Fig.~\ref{FigNeNa}(e) displaying the
$L$(Si\,{\sc viii})/$L$(Si\,{\sc ix}) radiance ratio.

Widing and Feldman (2001) found in active regions (AR) that the
confinement time of a plasma
is a decisive parameter for abundance variations.
A FIP bias of nearly ten was reached
after $\approx 6$~d. If these findings can be applied to plumes, we would
expect confinement times of a day or so\,---\,not too different
from plume and BP lifetimes.

\section{\small Relations of plumes to other solar phenomena}
\label{Relations}

Coronal plumes are rooted in the lower layers of the solar atmosphere. It is
essential to identify the relationship with features in these regions:
-- What is the association of plumes with the magnetic
network of the Sun?
-- What is the relationship between plumes and coronal BP?
-- Are magnetic field configurations, called ``rosettes'', around
plumes necessary for their formation and/or existence?
-- Is there any signature of plumes expected in
the Heliosphere?

% 6.1
\subsection{\small Chromospheric network}
\label{Network}

The reconnection processes in the complex magnetic network that are
thought to be required for the generation and acceleration of the SW
have been described as microflares by Axford and McKenzie (1992).
The fine structure of the network is of particular importance for the physics of
CH. Based on imaging results and the element composition of TR plasmas,
Feldman et al. (2001) concluded that it is likely that this plasma is confined
in magnetic structures and has no direct interface with the chromosphere.

High-frequency Alfv\'en waves produced in the network have been discussed by
Marsch and Tu (1997) in a SW acceleration and coronal heating model. In
addition to a steady coronal-funnel flow, solutions have been found with a
standing shock at a height of $h \approx 8$~Mm.
High densities and low outflow speeds are characteristic for the plasma
above the shock and might indicate plume conditions.
Such a process could operate at some stage of plume formation, however,
the higher electron and proton temperatures relative to the unshocked case
are in agreement with observations only, if an inversion occurs at greater
heights (cf., Sect.~\ref{Electron}).
Shock formation of slow magnetosonic waves in plumes is discussed by
Cuntz and Suess (2001), who showed that it is expected to take place below
$h = 0.3~R_\odot$. Numerical simulations show that the shock formation
can be suppressed, or can occur considerably higher when realistic compressive
viscosity is taken into account for slow magnetosonic waves with observed
amplitudes and periods (Ofman et al. 1999).

The relation of the chromospheric network to plumes has already been mentioned
in Sects.~\ref{Morphology} and \ref{Dynamics}.
Here we want to call attention to some
compressional wave observations in the lower solar atmosphere with dominant
periods of 25~min and an occasional downward direction in inter-network
regions, whereas only upward propagations prevailed within the network
(Gupta et al. 2009). Evidence of downward propagating waves in the TR had
earlier been reported by Judge et al. (1998).

In Sect.~\ref{Morphology}, it has been outlined that coronal plumes
arise from footpoints 2~Mm to 4~Mm wide in the network and expand
rapidly with height to diameters of 30~Mm (see Fig.~\ref{CH_QS_mag}).
Plumes are associated with
magnetic flux concentrations in the super-granular network boundaries.
Note in this context that the typical size of a super-granule agrees with the
typical diameter of \emph{beam} plumes at the base of the corona. However,
not all of the flux concentrations give
rise to coronal plumes. The plume footpoints are, in general, magnetically
rather complex\,---\,often related to BPs (see next section) and
rosettes (cf., Sect.~\ref{Magnetic}).
Plumes are thought to result from heating processes taking
place when unipolar magnetic flux concentrations reconnect
with oppositely directed fields of ephemeral
emergent loops (cf., e.g., Wang 1994; Wang and Sheeley 1995).
This scenario finds support in the high temperatures
observed at the footpoints of plumes yielding high-pressure plasma
that can diffuse along the magnetic field lines (van de Hulst 1950b).

% 6.2
\subsection{\small Bright points}
\label{BP}

The association of coronal BP with plumes is particularly intriguing,
because all possible combinations
have been found: plumes with BPs at their base, plumes without
BPs and BPs without plumes. Following a suggestion of Del~Zanna et al. (2003),
we will assume that BPs are typical features near plume footpoints only in the
early phases of plume formation. At that stage, high temperatures have been
observed at the base of plumes (see Sect.~\ref{Electron}),
whereas in a later phase\,---\,without BP\,---\,temperatures
of less than 1~MK prevail at all heights. BPs without plume, on
the other hand, have probably to be considered as precursors of plume formation.
Doppler velocity observations of BPs also gave complex results: blue and red
shifts of several kilometers per second or no shift are reported (Wilhelm et al.
2000; Madjarska et al. 2003; Popescu et al. 2004). The different findings
might not only be related to the stages of evolution of the BPs,
but also to differences in the formation temperatures of the emission
lines used in the studies.

As another observational fact it should be mentioned that compact 17~GHz radio
enhancements have been found with the Nobeyama Radioheliograph at the footpoint
of a plume with BP (characterized by Fe\,{\sc xii} emission),
but not in the plume structure above the base (Moran et al. 2001).

% 6.3
\subsection{\small Spicules, macrospicules and jets}
\label{Jets}

The mass supply to the corona and the SW\,---\,estimated
to be 1~\% of the spicular mass flux by Pneuman and Kopp (1978)\,---\,poses
a problem, because all TR studies (e.g., Doschek et al. 1976; Dere et al. 1989;
Brekke et al. 1997; Chae et al. 1998) demonstrated strong red shifts of
spectral lines and thus downflows of material, whereas the spectroscopic
signatures of outflows in spicules were only present in blue wings of spectral
lines, for instance, of N\,{\sc v} 124.3~nm,
indicating maximum LOS velocities of
150~km\,s$^{-1}$ without significant shift of the main profile (Wilhelm 2000).
Such profiles have also been observed with SUMER in other spectral lines (De
Pontieu et al. 2009), and near footpoints of AR loops
with EIS in Fe\,{\sc xiv} 27.4~nm (Hara et al. 2008).
Spicules seen in VUV emission lines with
formation temperatures between $T_{\rm F} = 30\,000$~K and 0.6~MK have
larger diameters at higher temperatures\,---\,interpreted as a signature of
evaporation (Budnik et al. 1998).
A direct relationship between spicule
activity and plumes could not be established (see also Fig.~\ref{FigVUV}).

\begin{figure}[!t]
% Figure 26
\begin{minipage}[b]{14.2cm}
\begin{minipage}[b]{10cm}
\includegraphics[width=10cm]{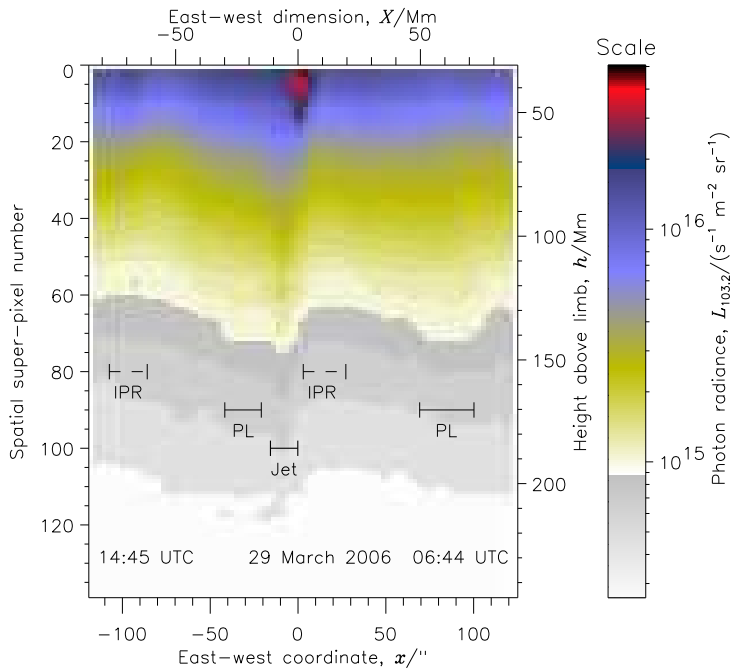} %{AAREV_JET.EPS}
\end{minipage}\hfill
\parbox[b]{4cm}{\caption{\label{Figjet} \small A jet seen in the
southern CH on 29 March 2006
during the total eclipse campaign embedded in plume and IPR structures.
The jet was recorded in O\,{\sc vi} line at
103.2~nm and 103.8~nm (not shown) between 10:26 and 11:23~UTC.
The ratio $L$(103.2~nm)/$L$(103.8~nm) was 2.3 in the centre of the jet at 70~Mm
height, just left of it in plume plasma the value was 2.35 and 2.4
to the right in an IPR. The same event is shown in Fig~\ref{Lybia} as green
rectangle. A super-pixel is equivalent to two detector pixels for this strong
line.}}
\end{minipage}
\end{figure}

The relation between jets and plumes is not clear at this stage.
From the rapid density decrease with height, van de Hulst (1950b)
concluded that ejected plasma could not produce polar rays (as he called
plumes). A fast jet with a speed of 400~km\,s$^{-1}$ in a plume assembly
displayed distinctly different characteristics from the surrounding plumes
(Moses et al. 1997). Similar observations of a jet related, in addition,
to a macrospicule obtained during a multi-spacecraft campaign in
November 2007 are reported by Kamio et al. (2010).
Lites et al. (1999) described a narrow jet embedded in a wider plume
from WL and EUV observations during the 1998 eclipse.
The outward propagation speed was
$\approx 200$~km\,s$^{-1}$. The acceleration of material
is favoured over a wave perturbation scenario in this case.
The acceleration could possibly be caused by explosive reconnection events
(cf., Dere et al. 1991; Innes et al. 1997).
Explosive events observed in chromospheric and
TR lines did not show any detectable signature in the coronal
Mg\,{\sc x} 62.5~nm line (Teriaca et al. 2002; Doyle et al. 2004).
However, the latter authors report their detection in the TRACE 17.1~nm
channel. Thus a
direct generation of the coronal plasma by the explosive event can be excluded;
however, an indirect process must operate in order to heat the plasma.

Correlated observations
of EUV and WL jets with EIT and LASCO near the limb and at several
solar radii showed that the bulk of the material travelled with a speed of
$\approx 250$~km\,s$^{-1}$ much lower than the injection speeds (Wang et al.
1998). The jets originated near BPs and, in many cases,
in the vicinity of plumes.
Detailed studies of six polar jets observed in H\,{\sc i} Ly\,$\alpha$ and
O\,{\sc vi} by UVCS (some of them also seen by LASCO and EIT) indicated a
relative decrease in line width during the jet brightenings comparable
to the plume/IPR ratio (Dobrzycka et al. 2002).
A pronounced plume-like brightening was observed in the northern CH during
the total eclipse on 29 March 2006 from five different sites
(Pasachoff et al. 2008). It was therefore
possible to determine an outward propagation speed of $\approx 65$~km\,s$^{-1}$
in the height range from $h$ = 0.07~$R_{\rm \odot}$ to 0.27~$R_{\rm \odot}$.
Based
on the available data, the authors could not decide whether the brightening was
of a wave-like nature or a material jet. More or less simultaneously, a jet was
observed in the southern CH by SUMER in the O\,{\sc vi} 103.2~nm,
103.8~nm lines. It is shown in Fig.~\ref{Figjet}
together with plumes in the neighbourhood of the jet, which has a
significantly different structure.

\begin{figure}[!t]
% Figure 27
\begin{minipage}[b]{14.2cm}
\begin{minipage}[b]{10.5cm}
\includegraphics[width=10.5cm]{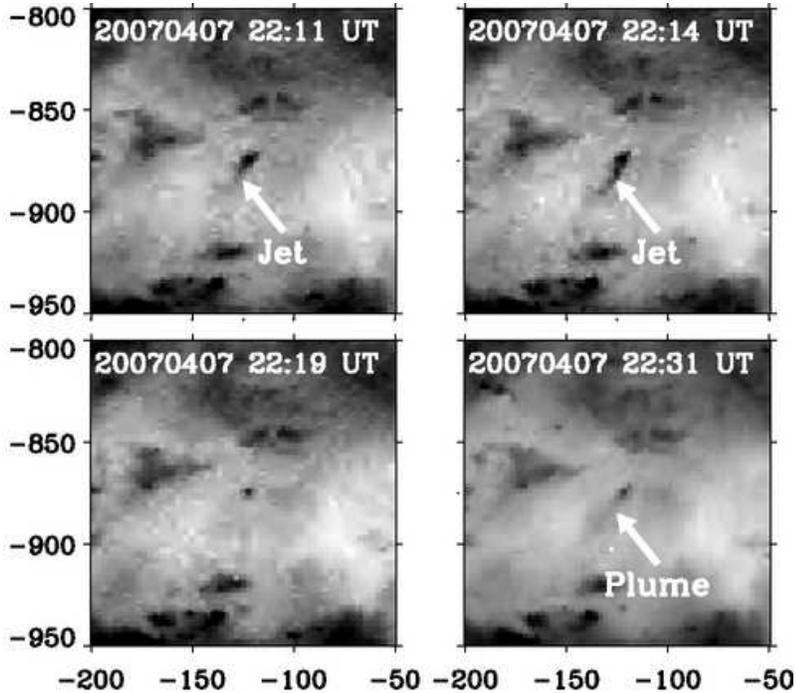} %{RAOUAFI-FG1.EPS}
\end{minipage}\hfill
\parbox[b]{3.2cm}{\caption{\label{EUVI_jet} \small
Observations made by EUVI on STEREO~A in the 17.1~nm passband on 7~April 2007
illustrating a jet seen in the negative prints. Its evolution into a coronal
plume occurred within 15~min. The $x$ and $y$ coordinate
scales are in seconds of arc from
the centre of the solar disk (from Raouafi et al. 2008).}}
\end{minipage}
\end{figure}

From XRT and EUVI observations in
April 2007, Raouafi et al. (2008) found evidence
that jets are precursors of coronal plume formation.
An example of a jet observed by EUVI aboard STEREO~A is recorded in
Fig.~\ref{EUVI_jet} over a time period of 20~min showing in four images
the transformation of a bright jet into plume haze
interpreted as material on newly opened field lines.
The base-heating model for plumes (Wang 1994) was applied to short-lived burst
to generate the jets in this model.
The association of plumes and jets is confirmed by the preliminary
results of the Hinode campaign in summer 2008, where we found
seven jets over a time interval of $\approx 3$~h. They lasted from 4~min to
17~min, and most of them were seen in regions with prominent plumes.
Fig.~\ref{XRT} shows the seven jets indicated as
solid red lines in the XRT Al\_poly image of a CH (see Sect.~\ref{Electron}).

Twelve jets where detected during the 66~h
LASCO-C2 observations in the right panel of Fig.~\ref{dtid}.
These features correspond to
those jets-like structures seen by St\,Cyr et al. (1997)
above the polar regions and analysed by Wang et al. (1998).
The conclusion was that the WL polar jets last for $\approx 2$~h in the C2
FOV, their occurrence frequency is three to four per day with an angular width
(measured with respect to their non-radial axis) of 2\deg~to 4\deg; in
agreement with the parameters deduced from the TID in Fig.~\ref{tidfive}.
The super-radial expansion factor of $\approx 2.2$ reduces the angular width to
1\deg~at the base of the corona (see DeForest et al. 2001b).
Recurrent WL polar jets reported by Wang et al. (1998)
look like transient events distinct from the quiescent emission
of WL polar plumes. The main argument is the specific
angular width and lifetime of these events. Some authors
(e.g., Raouafi et al. 2008) found evidence that jets are precursors of
coronal plume formation.
An TID analysis showed that jets observed by LASCO-C2 in WL agree in duration,
width and occurrence frequency with those seen in the VUV by EIT and UVCS
(Llebaria et al. 2002b).
Nearly all of the jets in WL appear to be associated with
bright plumes (at least in the projection plane of view).

% 6.4
\subsection{\small Fast solar wind and the Heliosphere}
\label{Wind}
The source region of the fast SW streams has to be sought in CHs
(see Sect.~\ref{Magnetic}). It is also firmly established that
funnel-type magnetic
structures in the low corona exist, in which the acceleration of the SW must
occur. It appears as if such structures are the building blocks for both IPRs
and plumes. In the context of this section, the main questions are where
the fast SW is produced and what signatures of the generation process might
be found in the Heliosphere. The discussion in Sect.~\ref{Outflow}
points to IPRs
as the main source region of the fast SW streams considering that the outflow
speeds of IPRs are approximately a factor of two higher than in plumes at a
height of $h \approx 1~R_\odot$. The higher densities in plumes might well be
more than compensated by the low filling factor of plumes in CHs
(cf., Sects.~\ref{Magnetic} and \ref{Densities}).

The strongest plume structures could be traced in the outer corona
beyond 0.1~ua (see Sect.~\ref{Magnetic}). If we were able to identify
plumes in the interplanetary medium, the problem might unambiguously
be solved whether or not they are relevant as contributors to the fast SW.
Hence, can we look at in situ data to ascertain
that plumes maintain their identity far out in the SW?
Fine structures in He$^{2+}$/H$^{+}$ density ratios
and flow speeds were observed in fast SW streams between 0.3~ua and 1.0~ua on
Helios (Thieme et al. 1990), suggesting that the ``ray-like structures''
seen in CHs expand while retaining an overall pressure balance with the
background. These structures\,---\,characterized by an anticorrelation between
gas and magnetic pressure\,---\,could possibly correspond to the interplanetary
manifestation of plumes. The flow tubes were traceable at different
heliocentric distances, because the two Helios probes were radially
aligned at that time, and the same plasma parcel could be recognized in the data
from both spacecraft. The flow tubes were becoming more and more
difficult to identify, as larger heliocentric distances were reached.
Stream interactions might have eliminated the speed
signatures inside 1~ua in the ecliptic plane. At high latitudes
this effect could happen at larger distances, and
SWOOPS might be able to detect such signals.
These structures have come to be known as PBSs. In their modeling attempts,
Velli et al. (1994) pursued the idea that plumes
expand into the SW under the assumption that they are hotter and more dense.
However, as demonstrated in Sects.~\ref{Densities} and \ref{Temperatures},
these assumptions could not be verified by observations.
Following the Helios result a series of papers dealt with
the problem of plume identification in the distant SW, but
controversial results have been obtained, for instance,
by Yamauchi et al. (2002, 2004) on the basis of Ulysses magnetometer data.

A different approach was taken by Reisenfeld et al. (1999) who examined the
correlations between the variations and magnitudes of plasma $\beta$ and
the $\alpha$-to-p ratio within PBSs  in Ulysses high-latitude data.
Since the element composition (cf., Sect.~\ref{Abundances})
is a robust plasma parameter during the SW
expansion, the authors concluded that at least the fraction of PBSs
with relatively high $\beta$ were indeed of solar origin, and
tentatively related them to coronal plumes. The conceptual process foresees an
expansion of the high-pressure plume structures at an altitude where the
low-$\beta$ regime ends. Based on a high $\alpha$-to-p ratio in plumes, a
positive correlation between this ratio and $\beta$ is then expected in,
agreement with the observations. However, the assumption of such a high ratio
is, in all likelihood, not justified.
The most distinct plume/IPR signature is
the low neon-to-magnesium ratio in plumes (see Sect.~\ref{Abundances}).
The FIP of
helium is even higher than that of neon and, consequently, a very low helium
abundance relative to low-FIP elements must be expected in plumes
(cf., also Wang 1996). Although the physics of helium and its ions in the
solar atmosphere is not fully understood, it should be pointed out
that helium is indeed less abundant than neon relative to
low-FIP elements in the SW (Geiss et al. 1995).

In order to explain the Ulysses observations, only a slight
modification of the assumptions is required: plume plasma, although on open
field lines, will not become part of the SW, at least not in great quantities.
Plume flux tubes would then be more or less void of
plasma at large distances from the Sun.
Therefore, the IPRs, the regions with higher plasma pressure,
will expand when $\beta$ increases,
and have, of course, a higher He$^{2+}$ abundance relative to H$^0$.
In Fig. 5 of Reisenfeld and co-workers, the plume and IPR assignments
just have to be reversed, and the lower portions of the newly defined plumes
filled with dense plasma. Figs.~1 and 2 of
Thieme et al. (1990) contain some examples of
PBSs measured near 0.63~ua at times (42.0, 43.4, 44.0)~d
with relatively low $\beta$, $\alpha$-to-p density ratio
and proton temperature as well as relatively
high magnetic fields that might qualify for remnants of a plume structures.

PBS have not been the only structures considered as possible candidates
of plume remnants. Microstreams\,---\,small features identified
in Ulysses data by a typical velocity profile\,---\,have been analysed
by Neugebauer et al. (1995), who concluded that these were
not to be identified with plumes.
The same result was reached by von Steiger et al. (1999), because no
significant depletion of the Ne/Mg abundance and charge-state deviation
in these structures could be detected with respect to the
surrounding fast SW.

Altogether, we may summarize that there is no clear indication for the presence
of plume plasma in the SW
(see, e.g., Poletto et al. 1996; DeForest et al. 2001a). Its
disappearance beyond $30~R_\odot$ must be ascribed to some interactions with
the plasma in the IPRs. It has been suggested, for instance, that
the differential flow speed between plumes and IPRs will lead to a
Kelvin--Helmholtz instability that mixes the plume with IPR plasmas,
removing the speed difference beyond  $\approx 10~R_\odot$
(Hardee and Clark 1995; Parhi et al. 1999). Alternatively, the shear could
have been reduced below these heights, as implied by the decreasing density
contrast illustrated in Fig.~\ref{Figden}, due to some other interaction
(Suess 2000). Plume signatures in temperature, composition and
ionization state may still exist unless there is an actual mixing of
plasma and magnetic field lines\,---\,something otherwise not expected to
occur in the fast SW.

Before concluding this section, a possible link between
plumes and the heliospheric current sheet should be mentioned as
envisaged by Veselovsky and Panassenko (2000). With an analytical model of
the global magnetic configuration in the heliosphere, and
by superimposing a multi-polar geometry onto an equatorial current sheet, they
argue that the shape and
physical conditions of plumes are dictated not only by local conditions, but
also by the global solar and heliospheric scenario.

% 6.5
\subsection{\small Density and magnetic-field fluctuations}

In situ data revealed the presence of interplanetary density and
magnetic-field fluctuations whose power spectra have, over some restricted
range of frequencies, a $1/f$ shape (see, e.g., Matthaeus and Goldstein 1986).
This had long been known; however, photospheric high-latitude magnetic field
spectra from MDI have recently been shown to display the same feature (see
Matthaeus et al. 2007). Bemporad et al. (2008) have analysed coronal
Ly\,$\alpha$ radiance fluctuations (that might be considered as a proxy for
density fluctuations) in UVCS data taken above a PCH at about
2~$R_\odot$ and came to the conclusion that coronal high-latitude
power spectra show the $1/f$ shape over about the same frequency interval where
it was detected in interplanetary data. These results appear to imply that
the 1/f noise persists throughout the solar atmosphere.
The origin of this phenomenon and its interpretation are not clear yet, but
a possible explanation invokes a cascade of scale-invariant reconnection
processes, originating somewhere in the solar atmosphere, which transfer
energy from smaller to larger scales and eventually
show up in the $1/f$ spectrum. Since
plumes can be identified by UVCS in the Ly\,$\alpha$ line
(see, e.g., Kohl et al. 1997), it might be possible to
check whether the $1/f$ spectrum is
ubiquitous over the CH by analysing Ly\,$\alpha$ fluctuations in
plumes, and
disentangle the plume from the IPR spectrum, possibly showing they
are different. Should plumes be the source of SW, or should the SW
originate from both plumes and IPRs, we might expect the
spectrum of plume density fluctuations to be similar to the in situ spectra.

There are problems in implementing such a project,
because we need to follow
a plume for a long enough time to be able to build up its spectrum, but the
LOS contribution from the IPR may mask any effect.
On the other hand, should we find different spectra, we might confirm
the different nature of the plume versus IPR plasmas. This could provide
a further means of checking the persistence of plume structures
in the interplanetary medium, if in situ spectra at different
heliocentric distances would become available, thus adding a further feature
to those adopted so far in this investigation (see Sect.~\ref{Wind}).

On the other hand, if the plumes and observations do not
cooperate in giving enough exposure time to get very low-frequency spectra,
it might be feasible to look at intermediate wave numbers and
frequencies to form a kind of $k$-$\omega$-diagram
(i.e., wave-number-frequency relation) over a modest range of scales and
periods. This could help distinguish whether the space-time structure of
plumes is more like wave with linear propagation,
or some kind of non-linear effect, such as turbulence or cascades.

% 7
\section{\small Classification}
\label{Classification}

Taking all the existing observations into account, it should be possible
to agree on a plume nomenclature and answer the questions:
-- Is there more than one type of plume?
-- How are polar rays and jets related to plumes?
-- Are polar plumes in PCHs comparable to coronal plumes in non-polar CHs?

The terminology related to our phenomenon has changed over the years:
``polar rays'' and ``brush-like plumes'' (van de Hulst 1950b);
``Polarstrahlen'' (Waldmeier 1955); ``polar rays'' (Saito 1965a);
``coronal polar plumes'' (Newkirk and Harvey 1968); ``polar plumes'' (Ahmad and
Withbroe 1977; Wang 1994) and ``polar rays'' (Koutchmy 1994). There can be
little doubt that all names are related to the same phenomenon:
field-aligned plasma density enhancements in the low and extended solar
corona. These enhancements can either be detected via Thomson scattering
of WL by electrons or via VUV emission lines from atoms and ions.
In the range from $R = 1~R_\odot$ to $R = 8~R_\odot$ mainly
considered in this report,
the density enhancements deduced with the help of both observational methods
yield comparable results (see Table~\ref{Tabden} and Fig.~\ref{Figden}). We call
them ``polar coronal plumes''. ``Coronal plumes'' have a wider meaning and
include plumes from non-polar CHs.
In VUV spectroheliograms polar coronal plumes appear as short spikes near
the polar limb. DeForest et al. (2001a)
have
claimed that individual high-altitude WL structures can be traced
to root structures in PCHs.

Although often difficult to detect outside the polar regions
because of the presence of bright foreground and background material,
plumes have been observed in low-latitude CHs (Wang and Muglach 2008).
These low-latitude plumes
appear to be completely analogous to their polar counterparts,
overlying regions of mixed polarity within the predominantly unipolar
CH and having characteristic lifetimes on the order of a day.

\begin{figure}[!t]
% Figure 28
\begin{minipage}[b]{14.2cm}
\begin{minipage}[b]{10cm}
\includegraphics[width=10cm]{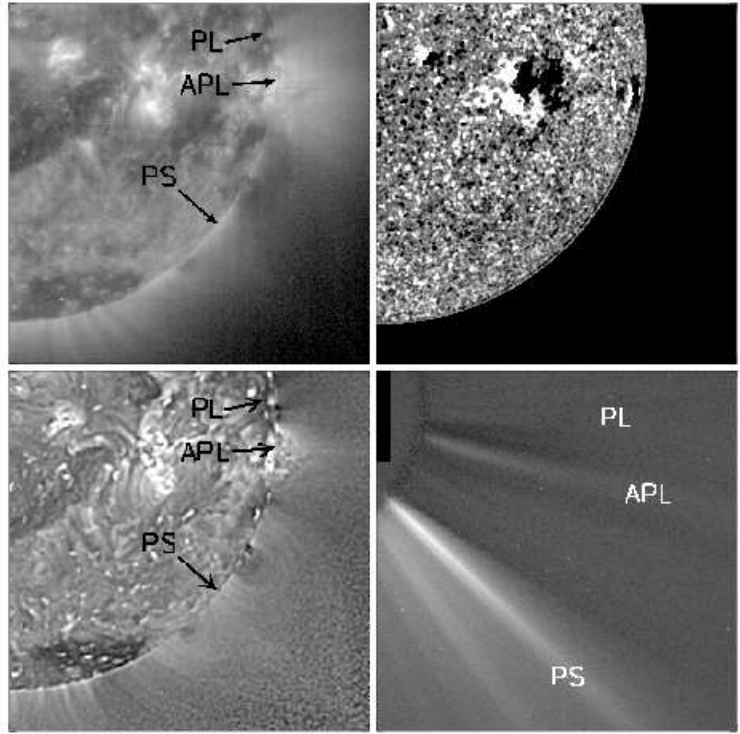} %{WANG_PLLIKE.EPS}
\end{minipage}\hfill
\parbox[b]{3.7cm}{\caption{\label{figAA} \small
Three kinds of plume-like structures observed at the southwest limb
on 6 February 2004: a low-latitude coronal plume (PL),
an active region plume (APL)
and a pseudostreamer (PS).
Top left: EIT Fe\,{\sc ix/x} 17.1~nm image at 19:00~UTC.
Bottom left: EIT Fe\,{\sc xii} 19.5~nm image at 19.13.
Top right: MDI magnetogram at 19:15.
Bottom right: LASCO-C2 image showing the
overlying WL corona beyond $R \approx 2~R_\odot$ at 19:31.
Coronal plumes in the southern PCH can also be seen in the left panels.}}
\end{minipage}
\end{figure}

Diffuse, plume-like structures are also seen above small active regions
that happen to emerge inside CHs.  On even larger spatial scales,
WL streamers that separate CHs of the same polarity
might be regarded as giant plumes; such ``pseudostreamers'' overlie
double arcades rooted in photospheric flux of the opposite polarity and
extend outwards into the heliosphere in the form of plasma sheets
without polarity reversals (Wang et al. 2007a, b). When the axis
of the double arcade is oriented perpendicular to the solar limb,
the pseudostreamer plasma sheet is seen edge-on and appears as a bright,
narrow stalk; when the axis is parallel to the limb, a fan-like structure
is observed.  In contrast, ordinary coronal plumes more often have a
cylindrical geometry, since the underlying minority-polarity flux is
generally concentrated within a very small area.  Examples of the
different kinds of plume-like structures are displayed in Fig.~\ref{figAA}.

% PLEASE DO NOT ROTATE THIS FIGURE!

According to Gabriel et al. (2009), there are at least two different
structures masquerading to give a similar 2D appearance.
The first is a beam plume; a quasi cylindrical
structure overlaying a photospheric bipolar loop or group of loops
(cf., Sect.~\ref{Dimension}).
The second structure has been
called network plume. These represent a very faint enhancement in the corona
(in volume luminosity) overlaying the network boundaries.
Its brightness becomes comparable to beam plumes only due to a large LOS
integration. It is likely that the network plumes are in reality made up of
many beam plumes, on a \emph{much} smaller scale
below the resolution limit of the imagers.
An individual microplume (cf., Sect.~\ref{Morphology}) appears fainter due
to the dilution by the normal coronal medium
within a pixel of the detector.

% 8
\section{\small Plume models and generation processes}
\label{Models}

The most important topic is, of course, how coronal or, in the restricted
sense, polar plumes are generated and maintained in the corona:
-- Plumes are ubiquitous in CHs, but not in quiet-Sun areas.
Can suitable constraints be defined and an agreement on the plume
generation process be reached?
-- Can forward modeling create plume structures in CHs?

\subsection{\small Plume formation and decay}
\begin{figure}[!t]
% Figure 29
\begin{minipage}[b]{14.2cm}
\begin{minipage}[b]{9.5cm}
\includegraphics[width=9.5cm]{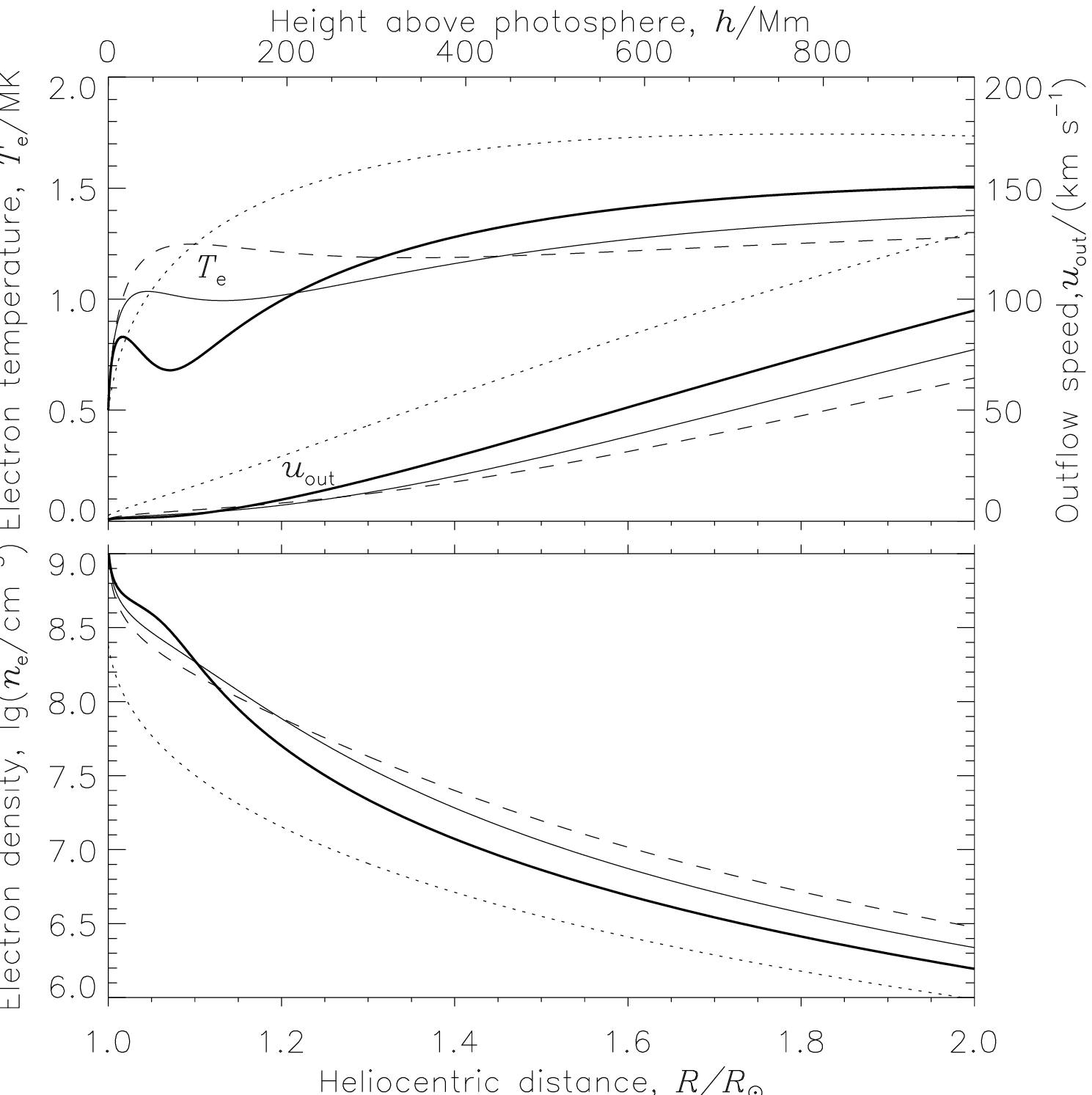} %{AAREV_MODEL.eps}
\end{minipage}\hfill
\parbox[b]{4.2cm}{\caption{\label{figX} \small
Three different ``plume'' SW solutions,
in which additional energy
$F_{\rm b0} = 400$~J\,m$^{-2}$ s$^{-1}$ is deposited
near the coronal base over a damping length $H_{\rm b}\,\ll\,1~R_\odot$.
Thick solid lines: $H_{\rm b} = 0.01~R_\odot$.
Thin solid lines: $H_{\rm b} = 0.02~R_\odot$.  Dashed lines:
$H_{\rm b} = 0.04~R_\odot$.  For comparison, dotted lines represent
the IPR case where $F_{\rm b0} = 0$.  The top panel shows
the variation of the flow speed, $u_{\rm out}$, and the electron
temperature, $T_{\rm e}$, with heliocentric
distance, $R$, while the bottom panel shows the corresponding density
variation, $n_{\rm e}(R)$ (after Wang 1994). At larger distances, the plume
streams are only slightly slower than the IPRs ones.}}
\end{minipage}
\end{figure}

The elevated densities in plumes can be explained if a strong
heating source
is present near their bases (Wang 1994). As plumes are found to overlie
areas of mixed magnetic polarity within the predominantly unipolar CHs,
this extra heating can be attributed to reconnection between the
small bipoles that continually emerge at the photosphere and the
neighbouring unipolar flux concentrations (Wang 1998; Xia et al. 2003;
Gabriel et al. 2009).  Plumes decay after the underlying minority-polarity
flux is cancelled on the $\approx 1$~d time scale of the super-granular
convection (Wang 1998; DeForest et al. 2001b; Wang and Muglach 2008).

The effect of two coronal heating sources,
one spread over a distance on the
order of $1~R_\odot$, the other concentrated very near the coronal base,
is illustrated by the SW solutions in Fig.~\ref{figX} (from Wang 1994).
Here, the single-fluid equations of mass, momentum and energy conservation
were solved, subject to the boundary condition that the downward heat flux
at the coronal base be balanced by the total radiation and enthalpy losses
from the TR.  The heating function was taken to be of the form
\begin{eqnarray}
F_{\rm PL} & = & F_{\rm global} + F_{\rm base}  \nonumber \\
& = & \frac{B}{B_0}\left[F_{\rm g0}\,\exp\left(\frac{R_\odot - R}
{H_{\rm g}}\right) +
F_{\rm b0}\,\exp\left(\frac{R_\odot - R}{H_{\rm b}}\right)\right] \quad ,
\label{heating}
\end{eqnarray}
with $F_{\rm g0} = F_{\rm b0} = 400$~J\,m$^{-2}$\,s$^{-1}$ and
(for simplicity) $B \propto (R_\odot/R)^2$. The three ``plume'' solutions
in Fig.~\ref{figX}
were obtained by setting $H_{\rm b}$, the damping length for the base heating,
equal to (0.01, 0.02, 0.04)~$R_\odot$, and $H_{\rm g} \approx 1~R_\odot$,
for the global heating, i.e., $H_{\rm b} << H_{\rm g}$.
An IPR solution is
also plotted for comparison, where the base heating is set to zero.
Not surprisingly, the strong low-level heating acts to steepen
the temperature gradient near the coronal base, giving rise to a
local temperature maximum and a large increase in the downward heat flux.
As $H_{\rm b}$ is reduced, the temperature ``kink'' moves inwards
and becomes more sharply localized near the footpoint; at the same time,
however, the plume temperature decreases.
Compared with the IPR solution, the plume models have higher footpoint
temperatures but lower temperatures at greater heights, as indeed suggested
by the observations plotted in Fig.~\ref{Figtem}; they are also characterized by
substantially smaller flow velocities in the low corona, in general agreement
with the Doppler measurements of Fig.~\ref{Figout}.
The base heating
has the desired effect of increasing the plume densities by a factor
of $\approx 4$ over the IPR solution.
Even though the velocities are very low near the plume base, the densities
are determined by the mass continuity equation, not by the hydrostatic
equilibrium condition; thus there is no contradiction between the
high densities and low temperatures of the plume plasma.
It may be noted that the
calculated densities in Fig.~\ref{figX} fall off more slowly with height than
the observed densities in Fig.~\ref{Figden}; this discrepancy is easily removed
by adopting a more rapid (and realistic) falloff rate for the magnetic
field.

Fig.~7 of Pinto et al. (2009), shows how a plume forms as the
base heating rate $F_{\rm b0}$ is suddenly raised from 0 to
400~J\,m$^{-2}$ s$^{-1}$.  The change in the heating rate
generates a wavefront that propagates upward along the flux tube at the
local sound speed.  The velocities increase during the first several hours,
but subsequently fall below the initial (IPR) values as the
densities continue to rise and the plasma above the coronal base cools.
The cooling is due to increased radiative losses and to the reduction
in the energy available per particle as the density increases.
The reverse process, in which the base heating is suddenly switched off,
is shown in Fig.~8 of Pinto and co-workers.
The velocities initially decrease, even becoming
slightly negative ($u_0 \approx -1$~km\,s$^{-1}$) near the coronal base;
the equilibrium profile, in which the speeds are everywhere higher than
the plume values, is attained only after $\approx 1$~d. We note that
it takes as long as $\approx 5$~h for the densities to drop by a factor
of two, a result that is consistent with the observed tendency for
EUV ``plume haze'' to linger well after the underlying bright point
has faded.

\subsection{\small Beam and network plumes}

Both beam plumes and the individual filaments of network plumes
(cf., Sect.~\ref{Dimension}) might have similar densities and
temperatures;
and extend to 1.5~$R_\odot$ to 2.0~$R_\odot$ in VUV images
(Gabriel et al. 2009).
They are both due to the interaction and reconnection of
emerging flux loops (with very different scales) with a monopolar
ambient field. Network plumes appear for a maximum of 2~d, due to
solar rotation and the evolution time of the network. The individual
filaments of network plumes might have a much shorter lifetime.
Jets are characterized by a basic monopole-bipole topology similar to
that of plumes, but the energy release is far more impulsive,
perhaps because the reconnection is driven
by the rapid emergence of small bipoles rather than by their slower
decay and dispersal in the supergranular flow field.

\subsection{\small Forward modeling}
\label{Forward}
\begin{figure}[!t]
% Figure 30
\centering
\includegraphics[width=\textwidth]{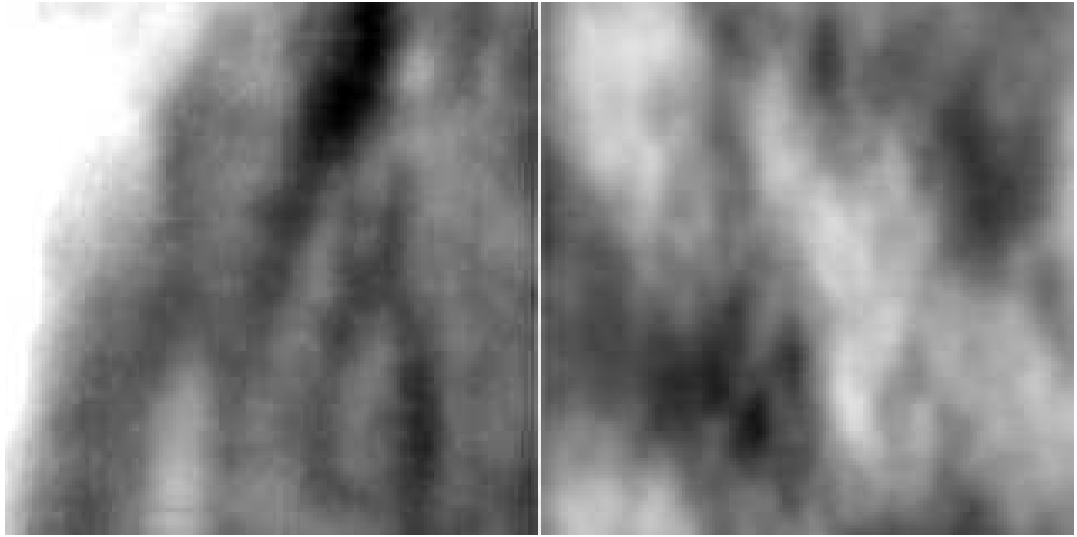} %{COMPARISON.PS}
\caption{\label{COMPAR} \small
Comparison between a zone of a real TID ($128 \times 128$~pixels wide) and a
TID obtained by simulation: (Left panel) enlarged detail of a real TID;
(right panel) enlarged detail of a TID generated from a fractal distribution
($D = 2.9$) of electrons over the polar cap.
The simulation includes 3D geometry,
Thompson scattering and projection onto the plane of view.}
\end{figure}

The ``forward modeling'' approach has been used for many years.
The purpose is to obtain simulated images from controlled
parameters with the same statistical characteristics (or at least
the same visual impression) as images of solar coronal plumes.
Wang and Sheeley (1995) simulated plumes and their
base areas in PCHs, observed in the Mg\,{\sc ix} 38.4~nm line with the
slitless spectrograph S082A on Skylab. They demonstrated that bright plumes
always show intense network features within their base areas, but the converse
does not hold. They also considered how the appearance of UV plumes
might depend on their orientation relative to the LOS.
From theoretical consideration about the formation of a coronal plume, they
deduced a typical time scale $\tau_{\rm p} = 14$~h for the energy transfer
from the network to the plume.

The purpose of simulations in WL is to reproduce their fluffy aspect in TID,
i.e., to mimic in space and time the visual behaviour of plumes,
as, for instance, shown in Fig.~\ref{COMPAR}.
Two different methods were used by Llebaria et al. (2004) to simulate
the image sequences and, from it, the TID.
The first one is based on a fractal affine model and the second on
an evolutive multi-scale model. These models required both the fractal
dimension of 2.9 for the 2D electron density distribution over the
polar cap. More recently,
Boursier and Llebaria (2008) introduced a parametric model based jointly on the
``hidden Markov trees'' and on the 2D wavelet transform, in order to control
both the localization of the electrons and the fractal dimension.
Simulations  of VUV plumes by Barbey et al. (2008)
reproduced the intermittent behaviour of plumes using a dynamic model.

The modeling activities have been successful in the sense that they
produce images with realistic plumes and TID structures and
their temporal evolution, although a convincing statistical
comparison between simulation and reality remains to be done.

% 10
\section{\small Conclusions}
\label{Conclusions}

In line with the definition of work in the introduction, ideally answers should
be given here to all questions asked in the proposal phase
of this study and discussed in
the various sections. However, the situation is not ideal: not all problems
could be solved and, in some cases, not all team members agreed on a proposed
solution. Consequently, three categories will have to be listed\,---\,accepted
results, controversial topics and open questions\,---\,before a concluding
statement can be made at the end of this section.

\begin{enumerate}
\item Agreed results:\\
-- The fore- and background problem of polar coronal plume observations can
be avoided during solar minimum periods,
taking into account the N-S asymmetry caused by the 7.25\deg~angle between
the solar equator and the ecliptic.
Careful consideration of the
actual PCH configuration is required as well.
The observations of non-polar plumes
need special conditions and precautions.\\
-- Polar plumes delineate magnetic field lines of the minimum corona in PCHs.
Plumes
expand super-radially\,---\,very fast below 30~Mm above the photosphere and more
slowly at greater altitudes. IPR funnels basically
show the same behaviour in the low
$\beta$-regime of the corona. \\
-- Plumes observed in WL and VUV results from plasma density enhancements in
CHs. \\
-- Sinogram analyses established a direct relationship between plumes observed
in different wavelength ranges. \\
-- The electron density ratio between plumes and IPRs is between three and
seven in the low corona and decreases at greater heights.\\
-- The electron temperature in plumes is $T_{\rm e} \leq 1$~MK. In IPRs it is
higher by $\approx$\,0.2~MK with a tendency of even higher values at greater
heights. This is compatible with
freeze-in temperatures derived from in situ SW observations.\\
-- The effective ion temperatures in plumes and IPRs are much higher than the
electron temperatures.\\
-- Effective ion temperatures in plumes are lower than in IPRs.\\
-- With stereoscopic observations, 3D reconstructions of more
or less cylindrical plumes\,---\,called here beam plumes\,---\,could
be achieved.\\
-- Beam plumes are, in many cases, related to coronal BPs during the beginning
of their lifetime that last
up to 2~d. They have a recurrence tendency over
much longer time scales.\\
-- Plumes show some evolution during their lifetime. \\
-- Compressional waves are frequently observed in plumes, but these slow
magnetosonic waves do not seem to carry large energy fluxes.\\
-- Non-compressional Alfv\'enic waves are thought to be of importance, at least,
in IPRs, but are difficult to detect.\\
-- Footpoints of beam plumes lie near magnetic flux concentrations
interacting with small magnetic dipoles. The reconnection activity generates
heat near the base of a plume and leads to jets that probably provide some of
the plume plasma.\\
-- The SW outflow velocity is higher in IPRs than in plumes above
a height of $h \approx 0.6~R_\odot$.
The filling factor of plumes in CHs is $\leq 0.1$.
The IPR contribution to the SW is thus more important than that of
plumes.\\
-- Plumes and IPRs have a distinctly different abundance composition, in the
sense that the ratio of low-FIP/high-FIP elements is much larger in plumes
than in IPRs.\\
-- Coronal plumes observed in low-latitude CHs have very similar
properties to plumes in PCHs. \\
-- Rosettes in the chromospheric network could be of importance for the plume
formation.\\
-- The ensemble of plumes appears to rotate rigidly within the CH boundaries.
The lifetime of a plume is too short for a definite answer.

\item Controversial topics:\\
-- Based on EIT observations, a second class of plumes appears to exist. In
contrast to beam plumes, they are composed of small
structures\,---\,microplumes\,---\,with footpoints along network lanes.
They are therefore called
network plumes and only visible when the LOS is directed in their long
dimension.\\
-- It is not excluded that beam plumes are also composed of microplumes in a
more compact fashion.\\
-- The indication that plumes can be described as fractals support the
microplume concept.\\
-- The outflow speed of the SW at low altitudes is slower in IPRs
than in plumes in some studies. Most of the results give, however,
higher speeds in IPRs at all heights.\\
-- Plume/IPR signatures in the fast SW at large distances from the Sun based on
compositional and magnetic variations appear to be present,
but are not sufficiently
pronounced for an unambiguous identification.

\item Open questions: \\
-- There are indications of temperature anisotropies of heavy ions both
in IPRs and in plumes, but the degree of anisotropy in both
regions needs further
investigations. \\
-- Although models of plumes and their formation are available, an exact
description of the
physical processes operating at the base and inside of plumes as well as their
interaction with the SW is still outstanding. \\
-- Is there any contribution of plume plasma to the fast SW streams at
all? \\
-- What produces the clear FIP effect signature between
plumes and IPRs?

\end{enumerate}

It is suggested that most of the plumes considered, as well as the
more transitory jets, arise from a similar mechanism. This is the
emergence of new bipolar loops at the base, interacting by reconnection
with an existing ambient, mainly monopolar field. The different spatial
and temporal scales observed could then be explained by the differing
parameters and reconnection rates. Many indications hint at a fine structure of
plumes that cannot be adequately resolved with present-day instrumentation.

{\bf Acknowledgements}
We dedicate this paper to the memory of Sir William Ian Axford who was very
interested in coronal plumes and their relation to the fast solar wind.
The team members thank the International Space Science Institute
for the opportunity to conduct this work
within the International Study Team programme
and the financial support. During two sessions in Bern, we enjoyed the
hospitality and the excellent working conditions at the institute.
We also want to thank the SOHO, TRACE, Hinode and STEREO teams. Without
their work we could not have conducted this study. We acknowledge
the analysis of the XRT data discussed here by Giulia Schettino and
Alphonse Sterling, the comments on a draft version of the
manuscript by Eckart Marsch and the review of the referee.
GP is grateful for support from the
Italian Space Agency (ASI/I015/07/0).
SI thanks Saku Tsuneta for fruitful discussions about the
magnetic fields in polar regions.

\begin{appendix}

\section{\small List of acronyms and abbreviations}
\begin{small}
AAS -- American Astronomical Society\\
APL -- active region plume\\
AOGS -- Asia Oceania Geoscience\\
AR -- active region\\
BP -- bright point\\
CDS -- Coronal Diagnostic Spectrometer\\
CH -- coronal hole\\
CHIANTI -- Atomic Database for Spectroscopic Diagnostics of Astrophysical
Plasmas\\
DEM -- differential emission measure\\
EKPol -- Liquid crystal polarimeter\\
EM -- emission measure\\
ECH -- equatorial coronal hole\\
EIS -- EUV Imaging Spectrometer\\
EIT -- EUV Imaging Telescope\\
EUV -- extreme UV (10~nm to 120~nm)\\
EUVI -- Extreme UV Imager\\
FIP -- first-ionization potential\\
FIT -- first-ionization time\\
FOV -- field of view\\
GI -- grazing-incidence\\
IPR -- inter-plume region\\
ISSI -- International Space Science Institute\\
LASCO -- Large Angle and Spectrometer Coronagraph\\
LOS -- line of sight\\
MDI -- Michelson Doppler Imager\\
MHD -- magnetohydrodynamic\\
MLSO -- Mauna Loa Solar Observatory\\
NI -- normal-incidence\\
PL -- coronal plume (in tables and diagrams)\\
PCH -- polar coronal hole\\
PBS -- pressure-balanced structure\\
PS -- pseudostreamer\\
QS -- quiet Sun\\
Secchi -- Sun Earth Connection Coronal and Heliospheric Investigation\\
SOHO -- Solar and Heliospheric Observatory\\
SW -- solar wind\\
SWOOPS -- Solar Wind Observations Over the Poles of the Sun\\
SWICS -- Solar Wind Ionization state and Composition Spectrometer\\
SUMER -- Solar UV Measurements of Emitted Radiation spectrometer\\
SP -- spectro-polarimeter\\
SOLIS -- Synoptic Optical Long-term Investigations of the Sun\\
SOT -- Solar Optical Telescope\\
STEREO -- Solar Terrestrial Relations Observatory\\
TID -- time-intensity diagram\\
TR -- transition region\\
TRACE -- Transition Region and Coronal Explorer\\
UV -- ultraviolet (10~nm to 380~nm)\\
UVCS -- UV Coronagraph Spectrometer\\
UCS -- UV Coronal Spectrometer\\
VUV -- vacuum UV (10~nm to 200~nm)\\
VHM/FGM -- Vector Helium and Fluxgate Magnetometers\\
VSM -- Vector Spectro-Magnetograph\\
WL -- white light (380~nm to 760~nm)\\
WLC -- WL Coronagraph\\
XRT -- X-Ray Telescope\\
3D (2.5D, 2D) -- three-(two and a half, two)-dimensional

\end{small}
\end{appendix}

\begin{small}

\end{small}
\end{document}